\documentclass[%
 aip,
 jmp,%
 amsmath,amssymb,
 reprint,%
]{revtex4-1}

\usepackage{graphicx}
\usepackage{dcolumn}
\usepackage{bm}
\usepackage[cp1251]{inputenc}
\usepackage{empheq}
\newcommand{\nc}{\newcommand}
\newcommand{\rnc}{\renewcommand}

\nc{\be}{\begin{equation}} \nc{\ee}{\end{equation}}
\nc{\bse}{\begin{equation*}} \nc{\ese}{\end{equation*}}

\nc{\ba}{\begin{array}} \nc{\ea}{\end{array}}
\nc{\bea}{\begin{eqnarray}} \nc{\eea}{\end{eqnarray}}
\nc{\bi}{\begin{itemize}} \nc{\ei}{\end{itemize}}
\nc{\bn}{\begin{enumerate}} \nc{\en}{\end{enumerate}}
\nc{\bt}{\begin{tabular}} \nc{\et}{\end{tabular}}
\nc{\bb}{\begin{equation}\begin{array}{|c|}\hline } \nc{\eb}{\\
\hline\end{array}\end{equation}}
\nc{\bw}{\begin{widetext}} \nc{\ew}{\end{widetext}}
\nc{\disp}{\displaystyle}

\newcommand{\lb}{\left[}
\newcommand{\rb}{\right]}
\newcommand{\lp}{\left(}
\newcommand{\rp}{\right)}
\newcommand{\lf}{\left\{}
\newcommand{\rf}{\right\}}

\newcommand{\rv}{\right|}
\newcommand{\ld}{\left.}
\newcommand{\rd}{\right.}

\nc{\Def}{\stackrel{def}{=}} \nc{\sign}{\mathrm{sign}\;}
\nc{\diag}{\mathrm{diag}\;}
\nc{\eq}{\equiv} \nc{\we}{\wedge} \nc{\ra}{\rightarrow}
\nc{\bfrac}{\disp\frac}
\nc{\bdel}{\bar{\del}}

\nc{\bfpa}{{\bm{\partial}}}                       
\nc{\pa}[1]{{\partial_{#1}}{}}                    
\nc{\pau}[1]{{\partial^{#1}}{}}                   
\nc{\alp}{\alpha} \nc{\bet}{\beta} \nc{\gam}{\gamma}
\nc{\del}{\delta} \nc{\eps}{\epsilon} \nc{\veps}{\varepsilon}
\nc{\zet}{\zeta}
\nc{\tet}{\theta} \nc{\vtet}{\vartheta} \nc{\iot}{\iota}
\nc{\kap}{\kappa} \nc{\vkap}{\varkappa} \nc{\lam}{\lambda}
\nc{\vpi}{\varpi}
\nc{\vrho}{\varrho} \nc{\sig}{\sigma} \nc{\vsig}{\varsigma}
\nc{\ups}{\upsilon}
\nc{\vphi}{\varphi}
\nc{\ome}{\omega}

\nc{\Gam}{\Gamma} \nc{\Del}{\Delta} \nc{\Tet}{\Theta}
\nc{\Lam}{\Lambda}
\nc{\Sig}{\Sigma} \nc{\Ups}{\Upsilon}
\nc{\Ome}{\Omega}

\nc{\BL}[1]{\mathbf{#1}}

\nc{\bfa}{\BL{a}} \nc{\bfb}{\BL{b}} \nc{\bfc}{\BL{c}}
\nc{\bfd}{\BL{d}} \nc{\bfe}{\BL{e}} \nc{\bff}{\BL{f}}
\nc{\bfg}{\BL{g}} \nc{\bfh}{\BL{h}} \nc{\bfi}{\BL{i}}
\nc{\bfj}{\BL{j}} \nc{\bfk}{\BL{k}} \nc{\bfl}{\BL{l}}
\nc{\bfm}{\BL{m}} \nc{\bfn}{\BL{n}} \nc{\bfo}{\BL{o}}
\nc{\bfp}{\BL{p}} \nc{\bfq}{\BL{q}} \nc{\bfr}{\BL{r}}
\nc{\bfs}{\BL{s}} \nc{\bft}{\BL{t}} \nc{\bfu}{\BL{u}}
\nc{\bfv}{\BL{v}} \nc{\bfw}{\BL{w}} \nc{\bfx}{\BL{x}}
\nc{\bfy}{\BL{y}} \nc{\bfz}{\BL{z}}

\nc{\bfA}{\BL{A}} \nc{\bfB}{\BL{B}} \nc{\bfC}{\BL{C}}
\nc{\bfD}{\BL{D}} \nc{\bfE}{\BL{E}} \nc{\bfF}{\BL{F}}
\nc{\bfG}{\BL{G}} \nc{\bfH}{\BL{H}} \nc{\bfI}{\BL{I}}
\nc{\bfJ}{\BL{J}} \nc{\bfK}{\BL{K}} \nc{\bfL}{\BL{L}}
\nc{\bfM}{\BL{M}} \nc{\bfN}{\BL{N}} \nc{\bfO}{\BL{O}}
\nc{\bfP}{\BL{P}} \nc{\bfQ}{\BL{Q}} \nc{\bfR}{\BL{R}}
\nc{\bfS}{\BL{S}} \nc{\bfT}{\BL{T}} \nc{\bfU}{\BL{U}}
\nc{\bfV}{\BL{V}} \nc{\bfW}{\BL{W}} \nc{\bfX}{\BL{X}}
\nc{\bfY}{\BL{Y}} \nc{\bfZ}{\BL{Z}}

\nc{\bfalp}{\bm{\alp}} \nc{\bfbet}{\bm{\bet}} \nc{\bfgam}{\bm{\gam}}
\nc{\bfdel}{\bm{\del}} \nc{\bfeps}{\bm{\eps}}
\nc{\bfveps}{\bm{\veps}} \nc{\bfzet}{{\bm{\zet}}}
\nc{\bfeta}{\bm{\eta}} \nc{\bftet}{\bm{\tet}}
\nc{\bfvtet}{\bm{\vtet}} \nc{\bfiot}{\bm{\iot}}
\nc{\bfkap}{\bm{\kap}} \nc{\bflam}{\bm{\lam}} \nc{\bfmu}{\bm{\mu}}
\nc{\bfnu}{\bm{\nu}} \nc{\bfxi}{\bm{\xi}} \nc{\bfpi}{\bm{\pi}}
\nc{\bfvpi}{\bm{\vpi}} \nc{\bfrho}{\bm{\rho}}
\nc{\bfvrho}{\bm{\vrho}} \nc{\bfsig}{\bm{\sig}}
\nc{\bfvsig}{\bm{\sig}} \nc{\bftau}{\bm{\tau}}
\nc{\bfups}{\bm{\ups}} \nc{\bfphi}{\bm{\phi}}
\nc{\bfvphi}{\bm{\vphi}} \nc{\bfchi}{\bm{\chi}}
\nc{\bfpsi}{\bm{\psi}} \nc{\bfome}{\bm{\ome}}

\nc{\bfGam}{\bm{\Gam}} \nc{\bfDel}{\bm{\Del}} \nc{\bfTet}{\bm{\Tet}}
\nc{\bfLam}{\bm{\Lam}} \nc{\bfXi}{\bm{\Xi}} \nc{\bfPi}{\bm{\Pi}}
\nc{\bfSig}{\bm{\Sig}} \nc{\bfUps}{\bm{\Ups}} \nc{\bfPhi}{\bm{\Phi}}
\nc{\bfPsi}{\bm{\Psi}} \nc{\bfOme}{\bm{\Ome}}

\DeclareFontFamily{OT1}{rsfs}{}
\DeclareFontShape{OT1}{rsfs}{m}{n}{<5> rsfs5 <7> rsfs7 <10>
rsfs10}{} \DeclareSymbolFont{mathrsfs}{OT1}{rsfs}{m}{n}
\DeclareSymbolFontAlphabet{\mathrsfs}{mathrsfs}

\nc{\rsA}{\mathrsfs{A}} \nc{\rsB}{\mathrsfs{B}}
\nc{\rsC}{\mathrsfs{C}} \nc{\rsD}{\mathrsfs{D}}
\nc{\rsE}{\mathrsfs{E}} \nc{\rsF}{\mathrsfs{F}}
\nc{\rsG}{\mathrsfs{G}} \nc{\rsH}{\mathrsfs{H}}
\nc{\rsI}{\mathrsfs{I}} \nc{\rsJ}{\mathrsfs{J}}
\nc{\rsK}{\mathrsfs{K}} \nc{\rsL}{\mathrsfs{L}}
\nc{\rsM}{\mathrsfs{M}} \nc{\rsN}{\mathrsfs{N}}
\nc{\rsO}{\mathrsfs{O}} \nc{\rsP}{\mathrsfs{P}}
\nc{\rsQ}{\mathrsfs{Q}} \nc{\rsR}{\mathrsfs{R}}
\nc{\rsS}{\mathrsfs{S}} \nc{\rsT}{\mathrsfs{T}}
\nc{\rsU}{\mathrsfs{U}} \nc{\rsV}{\mathrsfs{V}}
\nc{\rsW}{\mathrsfs{W}} \nc{\rsX}{\mathrsfs{X}}
\nc{\rsY}{\mathrsfs{Y}} \nc{\rsZ}{\mathrsfs{Z}}

\nc{\CA}[1]{\mathcal{#1}}
\nc{\caA}{\CA{A}} \nc{\caB}{\CA{B}} \nc{\caC}{\CA{C}}
\nc{\caD}{\CA{D}} \nc{\caE}{\CA{E}} \nc{\caF}{\CA{F}}
\nc{\caG}{\CA{G}} \nc{\caH}{\CA{H}} \nc{\caI}{\CA{I}}
\nc{\caJ}{\CA{J}} \nc{\caK}{\CA{K}} \nc{\caL}{\CA{L}}
\nc{\caM}{\CA{M}} \nc{\caN}{\CA{N}} \nc{\caO}{\CA{O}}
\nc{\caP}{\CA{P}} \nc{\caQ}{\CA{Q}} \nc{\caR}{\CA{R}}
\nc{\caS}{\CA{S}} \nc{\caT}{\CA{T}} \nc{\caU}{\CA{U}}
\nc{\caV}{\CA{V}} \nc{\caW}{\CA{W}} \nc{\caX}{\CA{X}}
\nc{\caY}{\CA{Y}} \nc{\caZ}{\CA{Z}}

\nc{\lag}{\rsL}      
\nc{\lagG}{\lag^G}   
\nc{\lagM}{\lag^M}   

\nc{\kro}[2]{\del^{#1}_{#2}}

\nc{\bfmet}{\bfg}                     
\nc{\met}[2]{g_{#1 #2}}               
\nc{\metu}[2]{g^{#1 #2}}              
\nc{\rmet}{\sqrt{-g}}                 
\nc{\bfcon}{\bfGam}                   
\nc{\con}[3]{\Gam^{#1}{}_{#2 #3}}     
\nc{\cond}[3]{\Gam_{#1,\, #2 #3}}        
\nc{\bftor}{\bfT}                     
\nc{\tor}[3]{T^{#1}{}_{#2 #3}}        
\nc{\tord}[3]{T_{#1,\, #2 #3}}        
\nc{\toru}[3]{T^{#1,\, #2 #3}}        
\nc{\bfsT}{{\stackrel{*}{\mathbf{T}}}{}}
\nc{\bfstor}{\bfsT}                   
\nc{\sT}{{\stackrel{*}{T}}{}}         
\nc{\stor}[3]{\sT^{#1}{}_{#2 #3}}     
\nc{\stord}[3]{\sT_{#1,\, #2 #3}}     
\nc{\storu}[3]{\sT^{#1,\, #2 #3}}     
\nc{\bfctor}{\bfK}                    
\nc{\ctor}[3]{K^{#1}{}_{#2 #3}}       
\nc{\ctord}[3]{K_{#1,\, #2 #3}}       
\nc{\ctoru}[3]{K^{#1,\, #2 #3}}       
\nc{\bfcur}{\bfR}                     
\nc{\cur}[4]{R^{#1}{}_{#2 #3 #4}}     
\nc{\curd}[4]{R_{#1 #2 #3 #4}}        
\nc{\curud}[4]{R^{#1 #2}{}_{#3 #4}}   
\nc{\ric}[2]{R_{#1 #2}}               
\nc{\ricu}[2]{R^{#1 #2}}              
\nc{\bfein}{\bfE}                     
\nc{\ein}[2]{E_{#1 #2}}               
\nc{\einu}[2]{E^{#1 #2}}              
\nc{\bfna}{{\bm{\nabla}}}             
\nc{\na}[1]{{\nabla_{#1}}{}}          
\nc{\nau}[1]{{\nabla^{#1}}{}}         
\nc{\bfsna}{{\stackrel{*}{{\bm{\nabla}}}}{}}         
\nc{\sna}[1]{{\stackrel{*}{\nabla}_{#1}}{}}          
\nc{\snau}[1]{{\stackrel{*}{\nabla}{}^{#1}}{}}       
\nc{\bfdome}{\bfd\bfome\,}                 
\nc{\dome}{dx\,}                           
\nc{\intOme}{\int\limits_{\Ome}}           
\nc{\intb}{\int\limits_{\Sig_1}^{\Sig_2}}  
\nc{\bfdsig}[1]{\bfd\bfsig_{#1}\,}         
\nc{\dsig}[1]{d\sig_{#1}\,}                
\nc{\intdOme}{\oint\limits_{\partial\Ome}} 
\nc{\intSig}{\int\limits_{\Sig}}           
\nc{\intSiga}{\int\limits_{\Sig_1}}        
\nc{\intSigb}{\int\limits_{\Sig_2}}        
\nc{\bfds}[2]{\bfd\bfs_{#1 #2}\,}          
\nc{\ds}[2]{ds_{#1 #2}\,}                  
\nc{\intdSig}{\oint\limits_{\partial\Sig}} 

\nc{\bfdx}[1]{\bfd\bfx^{#1}}          

\nc{\bfgfi}{\bfPhi}                   
\nc{\gfi}[1]{\Phi^{#1}}               
\nc{\gfiA}{\Phi^A}                    
\nc{\gfiB}{\Phi^B}                    
\nc{\bfjfi}{\bfphi}                   
\nc{\jfi}[1]{\phi^{#1}}               
\nc{\jfiA}{\jfi{a}}                    
\nc{\jfiB}{\jfi{b}}                    
\nc{\bfffi}{\bfvphi}                   
\nc{\ffi}[1]{\vphi^{#1}}               
\nc{\ffiA}{\ffi{a}}                    
\nc{\ffiB}{\ffi{b}}                    
\nc{\bfpara}{\bfxi}              
\nc{\para}[1]{\xi^{#1}}          
\nc{\dbfpara}{\del\bfpara}              
\nc{\dpara}[1]{\del\para{#1}}          
\nc{\dbfparK}{\del\bfpara_{\rsK}}              
\nc{\dparK}[1]{\del\para{#1}_{\rsK}}          
\nc{\dx}[1]{\del x^{#1}}
\nc{\dbfLam}{\del\bfLam} \nc{\dLam}[1]{\del\Lam^{#1}}
\nc{\dSig}{\del\Sig}
\nc{\dIdgfi}[1]{\bfrac{\del I}{\del\gfi{#1}}}
\nc{\dIdgfiA}{\dIdgfi{A}}
\nc{\DIDgfi}[1]{\bfrac{\Del I}{\Del\gfi{#1}}}
\nc{\DIDgfiA}{\DIDgfi{A}}
\nc{\DIDjfi}[1]{\bfrac{\Del I}{\Del\jfi{#1}}}
\nc{\DIDjfiA}{\DIDjfi{a}}
\nc{\DsIDjfi}[1]{\bfrac{\Del^* I}{\Del\jfi{#1}}}
\nc{\DsIDjfiA}{\DsIDjfi{a}}
\nc{\DIDnajfi}[2]{\bfrac{\Del I}{\Del(\na{#1}\jfi{#2})}}
\nc{\DIDnajfiA}[1]{\DIDnajfi{#1}{a}}
\nc{\DIDffi}[1]{\bfrac{\Del I}{\Del\ffi{#1}}}
\nc{\DIDffiA}{\DIDffi{a}}
\nc{\DIDnaffi}[2]{\bfrac{\Del I}{\Del(\na{#1}\ffi{#2})}}
\nc{\DIDnaffiA}[1]{\DIDnaffi{#1}{a}}
\nc{\DIDmet}[2]{\bfrac{\Del I}{\Del \met{#1}{#2}}}
\nc{\DIDtor}[3]{\bfrac{\Del I}{\Del \tor{#1}{#2}{#3}}}
\nc{\DsIDtor}[3]{\bfrac{\Del^* I}{\Del \tor{#1}{#2}{#3}}}
\nc{\DIDnator}[4]{\bfrac{\Del I}{\Del(\na{#1}\tor{#2}{#3}{#4})}}
\nc{\DIDcon}[3]{\bfrac{\Del I}{\Del\con{#1}{#2}{#3}}}
\nc{\DIMDmet}[2]{\bfrac{\Del I^M}{\Del \met{#1}{#2}}}
\nc{\DIMDtor}[3]{\bfrac{\Del I^M}{\Del \tor{#1}{#2}{#3}}}
\nc{\DIMDffi}[1]{\bfrac{\Del I^M}{\Del\ffi{#1}}}
\nc{\DIMDffiA}{\DIMDffi{a}}
\nc{\DIGDmet}[2]{\bfrac{\Del I^G}{\Del \met{#1}{#2}}}
\nc{\DIGDtor}[3]{\bfrac{\Del I^G}{\Del \tor{#1}{#2}{#3}}}

\nc{\K}{K}
\nc{\Kg}[2]{\K^{#1}|_{#2}} \nc{\KgA}[1]{\Kg{#1}{A}}
\nc{\KgB}[1]{\Kg{#1}{B}}
\nc{\Kj}[2]{\K^{#1}|_{#2}} \nc{\KjA}[1]{\Kj{#1}{a}}
\nc{\KjB}[1]{\Kj{#1}{b}}
\nc{\sKj}[2]{{}^*\K^{#1}|_{#2}} \nc{\sKjA}[1]{\sKj{#1}{a}}
\nc{\sKjB}[1]{\sKj{#1}{b}}
\nc{\Kf}[2]{\K^{#1}|_{#2}} \nc{\KfA}[1]{\Kf{#1}{a}}
\nc{\KfB}[1]{\Kf{#1}{b}}
\nc{\Km}[3]{\K^{#1}|^{#2 #3}} \nc{\Kt}[4]{\K^{#1}|_{#2}{}^{#3 #4}}
\nc{\sKt}[4]{{}^*\K^{#1}|_{#2}{}^{#3 #4}}
\rnc{\L}{L}
\nc{\Lg}[3]{\L^{#1 #2}|_{#3}} \nc{\LgA}[2]{\Lg{#1}{#2}{A}}
\nc{\LgB}[2]{\Lg{#1}{#2}{B}}
\nc{\Lj}[3]{\L^{#1 #2}|_{#3}} \nc{\LjA}[2]{\Lj{#1}{#2}{a}}
\nc{\LjB}[2]{\Lj{#1}{#2}{b}}
\nc{\Lf}[3]{\L^{#1 #2}|_{#3}} \nc{\LfA}[2]{\Lf{#1}{#2}{a}}
\nc{\LfB}[2]{\Lf{#1}{#2}{b}}
\nc{\Lm}[4]{\L^{#1 #2}|^{#3 #4}} \nc{\Lt}[5]{\L^{#1 #2}|_{#3}{}^{#4
#5}}
\nc{\gfia}[2]{\gfi{}_{#1}|^{#2}} \nc{\gfiaA}[1]{\gfia{#1}{A}}
\nc{\gfib}[3]{\gfi{}_{#1}{}^{#2}|^{#3}}
\nc{\gfibA}[2]{\gfib{#1}{#2}{A}} \nc{\gfic}[4]{\gfi{}_{#1}{}^{#2
#3}|^{#4}} \nc{\gficA}[3]{\gfic{#1}{#2}{#3}{A}}
\nc{\jfia}[2]{\jfi{}_{#1}|^{#2}} \nc{\jfiaA}[1]{\jfia{#1}{a}}
\nc{\jfib}[3]{\jfi{}_{#1}{}^{#2}|^{#3}}
\nc{\jfibA}[2]{\jfib{#1}{#2}{a}} \nc{\jfic}[4]{\jfi{}_{#1}{}^{#2
#3}|^{#4}} \nc{\jficA}[3]{\jfic{#1}{#2}{#3}{a}}
\nc{\ffia}[2]{\ffi{}_{#1}|^{#2}} \nc{\ffiaA}[1]{\ffia{#1}{a}}
\nc{\ffib}[3]{\ffi{}_{#1}{}^{#2}|^{#3}}
\nc{\ffibA}[2]{\ffib{#1}{#2}{a}} \nc{\ffic}[4]{\ffi{}_{#1}{}^{#2
#3}|^{#4}} \nc{\fficA}[3]{\ffic{#1}{#2}{#3}{a}}

\nc{\bfJpara}{\bfJ[\bfpara]} \nc{\Jpara}[1]{J^{#1}[\bfpara]}
\nc{\bfJdpara}{\bfJ[\dbfpara]} \nc{\Jdpara}[1]{J^{#1}[\dbfpara]}
\nc{\bfJdparK}{\bfJ[\dbfparK]} \nc{\JdparK}[1]{J^{#1}[\dbfparK]}

\nc{\QdxiSig}{Q[\dbfpara;\Sig]}
\nc{\U}[2]{U_{#1}{}^{#2}} \nc{\Uu}[2]{U^{#1 #2}} \nc{\Ud}[2]{U_{#1
#2}}
\nc{\M}[3]{M_{#1}{}^{#2 #3}} \nc{\Mu}[3]{M^{#1 #2 #3}}
\nc{\Md}[3]{M_{#1 #2 #3}}
\nc{\N}[4]{N_{#1}{}^{#2 #3 #4}} \nc{\Nu}[4]{N^{#1 #2 #3 #4}}
\nc{\Nud}[4]{N^{#1 #2 #3}{}_{#4}}
\nc{\Ia}[1]{I_{#1}} \nc{\Ib}[2]{I_{#1}{}^{#2}}
\nc{\bfpot}{\bftet}
\nc{\bfpotpara}{\bftet[\bfpara]} \nc{\potpara}[2]{\tet^{#1
#2}[\bfpara]}
\nc{\bfpotdpara}{\bftet[\dbfpara]} \nc{\potdpara}[2]{\tet^{#1
#2}[\dbfpara]}
\nc{\bfpotpdpara}{\bftet'[\dbfpara]} \nc{\potpdpara}[2]{\tet^{'#1
#2}[\dbfpara]}
\nc{\pota}[2]{\tet^{#1 #2}} \nc{\potb}[3]{\tet_{#1}{}^{#2 #3}}
\nc{\potbu}[3]{\tet^{#1 #2 #3}} \nc{\potc}[4]{\tet_{#1}{}^{#2 #3
#4}}
\nc{\ppota}[2]{\tet'^{#1 #2}} \nc{\dpota}[2]{\Del\tet^{#1 #2}}
\nc{\tpotb}[3]{\tilde{\tet}_{#1}{}^{#2 #3}}
\nc{\tpotbu}[3]{\tilde{\tet}^{#1 #2 #3}}
\nc{\tpotc}[4]{\tilde{\tet}_{#1}{}^{#2 #3 #4}}

\nc{\rsJdxi}[1]{\rsJ^{#1}[\dbfpara]}          
\nc{\bfrsJ}{{\bm{\rsJ}}} \nc{\bfrsJxi}{\bfrsJ[\bfpara]}
\nc{\bfrsJdxi}{\bfrsJ[\dbfpara]}
\nc{\bfJsdpara}{\stackrel{sym}{\bfJ}[\dbfpara]}          
\nc{\Jsdpara}[1]{\stackrel{sym}{J}{}^{#1}[\dbfpara]}          
\nc{\bfrsB}{{\bm{\rsB}}} \nc{\bfpotBdpara}{\bfrsB[\dbfpara]}
\nc{\potBdpara}[2]{\rsB^{#1 #2}[\dbfpara]}
\nc{\bfpotsdpara}{\stackrel{sym}{\bftet}[\dbfpara]}
\nc{\potsdpara}[2]{\stackrel{sym}{\tet^{#1 #2}}[\dbfpara]}
\nc{\bfUs}{\stackrel{sym}{\bfU}}
\nc{\Us}[2]{\stackrel{sym}{U}{}_{#1}{}^{#2}}
\nc{\bfJsdparK}{\stackrel{sym}{\bfJ}[\dbfparK]}          
\nc{\JsdparK}[1]{\stackrel{sym}{J^{#1}}[\dbfparK]}          

\nc{\A}[3]{A_{#1}{}^{#2 #3}} \nc{\B}[4]{B_{#1}{}^{#2 #3 #4}}
\nc{\C}[4]{C_{#1}{}^{#2 #3 #4}}
\nc{\youtaa}[2]{\ba{|c|c|}\hline #1 & #2\\ \hline \ea}
\nc{\youtab}[2]{\ba{|c|}\hline 1 \\ \hline 2\\ \hline \ea}
\nc{\youtba}[3]{\ba{|c|c|c|}\hline #1 & #2 & #3\\ \hline \ea}
\nc{\youtbb}[3]{\ba{|c|c|c}\hline #1 & #2\\ \hline #3\\ \cline{1-1}
\ea} \nc{\youtbc}[3]{\ba{|c|}\hline #1 \\ \hline #2\\ \hline #3\\
\hline \ea}
\nc{\yous}[1]{\hat{s}\lp\; #1 \;\rp} \nc{\youa}[1]{\hat{a}\lp\; #1
\;\rp}
\nc{\Na}[4]{a_{#1}{}^{#2 #3 #4}} \nc{\bfNb}{\bfb}
\nc{\Nb}[4]{b_{#1}{}^{#2 #3 #4}} \nc{\bfNc}{\bfc}
\nc{\Nc}[4]{c_{#1}{}^{#2 #3 #4}} \nc{\Nd}[4]{d_{#1}{}^{#2 #3 #4}}
\nc{\krob}[4]{\del^{{#1} {#2}}_{{#3} {#4}}}
\nc{\Dd}[6]{\Del^{#1 #2 #3}_{\underline{#4 #5 #6}}}
\nc{\Du}[6]{\Del^{\overline{#1 #2 #3}}_{#4 #5 #6}}
\nc{\dbfgfi}{\del\bfgfi}                 
\nc{\dgfi}[1]{\del\gfi{#1}}              
\nc{\dgfiA}{\dgfi{A}}                    
\nc{\dbfjfi}{\del\bfjfi}                 
\nc{\djfi}[1]{\del\jfi{#1}}              
\nc{\djfiA}{\djfi{a}}                    
\nc{\djfiB}{\djfi{b}}                    
\nc{\dbfffi}{\del\bfffi}                 
\nc{\dffi}[1]{\del\ffi{#1}}              
\nc{\dffiA}{\dffi{a}}                    
\nc{\dffib}{\dffi{b}}                    
\nc{\dmet}[2]{\del\met{#1}{#2}}          
\nc{\dmetu}[2]{\del\metu{#1}{#2}}        
\nc{\drmet}{\del\rmet}                   
\nc{\dlag}{\del\lag}                     
\nc{\dcon}[3]{\del\con{#1}{#2}{#3}}      
\nc{\dtor}[3]{\del\tor{#1}{#2}{#3}}      
\nc{\dcur}[4]{\del\cur{#1}{#2}{#3}{#4}}  
\nc{\bdbfgfi}{\bar{\del}\bfPhi}              
\nc{\bdgfiA}{\bar{\del}\Phi^A}               
\nc{\Dbrf}[4]{(\Del^{#1}{}_{#2})\ld^{#3}\rv_{#4}}  
\nc{\DbrfAB}[2]{\Dbrf{#1}{#2}{a}{b}}               
\nc{\Dbrfd}[4]{(\Del_{#1 #2})\ld^{#3}\rv_{#4}}  
\nc{\DbrfdAB}[2]{\Dbrfd{#1}{#2}{a}{b}}               
\nc{\Dbrfu}[4]{(\Del^{#1 #2})\ld^{#3}\rv_{#4}}  
\nc{\DbrfuAB}[2]{\Dbrfu{#1}{#2}{a}{b}}               
\nc{\Dbrj}[4]{(\Del^{#1}{}_{#2})\ld^{#3}\rv_{#4}} 
\nc{\DbrjAB}[2]{\Dbrj{#1}{#2}{a}{b}}              
\nc{\Dbrju}[4]{(\Del^{#1#2})\ld^{#3}\rv_{#4}} 
\nc{\DbrjuAB}[2]{\Dbrju{#1}{#2}{a}{b}}              
\nc{\Dbrjd}[4]{(\Del_{#1#2})\ld^{#3}\rv_{#4}} 
\nc{\DbrjdAB}[2]{\Dbrjd{#1}{#2}{a}{b}}              
\nc{\Dbrg}[4]{(\Del^{#1}{}_{#2})\ld^{#3}\rv_{#4}} 
\nc{\DbrgAB}[2]{\Dbrg{#1}{#2}{A}{B}}              
\nc{\Dbrm}[6]{(\Del^{#1}{}_{#2})\ld_{#3 #4}\rv^{#5 #6}} 
\nc{\Dbrt}[8]{(\Del^{#1}{}_{#2})\ld^{#3}{}_{#4 #5}\rv_{#6}{}^{#7 #8}\,} 
\nc{\Dbrtd}[8]{(\Del_{#1 #2})\ld^{#3}{}_{#4 #5}\rv_{#6}{}^{#7 #8}\,} 
\nc{\Dbrtu}[8]{(\Del^{#1 #2})\ld^{#3}{}_{#4 #5}\rv_{#6}{}^{#7 #8}\,} 
\nc{\Dbrc}[9]{(\Del^{#1}{}_{#2})\ld^{#3}{}_{#4 #5 #6}\rv_{#7}{}^{#8 #9}{}} 

\nc{\dparabfgfi}{\del_{\xi}\bfPhi}    
\nc{\dparagfiA}{\del_{\xi}\Phi^A}     
\nc{\pax}{\partial x}
\nc{\ten}{P}                          
\nc{\meta}[3]{g_{#1}|_{#2 #3}}                
\nc{\metb}[4]{g_{#1}{}^{#2}|_{#3 #4}}         
\nc{\tora}[4]{T_{#1}|^{#2}{}_{#3 #4}}         
\nc{\torb}[5]{T_{#1}{}^{#2}|^{#3}{}_{#4 #5}}  
\nc{\bfnator}{\bfna\bfT} \nc{\bfnanator}{\bfna\bfna\bfT}
\nc{\bfnaffi}{\bfna\bfvphi} \nc{\bfnanaffi}{\bfna\bfna\bfvphi}
\nc{\bfnajfi}{\bfna\bfphi} \nc{\bfnanajfi}{\bfna\bfna\bfphi}
\nc{\dslagdmet}[2]{\bfrac{\partial^*\lag}{\partial\met{#1}{#2}}}
\nc{\dlagdcur}[4]{\bfrac{\partial\lag}{\partial\cur{#1}{#2}{#3}{#4}}}
\nc{\dslagdtor}[3]{\bfrac{\partial^*\lag}{\partial\tor{#1}{#2}{#3}}}
\nc{\dlagdnator}[4]{\bfrac{\partial\lag}{\partial(\na{#1}
\tor{#2}{#3}{#4})}}
\nc{\dlagdnanator}[5]{\bfrac{\partial\lag}{\partial(\na{#1}\na{#2}
\tor{#3}{#4}{#5})}}
\nc{\dslagdjfi}[1]{\bfrac{\partial^*\lag}{\partial\jfi{#1}}}
\nc{\dslagdjfiA}{\dslagdjfi{a}}
\nc{\dlagdnajfi}[2]{\bfrac{\partial\lag}{\partial(\na{#1}\jfi{#2})}}
\nc{\dlagdnajfiA}[1]{\dlagdnajfi{#1}{a}}
\nc{\dlagdnanajfi}[3]{\bfrac{\partial\lag}{\partial(\na{#1}\na{#2}\jfi{#3})}}
\nc{\dlagdnanajfiA}[2]{\dlagdnanajfi{#1}{#2}{a}}
\nc{\dslagdffi}[1]{\bfrac{\partial^*\lag}{\partial\ffi{#1}}}
\nc{\dslagdffiA}{\dslagdffi{a}}
\nc{\dlagdnaffi}[2]{\bfrac{\partial\lag}{\partial(\na{#1}\ffi{#2})}}
\nc{\dlagdnaffiA}[1]{\dlagdnaffi{#1}{a}}
\nc{\dlagdnanaffi}[3]{\bfrac{\partial\lag}{\partial(\na{#1}\na{#2}\ffi{#3})}}
\nc{\dlagdnanaffiA}[2]{\dlagdnanaffi{#1}{#2}{a}}
\nc{\G}[4]{G_{#1}{}^{#2 #3 #4}}          
\nc{\Gu}[4]{G^{#1 #2 #3 #4}}             
\nc{\Gd}[4]{G_{#1 #2}{}^{#3 #4}}          
\nc{\bfsem}{\bft} \nc{\sem}[2]{t^{#1}{}_{#2}} \nc{\semu}[2]{t^{#1
#2}} \nc{\semd}[2]{t_{#1 #2}}
\nc{\bfsems}{\stackrel{sym}{\bft}}
\nc{\sems}[2]{\stackrel{sym}{t}{}^{#1}{}_{#2}}
\nc{\semsu}[2]{\stackrel{sym}{t}{}^{#1 #2}}
\nc{\semsd}[2]{\stackrel{sym}{t}{}_{#1 #2}}
\nc{\bfsemm}{\stackrel{met}{\bft}}
\nc{\semm}[2]{\stackrel{met}{t}{}^{#1}{}_{#2}}
\nc{\semmu}[2]{\stackrel{met}{t}{}^{#1 #2}}
\nc{\semmd}[2]{\stackrel{met}{t}{}_{#1 #2}}
\nc{\bfsema}{\stackrel{add}{\bft}}
\nc{\sema}[2]{\stackrel{add}{t}{}^{#1}{}_{#2}}
\nc{\semau}[2]{\stackrel{add}{t}{}^{#1 #2}}
\nc{\semad}[2]{\stackrel{add}{t}{}_{#1 #2}}
\nc{\bfsemi}{\stackrel{mod}{\bft}}
\nc{\semi}[2]{\stackrel{mod}{t}{}^{#1}{}_{#2}}
\nc{\semiu}[2]{\stackrel{mod}{t}{}^{#1 #2}}
\nc{\semid}[2]{\stackrel{mod}{t}{}_{#1 #2}}
\nc{\bfsemM}{\bfT} \nc{\semM}[2]{T^{#1}{}_{#2}} \nc{\semMu}[2]{T^{#1
#2}} \nc{\semMd}[2]{T_{#1 #2}}
\nc{\bfsemMs}{\stackrel{sym}{\bfT}}
\nc{\semMs}[2]{\stackrel{sym}{T}{}^{#1}{}_{#2}}
\nc{\semMsu}[2]{\stackrel{sym}{T}{}^{#1 #2}}
\nc{\semMsd}[2]{\stackrel{sym}{T}{}_{#1 #2}}
\nc{\bfsemMm}{\stackrel{met}{\bfT}}
\nc{\semMm}[2]{\stackrel{met}{T}{}^{#1}{}_{#2}}
\nc{\semMmu}[2]{\stackrel{met}{T}{}^{#1 #2}}
\nc{\semMmd}[2]{\stackrel{met}{T}{}_{#1 #2}}
\nc{\bfsemMa}{\stackrel{add}{\bfT}}
\nc{\semMa}[2]{\stackrel{add}{T}{}^{#1}{}_{#2}}
\nc{\semMau}[2]{\stackrel{add}{T}{}^{#1 #2}}
\nc{\semMad}[2]{\stackrel{add}{T}{}_{#1 #2}}
\nc{\bfsemMi}{\stackrel{mod}{\bfT}}
\nc{\semMi}[2]{\stackrel{mod}{T}{}^{#1}{}_{#2}}
\nc{\semMiu}[2]{\stackrel{mod}{T}{}^{#1 #2}}
\nc{\semMid}[2]{\stackrel{mod}{T}{}_{#1 #2}}
\nc{\bfspi}{\bfs} \nc{\spi}[3]{s^{#1}{}_{#2 #3}}
\nc{\spiu}[3]{s^{#1,\, #2 #3}} \nc{\spid}[3]{s_{#1,\, #2 #3}}
\nc{\spiud}[3]{s^{#1, #2}{}_{#3}}
\nc{\fj}[3]{{}^{(\jfi{})}f^{#1}{}_{#2 #3}}
\nc{\fju}[3]{{}^{(\jfi{})}f^{#1,\, #2 #3}}
\nc{\fjd}[3]{{}^{(\jfi{})}f_{#1,\, #2 #3}}
\nc{\fjud}[3]{{}^{(\jfi{})}f^{#1,\, #2}{}_{#3}}
\nc{\bfspia}{\stackrel{add}{\bfs}}
\nc{\spia}[3]{\stackrel{add}{s}{}^{#1}{}_{#2 #3}}
\nc{\spiau}[3]{\stackrel{add}{s}{}^{#1,\, #2 #3}}
\nc{\spiad}[3]{\stackrel{add}{s}{}_{#1,\, #2 #3}}
\nc{\spiaud}[3]{\stackrel{add}{s}{}^{#1, #2}{}_{#3}}
\nc{\bfspii}{\stackrel{mod}{\bfs}}
\nc{\spii}[3]{\stackrel{mod}{s}{}^{#1}{}_{#2 #3}}
\nc{\spiiu}[3]{\stackrel{mod}{s}{}^{#1,\, #2 #3}}
\nc{\spiid}[3]{\stackrel{mod}{s}{}_{#1,\, #2 #3}}
\nc{\spiiud}[3]{\stackrel{mod}{s}{}^{#1, #2}{}_{#3}}
\nc{\bfspiM}{\bfS} \nc{\spiM}[3]{S^{#1}{}_{#2 #3}}
\nc{\spiMu}[3]{S^{#1,\, #2 #3}} \nc{\spiMd}[3]{S_{#1,\, #2 #3}}
\nc{\spiMud}[3]{S^{#1, #2}{}_{#3}}
\nc{\bfspiMa}{\stackrel{add}{\bfS}}
\nc{\spiMa}[3]{\stackrel{add}{S}{}^{#1}{}_{#2 #3}}
\nc{\spiMau}[3]{\stackrel{add}{S}{}^{#1,\, #2 #3}}
\nc{\spiMad}[3]{\stackrel{add}{S}{}_{#1,\, #2 #3}}
\nc{\spiMaud}[3]{\stackrel{add}{S}{}^{#1, #2}{}_{#3}}
\nc{\bfspiMi}{\stackrel{mod}{\bfS}}
\nc{\spiMi}[3]{\stackrel{mod}{S}{}^{#1}{}_{#2 #3}}
\nc{\spiMiu}[3]{\stackrel{mod}{S}{}^{#1,\, #2 #3}}
\nc{\spiMid}[3]{\stackrel{mod}{S}{}_{#1,\, #2 #3}}
\nc{\spiMiud}[3]{\stackrel{mod}{S}{}^{#1, #2}{}_{#3}}
\nc{\bfbel}{\bfb} \nc{\bel}[3]{b^{#1}{}_{#2 #3}} \nc{\belu}[3]{b^{#1
#2 #3}} \nc{\beld}[3]{b_{#1 #2 #3}} \nc{\belud}[3]{b^{#1 #2}{}_{#3}}
\nc{\bfbela}{\stackrel{add}{\bfbel}}
\nc{\bela}[3]{\stackrel{add}{b}{}^{#1}{}_{#2 #3}}
\nc{\belau}[3]{\stackrel{add}{b}{}^{#1 #2 #3}}
\nc{\belad}[3]{\stackrel{add}{b}{}_{#1 #2 #3}}
\nc{\belaud}[3]{\stackrel{add}{b}{}^{#1 #2}{}_{#3}}
\nc{\bfbeli}{\stackrel{mod}{\bfbel}}
\nc{\beli}[3]{\stackrel{mod}{b}{}^{#1}{}_{#2 #3}}
\nc{\beliu}[3]{\stackrel{mod}{b}{}^{#1 #2 #3}}
\nc{\belid}[3]{\stackrel{mod}{b}{}_{#1 #2 #3}}
\nc{\beliud}[3]{\stackrel{mod}{b}{}^{#1 #2}{}_{#3}}
\nc{\bfbelM}{\bfB} \nc{\belM}[3]{B^{#1}{}_{#2 #3}}
\nc{\belMu}[3]{B^{#1 #2 #3}} \nc{\belMd}[3]{B_{#1 #2 #3}}
\nc{\belMud}[3]{B^{#1 #2}{}_{#3}}
\nc{\bfbelMa}{\stackrel{add}{\bfbelM}}
\nc{\belMa}[3]{\stackrel{add}{B}{}^{#1}{}_{#2 #3}}
\nc{\belMau}[3]{\stackrel{add}{B}{}^{#1 #2 #3}}
\nc{\belMad}[3]{\stackrel{add}{B}{}_{#1 #2 #3}}
\nc{\belMaud}[3]{\stackrel{add}{B}{}^{#1 #2}{}_{#3}}
\nc{\bfbelMi}{\stackrel{mod}{\bfbel}}
\nc{\belMi}[3]{\stackrel{mod}{B}{}^{#1}{}_{#2 #3}}
\nc{\belMiu}[3]{\stackrel{mod}{B}{}^{#1 #2 #3}}
\nc{\belMid}[3]{\stackrel{mod}{B}{}_{#1 #2 #3}}
\nc{\belMiud}[3]{\stackrel{mod}{B}{}^{#1 #2}{}_{#3}}
\nc{\ogG}{\rv_{\lag = \lagG}} \nc{\ogM}{\rv_{\lag = \lagM}}

\nc{\bfCar}{\rsC} \nc{\Car}[3]{\rsC^{#1}{}_{#2 #3}}
\nc{\Caru}[3]{\rsC^{#1 #2 #3}} \nc{\Carud}[3]{\rsC^{#1 #2}{}_{#3}}
\nc{\bfEin}{\rsE} \nc{\Ein}[2]{\rsE^{#1}{}_{#2}}
\nc{\Einu}[2]{\rsE^{#1 #2}} \nc{\Eind}[2]{\rsE_{#1 #2}}

\nc{\bfpaffi}{\bfpa\bfffi} \nc{\bfpapaffi}{\bfpa\bfpa\bfffi}

\nc{\dIdpaffi}[2]{\bfrac{\del I}{\del(\pa{#1}\ffi{#2})}}
\nc{\dIdpaffiA}[1]{\dIdpaffi{#1}{a}}
\nc{\dlagdpaffi}[2]{\bfrac{\partial\lag}{\partial(\pa{#1}\ffi{#2})}}
\nc{\dlagdpaffiA}[1]{\dlagdpaffi{#1}{a}}
\nc{\dlagdpapaffi}[3]{\bfrac{\partial\lag}{\partial(\pa{#1}\pa{#2}\ffi{#3})}}
\nc{\dlagdpapaffiA}[2]{\dlagdpapaffi{#1}{#2}{a}}
\nc{\dIdmet}[2]{\bfrac{\del I}{\del \met{#1}{#2}}}

\nc{\bfMin}{\bfeta}                      
\nc{\Min}[2]{\eta_{#1 #2}}               
\nc{\Minu}[2]{\eta^{#1 #2}}              

\nc{\bftmet}{\tilde{\bfg}}                      
\nc{\tmet}[2]{\tilde{g}{}_{#1 #2}}               
\nc{\tmetu}[2]{\tilde{g}{}^{#1 #2}}              

\nc{\bftna}{\tilde{{\bfna}}}                     
\nc{\tna}[1]{{\tilde{\nabla}{}_{#1}}{}}          
\nc{\tnau}[1]{{\tilde{\nabla}{}^{#1}}{}}         

\nc{\DIDtnaffi}[2]{\bfrac{\Del I}{\Del(\tna{#1}\ffi{#2})}}
\nc{\DIDtnaffiA}[1]{\DIDtnaffi{#1}{a}}
\nc{\dlagdtnaffi}[2]{\bfrac{\partial\lag}{\partial(\tna{#1}\ffi{#2})}}
\nc{\dlagdtnaffiA}[1]{\dlagdtnaffi{#1}{a}}
\nc{\dlagdtnatnaffi}[3]{\bfrac{\partial\lag}{\partial(\tna{#1}\tna{#2}\ffi{#3})}}
\nc{\dlagdtnatnaffiA}[2]{\dlagdtnatnaffi{#1}{#2}{a}}

\nc{\bftcur}{\tilde{\bfR}}                     
\nc{\tcur}[4]{\tilde{R}{}^{#1}{}_{#2 #3 #4}}     
\nc{\tcurd}[4]{\tilde{R}{}_{#1 #2 #3 #4}}        
\nc{\tcurud}[4]{\tilde{R}{}^{#1 #2}{}_{#3 #4}}   
\nc{\bfttor}{\tilde{\bfT}}                     
\nc{\ttor}[3]{\tilde{T}{}^{#1}{}_{#2 #3}}        
\nc{\ttord}[3]{\tilde{T}{}_{#1,\, #2 #3}}        
\nc{\ttoru}[3]{\tilde{T}{}^{#1,\, #2 #3}}        

\nc{\bffsem}{{}^{(\vphi)}\bft}
\nc{\fsem}[2]{{}^{(\vphi)}t^{#1}{}_{#2}}
\nc{\fsemu}[2]{{}^{(\vphi)}t^{#1 #2}}
\nc{\fsemd}[2]{{}^{(\vphi)}t_{#1 #2}}
\nc{\bfmsem}{{}^{(R)}\bft} \nc{\msem}[2]{{}^{(R)}t^{#1}{}_{#2}}
\nc{\msemu}[2]{{}^{R)}t^{#1 #2}} \nc{\msemd}[2]{{}^{(R)}t_{#1 #2}}
\nc{\bftsem}{{}^{(T)}\bft} \nc{\tsem}[2]{{}^{(T)}t^{#1}{}_{#2}}
\nc{\tsemu}[2]{{}^{T)}t^{#1 #2}} \nc{\tsemd}[2]{{}^{(T)}t_{#1 #2}}
\nc{\bfjsem}{{}^{(\phi)}\bft}
\nc{\jsem}[2]{{}^{(\phi)}t^{#1}{}_{#2}}
\nc{\jsemu}[2]{{}^{(\phi)}t^{#1 #2}} \nc{\jsemd}[2]{{}^{(\phi)}t_{#1
#2}}
\nc{\bffspi}{{}^{(\vphi)}\bfs}
\nc{\fspi}[3]{{}^{(\vphi)}s^{#1}{}_{#2 #3}}
\nc{\fspiu}[3]{{}^{(\vphi)}s^{#1,\, #2 #3}}
\nc{\fspid}[3]{{}^{(\vphi)}s_{#1,\, #2 #3}}
\nc{\fspiud}[3]{{}^{(\vphi)}s^{#1, #2}{}_{#3}}
\nc{\bfmspi}{{}^{(R)}\bfs} \nc{\mspi}[3]{{}^{(R)}s^{#1}{}_{#2 #3}}
\nc{\mspiu}[3]{{}^{(R)}s^{#1,\, #2 #3}}
\nc{\mspid}[3]{{}^{(R)}s_{#1,\, #2 #3}}
\nc{\mspiud}[3]{{}^{(R)}s^{#1, #2}{}_{#3}}
\nc{\bftspi}{{}^{(T)}\bfs} \nc{\tspi}[3]{{}^{(T)}s^{#1}{}_{#2 #3}}
\nc{\tspiu}[3]{{}^{(T)}s^{#1,\, #2 #3}}
\nc{\tspid}[3]{{}^{(T)}s_{#1,\, #2 #3}}
\nc{\tspiud}[3]{{}^{(T)}s^{#1, #2}{}_{#3}}
\nc{\bfjspi}{{}^{(\phi)}\bfs} \nc{\jspi}[3]{{}^{(\phi)}s^{#1}{}_{#2
#3}} \nc{\jspiu}[3]{{}^{(\phi)}s^{#1,\, #2 #3}}
\nc{\jspid}[3]{{}^{(\phi)}s_{#1,\, #2 #3}}
\nc{\jspiud}[3]{{}^{(\phi)}s^{#1, #2}{}_{#3}}
\nc{\bffbel}{{}^{(\vphi)}\bfb}
\nc{\fbel}[3]{{}^{(\vphi)}b^{#1}{}_{#2 #3}}
\nc{\fbelu}[3]{{}^{(\vphi)}b^{#1 #2 #3}}
\nc{\fbeld}[3]{{}^{(\vphi)}b_{#1 #2 #3}}
\nc{\fbelud}[3]{{}^{(\vphi)}b^{#1 #2}{}_{#3}}
\nc{\bfmbel}{{}^{(R)}\bfb} \nc{\mbel}[3]{{}^{(R)}b^{#1}{}_{#2 #3}}
\nc{\mbelu}[3]{{}^{(R)}b^{#1 #2 #3}} \nc{\mbeld}[3]{{}^{(R)}b_{#1 #2
#3}} \nc{\mbelud}[3]{{}^{(R)}b^{#1 #2}{}_{#3}}
\nc{\bftbel}{{}^{(T)}\bfb} \nc{\tbel}[3]{{}^{(T)}b^{#1}{}_{#2 #3}}
\nc{\tbelu}[3]{{}^{(T)}b^{#1 #2 #3}} \nc{\tbeld}[3]{{}^{(T)}b_{#1 #2
#3}} \nc{\tbelud}[3]{{}^{(T)}b^{#1 #2}{}_{#3}}
\nc{\bfjbel}{{}^{(\phi)}\bfb} \nc{\jbel}[3]{{}^{(\phi)}b^{#1}{}_{#2
#3}} \nc{\jbelu}[3]{{}^{(\phi)}b^{#1 #2 #3}}
\nc{\jbeld}[3]{{}^{(\phi)}b_{#1 #2 #3}}
\nc{\jbelud}[3]{{}^{(\phi)}b^{#1 #2}{}_{#3}}

\begin{document}


\preprint{AIP/123-QED}

\title[Differential Identities and Conservation Laws in Metric-Torsion Theories of Gravitation. II.]
{Covariant Differential Identities and Conservation Laws in
Metric-Torsion Theories of Gravitation. II. Manifestly Generally
Covariant Theories}

\thanks{Submitted to Journal of Mathematical Physics.}

\author{Robert R. Lompay}
\affiliation{Department of Physics, Uzhgorod National University,
Voloshyna str., 54, Uzhgorod, 88000, UKRAINE}
\email{rlompay@gmail.com}

\author{Alexander N. Petrov}
\affiliation{Moscow M. V. Lomonosov State University, Sternberg
Astronomical institute, Universitetskii pr., 13, Moscow, 119992,
RUSSIA} \email{alex.petrov55@gmail.com}
\date{\today}


\begin{abstract}
The present paper continues the work of the authors [J. Math. Phys.
{\bf 54},  062504 (2013)] where manifestly covariant differential
identities and conserved quantities in generally covariant metric-torsion
theories of gravity of the most general type have been
constructed. Here, we study these theories presented more
concretely, setting that their Lagrangians $\lag$ are
\emph{manifestly} generally covariant scalars: algebraic functions
of contractions of tensor functions and their covariant derivatives.
It is assumed that Lagrangians depend on metric tensor $\bfmet$,
curvature tensor $\bfcur$, torsion tensor $\bftor$ and its first
$\bfnator$ and second $\bfnanator$ covariant derivatives, besides,
on an arbitrary set of other tensor (matter) fields $\bfffi$ and
their first $\bfnaffi$ and second $\bfnanaffi$ covariant
derivatives: $\lag = \lag (\bfmet,\bfcur;
\;\bftor,\bfnator,\bfnanator; \;\bfffi,\bfnaffi,\bfnanaffi)$. Thus,
both the standard minimal coupling with the Riemann-Cartan geometry
and non-minimal coupling with the curvature and torsion tensors are
considered.

The studies and results are as follow. (a) A physical interpretation
of the Noether and Klein identities is examined. It was found that
they are the basis for constructing equations of balance of
energy-momentum tensors of various types (canonical, metrical and
Belinfante symmetrized). The equations of balance are presented. (b)
Using the generalized equations of balance, new (generalized)
manifestly generally covariant expressions for canonical
energy-momentum and spin tensors of the matter fields are
constructed. In the cases, when the matter Lagrangian contains both
the higher derivatives and non-minimal coupling with curvature and
torsion, such generalizations are non-trivial. (c) The Belinfante
procedure is generalized for an arbitrary Riemann-Cartan space. (d)
A more convenient in applications generalized expression for the
canonical superpotential is obtained. (e) A total system of
equations for the gravitational fields and matter sources are
presented in the form more naturally generalizing the
Einstein-Cartan equations with matter. This result, being a one of
more important results itself, is to be also a basis for
constructing physically sensible conservation laws and their
applications.
\end{abstract}

\maketitle

%
\section{Introduction}\label{sec_02_00-00}

The present work is the second one of the series of works related to
constructing manifestly covariant differential identities and
conserved quantities, and their study in generally covariant metric-torsion
theories of gravity. In the first work of the series
\cite{Lompay_Petrov_2013_a} (we will call it as the Paper~I), in an
arbitrary Riemann-Cartan space $\caC(1,D)$ the next manifestly
covariant expressions and relations have been obtained: (a) the
generalized Noether current $\bfJdpara$; (b) the system of
differential Klein and Noether identities; (c) the generalized
superpotential $\bfpotdpara$, with the use of which the generalized
Noether current is presented; (d) the generalized symmetrized
Noether current $\bfJsdpara$ as a result of an application of the
generalized Belinfante procedure to the generalized Noether current.

For the sake of a definiteness, let us repeat the definitions, which
we use. A theory is called as \emph{generally covariant} if it is
invariant with respect to general diffeomorphisms. At the same time,
a form of its presentation can be arbitrary. Because \emph{gauge
covariant} theories that are invariant with respect to internal gauge
transformations are very similar to the generally covariant ones
(have the same structure of currents, etc), for the sake of an
universality we call theories of both these types as a
\emph{gauge-invariant theories}. On the other hand, the usual gauge
theories with an internal gauge group we call separately as the
gauge theories of Utiyama-Yang-Mills type.

Thus, in the Paper~I, the quantities and relations in the theories
of the most general type have been constructed. In the present work,
we concretize them. We apply the developed formalism for the study
of manifestly generally covariant theories, which are a more
interesting and important example of diffeomorphically invariant
theories. We call a theory as \emph{manifestly generally covariant}
if its Lagrangian $\lag$ is a generally covariant scalar constructed
as algebraic scalar function of manifestly covariant objects that
are transformed following the linear \emph{homogeneous}
representations of the diffeomorphism group. This means that $\lag$
is an algebraic function of scalar contractions of tensor (and/or
spinor) field functions and their covariant derivatives; besides
manifest dependence on field variables, $\lag$ can also depend on
curvature and torsion tensors independently. It seems that almost
all the physically interesting theories are manifestly generally
covariant or can be presented in such a form. Exceptions are, e.g.,
topological theories of the Chern-Simons type (see  reviews
\cite{Zanelli_2012, Zanelli_2011, Zanelli_2008_b, Marino_2005,
Kaul_Govindarajan_Ramadevi_2005, Dunne_1999,
Jackiw_Grieshammer_Schnetz_Fischer_Simburger_1997, Jackiw_1990,
Jackiw_1983} and references therein), Lagrangians of which are
presented by a secondary characteristic class of a topological
invariant (Chern-Simons form). Such Lagrangians explicitly contain
connections that are transformed following a linear
\emph{non-homogeneous} representation. At the same time, under gauge
transformations the Lagrangians themselves change to a total
divergence. In the Chern-Simons theories, conserved quantities were
constructed in the works \cite{Banados_1995,
Borowiec_Ferraris_Francaviglia_1998,
Borowiec_Ferraris_Francaviglia_2003,
Allemandi_Francaviglia_Raiteri_2003,
Borowiec_Ferraris_Francaviglia_Palese_2003, Sardanashvily_2003_a,
Sardanashvily_2003_b, Aros_2006}. Notice also that Lagrangians in
such theories can be presented in the exactly gauge invariant form
by expanding the Chern-Simons form to the transgression form
\cite{Banados_Mendez_1998, Borowiec_Ferraris_Francaviglia_2003,
Mora_2005, Mora_Olea_Troncoso_Zanelli_2004,
Mora_Olea_Troncoso_Zanelli_2006, Izaurieta_Rodriguez_Salgado_2005,
Izaurieta_Rodriguez_Salgado_2006, Aros_2006,
Borowiec_Fatibene_Ferraris_Francaviglia_2006, Zanelli_2007,
Mora_2007}.

Earlier, manifestly generally covariant theories both in Riemannian
spacetime (see, for example, Refs.~\cite{Belinfante_1939, Belinfante_1940,
Rosenfeld_1940, Rosenfeld_1940_en, Trautman_1962_b, Trautman_1965,
Trautman_1966_en, Logunov_Folomeshkin_1977_a_en,
Logunov_Folomeshkin_1977_b_en, Logunov_Folomeshkin_1977_c_en,
Logunov_Folomeshkin_1977_d_en, Szabados_1991, Szabados_1992,
Obukhov_Rubilar_2006_b, Deruelle_Katz_Ogushi_2004,
Katz_Livshits_2008, Petrov_2004_b_en, Petrov_2008, Petrov_2009_a,
Petrov_2010_a, Petrov_2011, Petrov_Lompay_2013,
Baykal_Delice_2011}), and in the Riemann-Cartan space (see, for
example,  Refs.~\cite{Trautman_1972_a, Trautman_1972_b, Trautman_1973_b,
Trautman_1980, Hehl_1973, Hehl_1974, Hehl_Heyde_Kerlick_Nester_1976}) were studied
already. In particular, in the works by Trautman
\cite{Trautman_1972_a, Trautman_1972_b, Trautman_1973_b,
Trautman_1980} and by Hehl at al
\cite{Hehl_1973, Hehl_1974, Hehl_Heyde_Kerlick_Nester_1976} the simplest theory of gravity
with torsion, the Einstein-Cartan theory (ECT), with matter,
presented by the Lagrangian $\lag = \lagG + \lagM = -\frac1{2k} R+
\lagM(\bfmet; \; \bfffi, \bfnaffi)$ is examined. Their main results
are: (a) clarification of the role of the \emph{canonical}
energy-momentum tensor (EMT) of matter as a source of the metric
field; (b) determination of the connection between the variation
derivative $\Del I^M/\Del\bftor$ with respect to the torsion field
$\bftor$ and the Belinfante tensor $\bfbelM$, induced by the spin
tensor (ST) $\bfspiM$ in the matter sector of the system related to
the action functional $I^M$; (c) clarification of the role of the
\emph{canonical} ST of matter as a source of the torsion field; (d)
in the Riemann-Cartan space, construction of the \emph{universal}
balance equation for the canonical EMT of matter.

In the present work, we consider significantly more general
manifestly generally covariant theories, the total Lagrangians of
which  $\lag = \lagG + \lagM$ have the form: $\lag = \lag
(\bfmet,\bfcur; \;\bftor,\bfnator,\bfnanator;
\;\bfffi,\bfnaffi,\bfnanaffi)$ that, besides of the second
derivatives of matter fields $\bfffi$, includes a dependence on
non-minimal coupling both with the curvature $\bfcur$ and with the
torsion $\bftor$. By this, we significantly generalize the results
of earlier works. At first, following the recommendations of the
Paper~I, we recalculate the elements necessary for constructing
currents and superpotentials of various types. Basing on this,
generalized covariant dynamical quantities are constructed. They are
total canonical both ST $\bfspi$ and EMT $\bfsem$, so-called
modified canonical both ST $\bfspii$ and EMT $\bfsemi$, at last,
symmetrized EMT $\bfsems$ and metric EMT $\bfsemm$. Connections
between these quantities are clarified, besides, for each of the
types of the dynamic characteristics equations of balance are
presented. Next, the correspondent currents and superpotentials are
constructed. The generalized equations of balance are also the basis
for constructing the total system of the equations of the
aforementioned theories generalizing the equations of the ECT with
matter.

The main original results of the present work related to the
manifestly generally covariant metric-torsion theories of gravity are:
\bi
\item a physical interpretation of the Klein and Noether identities, which are a basis for constructing equations of balance for EMTs
of various types;
\item a construction of manifestly generally
covariant expressions for canonical EMT and ST of matter fields. In
the more complicated cases, when $\lagM$ has a generalized
dependence as remarked above, canonical EMT and ST of matter fields
\emph{cannot be obtained} with the use of the standard procedure,
namely, applying the 1-st Noether theorem in the Minkowski space and
covariantization of the expressions. In the
complicated cases, the generalized equations of balance are crucial
for constructing canonical EMT and ST;
\item a \emph{nontrivial} generalization
of the Belinfante procedure applied to the canonical  EMT $\bfsem$
for constructing symmetrized EMT $\bfsems$ \emph{in an arbitrary
Riemann-Cartan space};
\item a construction of a more simple and convenient in applications generalized expression for the
canonical superpotential;
\item a presentation of a total system of
equations for the gravitational fields and matter sources in the
form, which more naturally generalizes the Einstein-Cartan equations
with matter. This result is to be a basis for constructing
physically sensible conservation laws and their applications in the
Paper~III of the series.
\ei

The most of the calculations are very cumbersome and intricate.
Therefore, to give a possibility to a reader to repeat them, many
steps are presented in the main text. Besides, the more important
formulae are given in boxes.  It is important also that initial
identities are analyzed under different assumptions, when either the
total set of field equations hold, or a part of field equations
(say, the gravitational ones only, or the matter ones only) hold. In
future, this will be useful for studying both gravitational theories
with sources of the general type and field theories on a given
geometrical background. At last, one has to note that, in spite of
the present work (Paper~II) is the second work of the series,
developing the Paper~I, it presents a quite independent research.

The paper is organized as follows.  In Sec. \ref{sec_02_00b-00},
necessary formulae for the current and superpotentials, and the
Klein-Noether identities obtained in the Paper~I are given. In the
present work, namely for all of them concrete expressions in the
framework of the manifestly generally covariant theories are
constructed. Also problems, which are elaborated in the present work
are formulated.

All the next studies are related to the theories with the
generalized manifestly covariant Lagrangians $\lag$ of the described
above type. In Sec. \ref{sec_02_01-00}, corresponding to the formulae
\eqref{sec_02_00-08} - \eqref{sec_02_00-10}, the covariant tensors
$\bfK$ and $\bfL$ are constructed. They are necessary to construct
the tensors $\bfU$, $\bfM$ and $\bfN$ determining the generalized
Noether current $\bfJdpara$.

In Sec. \ref{sec_02_02-00}, the manifestly covariant expressions for
the tensors $\bfU$, $\bfM$ and  $\bfN$ themselves are carried out.
The generalized covariant expressions for the total canonical EMT $\bfsem$ and ST $\bfspi$ are found. A contribution initiated both by
the curvature tensor $\bfcur$ and by the non-minimal coupling with
the torsion tensor $\bftor$ is taken into account. It is shown that
in the manifestly covariant theories the tensor $\bfN$ does not
vanish only, when the Lagrangian $\lag$ contains the curvature
tensor $\bfcur$ explicitly. Also additional (with respect to the
standard ones) symmetries of $\bfN$ are clarified.

In Sec. \ref{sec_02_03-00}, a structure of variational derivatives
$\Del I/ \Del\bftor$ and $\Del I/ \Del\bfmet$ of the action
functional $I$ is analyzed. This gives a basis to clarify a physical
sense both of the Noether identity and of all the Klein identities.
It is shown that the results by Trautman and by Hehl at al:
\eqref{sec_02_00-12} and \eqref{sec_02_00-13} are held in a more
general case, when the Lagrangian has the form: $\lag = \lag
(\bfmet, \bfcur; \; \bfffi, \bfnaffi, \bfnanaffi )$. However, in a
case of the next generalization, when the Lagrangian contains a
non-minimal coupling with the torsion, the results become more
complicated: \emph{additional} terms appear. Therefore, one needs to
modify the basic dynamical characteristics. We introduce such a
modification in an explicit form and construct \emph{modified}
total canonical ST $\bfspii$ and EMT $\bfsemi$. The use of these
quantities permits to conserve a connection between $\Del I^M/
\Del\bftor$ and $\bfbelM$, and the equations of balance in the
standard form. Besides, in this section, manifestly covariant
equations of balance for both the total symmetrized EMT $\bfsems$
and the total canonical EMT $\bfsem$ are carried out. It is shown
also that symmetrized EMT $\bfsems$ and the metrical EMT $\bfsemm$
are equivalent if the matter equations hold. In the general case we
prove that the generalized symmetrized Noether current $\bfJsdpara$ is
determined by the symmetrized EMT $\bfsems$ only. Then, it turns out
that surface terms in the functional action do not influence to
constructing both $\bfsems$ and $\bfJsdpara$ (see the Paper~I, Sec. V).

Sec. \ref{sec_02_04-00} is devoted to calculating the superpotential
$\bfpotdpara$ \eqref{sec_02_00-06} and clarifying the role of the
dynamical characteristics EMT and ST in the structure of the
generalized current $\bfJdpara$ \eqref{sec_02_00-05}. The obtained
manifestly covariant formula for the superpotential is quite simple,
it is expressed only through the Belinfante tensor $\bfbel$ and a
tensor $\bfG$, the last exists only if the Lagrangian $\lag$
depends on the curvature tensor $\bfcur$ explicitly.

In Sec. \ref{sec_02_05-00}, the general structure of the equations
of motion of the gravitational fields is examined. The point of
view, which is beginning from Lorentz is discussed. Following it, in
the background independent field theories, the total EMT and ST are equal
to zero identically. The Einstein arguments against are given. We
show that the equations of balance for the pure gravitational part
hold identically and have a clear geometrical sense: they generalize
twice contracted the Bianchi identities onto the case of an
arbitrary metric-torsion theory of gravity in the Riemann-Cartan
space. Basing on this result, we suggest a more preferable
(decomposed) form for the equations of the gravitational fields,
where the pure gravitational part is placed on  the left hand side
of the equations, whereas the other (matter) part is transformed to
the right hand side. This generalizes the form of the equations in
the ECT with matter as well as the Einstein equations themselves.
Their structure is more natural: the modified canonical EMT of
matter $\bfsemMi$ is a source of the metric field $\bfmet$, whereas
the modified canonical ST of matter $\bfspiMi$ is a source of the
torsion field $\bftor$. Such a presentation of the equations is
interesting and important itself. However, besides of that, it is
the basis for constructing physically sensible conservation laws in
the next Paper~III of the series. By this, one concludes also that
the total dynamical characteristics of the physical system are not
equal to zero identically, adding the Einstein arguments.

A calculation of auxiliary quantities is presented in Appendixes. In
Appendix \ref{app_02_a-00}, universal tensors $\{
\Dd\alp\bet\gam\lam\mu\nu \}$ and $\{ \Du\alp\bet\gam\lam\mu\nu \}$,
and various related identities are carried out. The use of them
permits significantly to simplify a presentation of many formulae.
In Appendix \ref{app_02_b-00}, manifestly covariant expressions for
general variations of various quantities, which appear under
calculation of the functional variation of the action functional,
are found. In Appendix \ref{app_02_c-00}, the general theory of the
Belinfante-Rosenfeld symbols, which permits to present the main
relations of the Riemann-Cartan geometry more generally and
economically, is developed. At last, in Appendix \ref{app_02_d-00},
the general identity, which is a central one in constructing
modified canonical EMT $\bfsemi$ and ST $\bfspii$, is proved.

%
\section{Preliminary formulae and a statement of tasks}\label{sec_02_00b-00}

In this Section, we present the main results of the Paper~I, which
are necessary below. Also, here, we formulate the goals of the
present paper. In the Paper~I, an arbitrary generally covariant
theory of tensor fields $\bfgfi$, including both gravitational and
matter ones, with the action functional
\be\label{sec_02_00-15} \ba{l} I[\bfgfi;\Sig_{1,2}] =
\intb\dome\rmet\lag,\ea \ee
is studied. In \eqref{sec_02_00-15}, the integration is provided
over an arbitrary $(D+1)$-dimensional volume in the Riemann-Cartan
space $\caC(1,D)$, restricted by two spacelike $D$-dimensional
hypersurfaces  $\Sig_1$ and $\Sig_2$; the Lagrangian $\lag$ is a
local function of the field variables $\bfPhi=\{\Phi^A(x);
A=\overline{1,N}\}$ and their derivatives up to a second order. One
of the main results of the Paper~I is a construction of the
manifestly covariant expression for the generalized Noether current:
\be\label{sec_02_00-05} \Jdpara\mu = \U\alp\mu \dpara\alp +
\M\alp\bet\mu \na\bet\dpara\alp + \N\alp\bet\gam\mu
\na{(\gam}\na{\bet)}\dpara\alp. \ee
The displacement vectors $\dbfpara$ are induced by diffeomorphisms;
the tensors $\bfU$, $\bfM$ and $\bfN$ are presented by expressions:
\begin{empheq}{flalign}
\U\alp\mu & \Def \lag\kro\mu\alp + \KgA\mu
\gfiaA\alp\nonumber\\
 & + \LgA\kap\mu \lp \na\kap\gfiaA\alp
 + \bfrac{1}{2} \cur\veps\alp\kap\lam \gfibA\veps\lam \rp;\label{sec_02_00-08}\\
\M\alp\bet\mu & \Def \KgA\mu \gfibA\alp\bet +
\LgA\bet\mu \gfiaA\alp\nonumber\\
 & +\LgA\kap\mu \lp \na\kap\gfibA\alp\bet -
\bfrac{1}{2} \tor\bet\kap\lam \gfibA\alp\lam \rp;\label{sec_02_00-09}\\
\N\alp\bet\gam\mu & \Def \LgA{(\gam|}{\mu}
\gfibA{\alp}{|\bet)}.\label{sec_02_00-10}
\end{empheq}
Important definitions and relations in the Riemann-Cartan geometry
are given in the Paper~I. Now, recall necessary notations only. The
torsion tensor $\bftor $ and the curvature tensor $\bfcur $ are
presented as
\be\label{app_02_b-10} \tor\lam\mu\nu = -2\con\lam{[\mu}{\nu]}; \ee
\be\label{app_01_a-02b} \cur\kap\lam\mu\nu = \pa\mu\con\kap\lam\nu -
\pa\nu\con\kap\lam\mu + \con\kap\alp\mu \con\alp\lam\nu -
\con\kap\alp\nu \con\alp\lam\mu. \ee
Here, the connection $\bfcon \Def \{\con\lam{\mu}{\nu}\}$ is defined
by a metric compatible condition
\be\label{app_01_a-02} \na\lam\met\mu\nu = \pa\lam\met\mu\nu -
\con\alp\mu\lam \met\alp\nu - \con\alp\nu\lam \met\mu\alp = 0, \ee
where the standard covariant derivative $\bfna \Def \{ \na\lam \}$
is used. The modified covariant derivative $\bfsna$ is
\be\label{app_01_a-07} \sna\lam = \na\lam + \tor\alp\lam\alp. \ee
Quantities presented by the notations $\{ \gfiaA\alp \}$ and $\{
\gfibA\alp\bet \}$ are defined by the transformation properties of
the fields $\bfPhi$ under diffeomorphisms:
\be \dparagfiA(x) = \gfiaA\alp \dpara\alp (x) + \gfibA\alp\bet
\na\bet\dpara\alp (x). \ee
The tensors $\bfK$ and $\bfL$ are defined as a result of a
comparison of the variation of the action functional
\eqref{sec_02_00-15}:
\be\label{sec_02_00-16} \ba{l}
 \del_{\gfi{}} I[\bfgfi;\Sig_{1,2}] \Def
I[\bfgfi+\dbfgfi;\Sig_{1,2}] - I[\bfgfi;\Sig_{1,2}]\\
\quad = \intb\dome \del_\gfi{}\lp\rmet\lag\rp \ea \ee
with the formula
\be\label{sec_02_01-65} \ba{l} \del_{\gfi{}} I[\bfgfi;\Sig_{1,2}] =
\intb\dome\rmet\DIDgfiA\dgfiA\\
\quad + \intb\dome\rmet \sna\mu \lf \KgA\mu \dgfiA + \LgA\bet\mu
\na\bet \dgfiA \rf. \ea \ee
Hereinafter, instead of the usual variational derivative we use a
quantity proportional it
\be\label{sec_02_01-65V} \frac{\Del I}{\Del\gfiA} =
\frac{1}{\rmet}\frac{\del I}{\del\gfiA} \ee
--- covariant variational derivative.

The tensors $\bfU$, $\bfM$ and $\bfN$ are not independent, they
satisfy the system of the Klein-Noether differential identities:
\bw
\begin{empheq}[left=\empheqlbrace]{flalign}
\sna\mu \Ib\alp\mu \eq \Ia\alp;\label{sec_02_00-01}\\
\ba{rl} \lp \U\alp\bet - \bfrac{1}{3} \N\lam\bet\rho\sig
\cur\lam\alp\rho\sig \rp +  \sna\mu \lp \M{\alp}{[\bet}{\mu]} -
\bfrac{2}{3} \sna\lam \N{\alp}{\lam}{[\bet}{\mu]} +
\bfrac{1}{3} \N{\alp}{[\bet|}{\rho}{\sig} \tor{|\mu]}{\rho}{\sig} \rp \\
+ \bfrac{1}{2} \lp \M{\alp}{[\rho}{\sig]} - \bfrac{2}{3} \sna\lam
\N{\alp}{\lam}{[\rho}{\sig]} + \bfrac{1}{3}
\N{\alp}{[\rho|}{\kap}{\lam} \tor{|\sig]}{\kap}{\lam} \rp
\tor\bet\rho\sig & \eq -\Ib\alp\bet;
\ea\label{sec_02_00-02}\\
\M{\alp}{(\bet}{\gam)} + \sna\mu \N\alp\bet\gam\mu + \N{\alp}{(\bet|}{\mu}{\nu} \tor{|\gam)}{\mu}{\nu} \eq 0;\label{sec_02_00-03}\\
\N{\alp}{(\bet}{\gam}{\del)} \eq 0, \label{sec_02_00-04}
\end{empheq}
\ew
where
\be\label{sec_02_04-61} \Ia\alp \Def \DIDgfiA\gfiaA\alp; \ee
\be\label{sec_02_04-62} \Ib\alp\bet \Def \DIDgfiA\gfibA\alp\bet \ee
are equal to zero if the equations of motion $\Del I/ \Del\gfiA = 0$
hold.

The analysis of the identities \eqref{sec_02_00-01} --
\eqref{sec_02_00-04} lead to the boundary Klein theorem (the 3-rd
Noether theorem), which states that the current \eqref{sec_02_00-05}
can be presented in the form:

\be\label{sec_02_05-19} \Jdpara\mu = -\Ib\alp\mu \dpara\alp + \lp
\sna\nu \potdpara\mu\nu + \frac{1}{2} \potdpara\rho\sig
\tor\mu\rho\sig \rp, \ee
where the generalized canonical superpotential is
\bw
\be\label{sec_02_00-06} \potdpara\mu\nu = \lf -\M{\alp}{[\mu}{\nu]}
+ \bfrac{2}{3} \lp \sna\lam \N{\alp}{\lam}{[\mu}{\nu]} +
\bfrac{1}{2} \tor{[\mu}{\rho}{\sig} \N{\alp}{\nu]}{\rho}{\sig} \rp
\rf \dpara\alp + \lf -\bfrac{4}{3} \N{\alp}{\bet}{[\mu}{\nu]} \rf
\na\bet \dpara\alp. \ee
\ew
With the use of the generalized Belinfante procedure the generalized
symmetrized Noether current
\be\label{sec_02_00-14} \ba{l} \Jsdpara\mu  \Def \Jdpara\mu - \lp
\sna\nu \potBdpara\mu\nu + \frac{1}{2} \potBdpara\rho\sig
\tor\mu\rho\sig
\rp\\
\quad = \Us\alp\mu \dpara\alp \ea \ee
has been constructed. It turns out that the generalized Belinfante
tensor $\bfpotBdpara$, determining the procedure, coincides with the
generalized canonical superpotential $\bfpotdpara$
\eqref{sec_02_00-06}. Thus the current $\bfJsdpara$
\eqref{sec_02_00-14}, by \eqref{sec_02_05-19}, is proportional to
the operators of the equations of motion, that is proportional to
the variational derivatives of the action.  This means that it does
not depend on divergences in the Lagrangian (in the other words, it
does not depend on surface terms in the action functional
\eqref{sec_02_00-15}) and is equal to zero on the equations of
motion.

In the present paper, we consider more concrete theories presented
by the action \eqref{sec_02_00-15}, examining Lagrangians in a
manifestly covariant form:
\be\label{sec_02_01-06} \lag = \lag (\bfmet,\bfcur; \;
\bftor,\bfnator,\bfnanator; \; \bfffi,\bfnaffi,\bfnanaffi). \ee
Here, the total set of the fields $\bfgfi$ is presented by the
metric tensor $\bfmet$, by the torsion tensor $\bftor$
and by a set of the matter fields $\bfffi \Def \{\ffiA(x); \;
a=\overline{1,n} \}$, which are considered as tensorial ones also.
Lagrangians of the type \eqref{sec_02_01-06} include, together with
the minimal coupling, the non-minimal coupling related both to the
curvature and to the torsion. The main task of the present paper is
to present relations and conserved quantities (currents and
superpotentials) constructed in the Paper~I in a maximally concrete
form that follows from the concrete structure of the Lagrangian
\eqref{sec_02_01-06}.

Recall, see formulae \eqref{sec_02_00-05} and \eqref{sec_02_00-06},
that for constructing the generalized current $\bfJdpara$ and
superpotential $\bfpotdpara$ one needs the tensors $\bfU$
\eqref{sec_02_00-08}, $\bfM$ \eqref{sec_02_00-09} and $\bfN$
\eqref{sec_02_00-10}. For constructing the last the other tensors,
$\bfK$ and $\bfL$, defined in \eqref{sec_02_01-65} have to be
calculated. To do this one has to compare \eqref{sec_02_01-65} with
\eqref{sec_02_00-16}, for which one has to know the variation
$\del_\gfi{}(\rmet\lag)$. Because the fields $\bftor$ and $\bfffi$
are included in the Lagrangian in a similar way, for simplification
of the calculations we unite them into the unique set $\bfjfi$:
\be\label{sec_02_01-16} \bftor,\bfffi \quad \rightarrow \quad \bfjfi
\Def \{\jfiA\} \Def \{\bftor,\bfffi\}. \ee
If necessary one can decompose the set $\bfjfi$ again. Now, the
Lagrangian \eqref{sec_02_01-06} is presented as
\be\label{sec_02_01-07} \lag = \lag (\bfmet,\bfcur;
\;\bfjfi,\bfnajfi,\bfnanajfi). \ee
One has to keep in mind that the torsion $\bftor$ is included in the
Lagrangian \eqref{sec_02_01-06} not only \emph{explicitly} as
arguments $\bftor$, $\bfnator$ and $\bfnanator$, but \emph{not
explicitly} also over the connection $\bfcon$, which is used for
constructing the covariant derivative $\bfna$ and the curvature
tensor $\bfcur$.

As it was remarked in Introduction, already in the works by Trautman
\cite{Trautman_1972_a, Trautman_1972_b, Trautman_1973_b,
Trautman_1980} and by Hehl at al
\cite{Hehl_1973, Hehl_1974, Hehl_Heyde_Kerlick_Nester_1976} the construction of the
conservation laws and conserved quantities in the framework of the
manifestly covariant theories has been carried out.  Theories with
the Lagrangians of the type
\be\label{sec_02_00-11} \lag = \lag(\bfmet; \; \bfffi, \bfnaffi).\
\ee
were considered. It was shown that the general relations
\be\label{sec_02_00-12} \DIDtor\alp\bet\gam = \frac12
\belud\gam\bet\alp;\ee
\be\label{sec_02_00-13} \ba{r} \sna\mu \sem\mu\nu = - \sem\mu\lam
\tor\lam\mu\nu +  \bfrac12 \spiu\pi\rho\sig \curd\rho\sig\pi\nu\\
\mbox{(on the $\bfffi$-equations)} \ea \ee
take a place. The first of them shows that the variational
derivative of the action functional $I$ with respect to the torsion
$\bftor$ is equal to the half of the Belinfante tensor $\bfbel \Def
\{ \belu\gam\bet\alp \}$ induced by the canonical ST $\bfspi \Def \{
\spi\pi\rho\sig \}$. The second one is the equation of balance for
the canonical EMT $\bfsem \Def \{ \sem\mu\nu \}$. Of course, the
study of the theories with the Lagrangians of the type
\eqref{sec_02_01-06}, generalizing \eqref{sec_02_00-12} and
\eqref{sec_02_00-13}, has not to lead to contradictions with them.

%
\section{Calculation of the tensors $\bfK$ and $\bfL$}\label{sec_02_01-00}

Variate the lagrangian \eqref{sec_02_01-07}:
\be\label{sec_02_01-13} \del\lp \rmet\lag \rp = \lp\drmet\rp \lag +
\rmet\;\dlag. \ee
The variation of the first term is defined by the well known
relation $\drmet = \rmet \metu\bet\gam \dmet\bet\gam/2$. The second
one, taking into account the above, can be presented in the form:
\be\label{sec_02_01-08} \ba{l} \dlag = \dslagdmet\bet\gam
\dmet\bet\gam + \dlagdcur\kap\lam\mu\nu \dcur\kap\lam\mu\nu
 + \dslagdjfiA \djfiA\\
\quad + \dlagdnajfiA\mu \del(\na\mu\jfiA) + \dlagdnanajfiA\mu\nu
\del(\na\mu \na\nu \jfiA). \ea \ee
Hereinafter, $\pa{}^*\lag/\pa{}\met\bet\gam$ means \emph{explicit}
derivative with respect to $\met\bet\gam$, that is the
differentiation is provided only with respect to $\met\bet\gam$,
which do not included into $\bfcur$ and $\bfna$; analogously,
$\pa{}^*\lag/\pa{}\jfiA$ means differentiation only with respect to
$\jfiA$, which do not included into $\bfnajfi$ and $\bfnanajfi$.
Substituting the expressions for variations $\dcur\kap\lam\mu\nu$
\eqref{sec_02_01-09}, $\del(\na\mu\jfiA)$ \eqref{app_02_b-09} and
$\del(\na\mu\na\nu\jfiA)$ \eqref{sec_02_01-10} into
\eqref{sec_02_01-08}, providing the differentiation by parts and
grouping similar terms, obtain
\bw
\be\label{sec_02_01-11} \ba{rl} \dlag & = \lf \dslagdmet\bet\gam \rf
\dmet\bet\gam + \lf \DIDcon\kap\lam\tau \rf \dcon\kap\lam\tau + \lf
\DsIDjfiA \rf \djfiA\\
 &  + \sna\mu \lb \lf \G\kap\lam\mu\nu + \dlagdnanajfiA\mu\nu
\DbrjAB\lam\kap \jfiB\rf \dcon\kap\lam\nu + \lf \DIDnajfiA\mu \rf
\djfiA + \lf \dlagdnanajfiA\mu\alp \rf \na\alp \djfiA \rb, \ea \ee
\ew
where the notations
\be\label{sec_02_02-02} \ba{l}
\DIDcon\kap\lam\tau \Def \lp \sna\nu \G\kap\lam\tau\nu + \bfrac12 \G\kap\lam\rho\sig \tor\tau\rho\sig \rp\\
 \\
\qquad + \DIDnajfiA\tau \DbrjAB\lam\kap \jfiB\\
 \\
\qquad + \dlagdnanajfiA\tau\nu \lb \DbrjAB\lam\kap \na\nu \jfiB -
\kro\lam\nu \na\kap \jfiA \rb; \ea \ee
\be\label{sec_02_01-35} \G\kap\lam\mu\nu \Def
2\dlagdcur\kap\lam\mu\nu; \ee

\be \DsIDjfiA \Def \dslagdjfiA - \sna\mu \lp \dlagdnajfiA\mu \rp +
\sna\nu\sna\mu \lp \dlagdnanajfiA\mu\nu \rp; \ee
\be \DIDnajfiA\mu \Def \dlagdnajfiA\mu - \sna\nu \lp
\dlagdnanajfiA\nu\mu \rp \ee
are used, and  $\{ \DbrjAB\lam\kap \}$ are the Belinfante-Rosenfeld
symbols (see Appendix \ref{app_02_c-01}). Next, substituting the
expression $\{ \dcon\kap\lam\tau \}$ \eqref{sec_02_01-12} into
\eqref{sec_02_01-11}, providing in the term $\{ \Del I/\Del
\con\kap\lam\tau \}\, \dcon\kap\lam\tau$ differentiation by parts
and again grouping similar terms, one obtains
\bw
\be\label{sec_02_01-15} \ba{l}
\dlag = \lf \dslagdmet\bet\gam -\sna\mu \lp \DIDcon\kap\lam\tau \rp \metu\kap\pi \Dd\mu\bet\gam\pi\lam\tau \rf \dmet\bet\gam + \lf \DIDcon\kap\lam\tau \metu\kap\pi \Dd\mu\bet\gam\pi\lam\tau \met\mu\veps \rf \dtor\veps\bet\gam  + \lf \DIDjfiA \rf \djfiA\\
\quad + \sna\mu \lb \lf \DIDcon\kap\lam\tau \metu\kap\pi \Dd\mu\bet\gam\pi\lam\tau \rf \dmet\bet\gam + \lf \lp \G\kap\lam\mu\nu + \dlagdnanajfiA\mu\nu \DbrjAB\lam\kap \jfiB \rp \metu\kap\pi \Dd\alp\bet\gam\pi\lam\nu \rf \na\alp \dmet\bet\gam \rd\\
\quad \ld + \lf \lp \G\kap\lam\mu\nu + \dlagdnanajfiA\mu\nu
\DbrjAB\lam\kap \jfiB \rp \metu\kap\pi \Dd\alp\bet\gam\pi\lam\nu
\met\alp\veps \rf \dtor\veps\bet\gam + \lf \DIDnajfiA\mu \rf \djfiA +
\lf \dlagdnanajfiA\mu\alp \rf \na\alp \djfiA \rb. \ea \ee
\ew
At last, substituting \eqref{sec_02_01-15} into \eqref{sec_02_01-13}
and recalling that due to the convention \eqref{sec_02_01-16} the
set of fields $\bfjfi$ consists of the torsion field $\bftor$ and a
set of matter fields $\bfffi$, one obtains the search expression for
the functional variation of the action:

\bw
\be\label{sec_02_01-66} \ba{l}
\del_{\gfi{}} I = \disp\intb\dome\rmet \lb \lf \frac12 \lag \metu\bet\gam + \dslagdmet\bet\gam -\sna\mu \lp \DIDcon\kap\lam\tau \rp \metu\kap\pi \Dd\mu\bet\gam\pi\lam\tau \rf \dmet\bet\gam \rd\\
\quad + \lf \dslagdtor\veps\bet\gam - \sna\mu \lp \dlagdnator\mu\veps\bet\gam \rp + \sna\nu \sna\mu \lp \dlagdnanator\mu\nu\veps\bet\gam \rp + \DIDcon\kap\lam\tau \metu\kap\pi \Dd\mu\bet\gam\pi\lam\tau \met\mu\veps \rf \dtor\veps\bet\gam\\
\quad \ld + \lf \dslagdffiA - \sna\mu \lp \dlagdnaffiA\mu \rp + \sna\nu \sna\mu \lp \dlagdnanaffiA\mu\nu \rp \rf \dffiA \rb \\
+ \disp\intb\dome\rmet \sna\mu \lb \lf \DIDcon\kap\lam\tau \metu\kap\pi \Dd\mu\bet\gam\pi\lam\tau \rf \dmet\bet\gam \rd\\
\quad + \lf \lp \G\kap\lam\mu\nu + \dlagdnanator\mu\nu\tau\tet\phi \Dbrt\lam\kap\tau\tet\phi\ome\rho\sig \tor\ome\rho\sig  + \dlagdnanaffiA\mu\nu \DbrfAB\lam\kap \ffiB \rp \metu\kap\pi \Dd\alp\bet\gam\pi\lam\nu \rf \na\alp \dmet\bet\gam \\
\quad +\lf \lp \G\kap\lam\mu\nu + \dlagdnanator\mu\nu\tau\tet\phi \Dbrt\lam\kap\tau\tet\phi\ome\rho\sig \tor\ome\rho\sig +\dlagdnanaffiA\mu\nu \DbrfAB\lam\kap \ffiB \rp \metu\kap\pi \Dd\alp\bet\gam\pi\lam\nu \met\alp\veps \rd\\
\quad \ld\ld + \DIDnator\mu\veps\bet\gam \rf \dtor\veps\bet\gam +
\lf \dlagdnanator\mu\alp\veps\bet\gam \rf \na\alp \dtor\veps\bet\gam
+ \lf \DIDnaffiA\mu \rf \dffiA + \lf \dlagdnanaffiA\mu\alp \rf
\na\alp \dffiA \rb, \ea \ee
%
%
where
\be\label{sec_02_01-17} \ba{l} \DIDcon\kap\lam\tau \Def
\lp \sna\nu \G\kap\lam\tau\nu + \bfrac12 \G\kap\lam\rho\sig \tor\tau\rho\sig \rp\\
\quad + \DIDnator\tau\veps\bet\gam
\Dbrt\lam\kap\veps\bet\gam\ome\rho\sig \tor\ome\rho\sig
 + \dlagdnanator\tau\nu\veps\bet\gam \lb \Dbrt\lam\kap\veps\bet\gam\ome\rho\sig \na\nu \tor\ome\rho\sig - \kro\lam\nu \na\kap \tor\veps\bet\gam \rb\\
\quad + \DIDnaffiA\tau \DbrfAB\lam\kap \ffiB
 + \dlagdnanaffiA\tau\nu
\lb \DbrfAB\lam\kap \na\nu \ffiB - \kro\lam\nu \na\kap \ffiA \rb;
\ea \ee
\be\label{sec_02_01-18} \DIDnator\mu\veps\bet\gam \Def
\dlagdnator\mu\veps\bet\gam - \sna\nu \lp
\dlagdnanator\nu\mu\veps\bet\gam \rp; \ee
\be\label{sec_02_01-19} \DIDnaffiA\mu \Def \dlagdnaffiA\mu - \sna\nu
\lp \dlagdnanaffiA\nu\mu \rp. \ee
On the other hand, for the Lagrangian \eqref{sec_02_01-06} the
formula \eqref{sec_02_01-65} acquires the form:
\be \ba{l}
\del_{\gfi{}} I = \disp\intb\dome\rmet \lb \DIDmet\bet\gam \dmet\bet\gam + \DIDtor\veps\bet\gam \dtor\veps\bet\gam + \DIDffiA \dffiA \rb\\
\quad + \disp\intb\dome\rmet \sna\mu \lb \lp \Km\mu\bet\gam \dmet\bet\gam + \Lm\alp\mu\bet\gam \na\alp \dmet\bet\gam \rp + \lp \Kt\mu\veps\bet\gam \dtor\veps\bet\gam + \Lt\alp\mu\veps\bet\gam \na\alp \dtor\veps\bet\gam \rp \rd\\
\quad \ld + \lp \KfA\mu \dffiA + \LfA\alp\mu \na\alp \dffiA \rp \rb.
\ea \ee
Comparing \eqref{sec_02_01-66} with the last expression, one can recognize
expressions and quantities interesting in our study:
\begin{empheq}[left=\empheqlbrace]{align}
\DIDmet\bet\gam & = \bfrac12 \lag \metu\bet\gam + \dslagdmet\bet\gam -\sna\mu \lp \DIDcon\kap\lam\tau \metu\kap\pi \rp \Dd\mu{(\bet}{\gam)}\pi\lam\tau;\label{sec_02_04-15} \\
\DIDtor\veps\bet\gam & = \DsIDtor\veps\bet\gam + \lp \DIDcon\kap\lam\tau \metu\kap\pi \rp \Dd\mu{[\bet}{\gam]}\pi\lam\tau \met\mu\veps;\label{sec_02_04-10}\\
\DIDffiA & = \dslagdffiA - \sna\mu \lp \dlagdnaffiA\mu \rp + \sna\nu
\sna\mu \lp \dlagdnanaffiA\mu\nu \rp;\label{sec_02_06-28}
\end{empheq}
\begin{empheq}[left=\empheqlbrace]{align}
\Km\mu\bet\gam & = \DIDcon\kap\lam\tau \metu\kap\pi \Dd\mu{(\bet}{\gam)}\pi\lam\tau; \label{sec_02_01-26}\\
\Lm\alp\mu\bet\gam & = \lp \G\kap\lam\mu\nu +
\dlagdnanator\mu\nu\tau\tet\phi
\Dbrt\lam\kap\tau\tet\phi\ome\rho\sig \tor\ome\rho\sig  +
\dlagdnanaffiA\mu\nu \DbrfAB\lam\kap \ffiB \rp \metu\kap\pi
\Dd\alp{(\bet}{\gam)}\pi\lam\nu;\label{sec_02_01-40}
\end{empheq}
\begin{empheq}[left=\empheqlbrace]{align}
\Kt\mu\veps\bet\gam & = \lp \G\kap\lam\mu\nu + \dlagdnanator\mu\nu\tau\tet\phi \Dbrt\lam\kap\tau\tet\phi\ome\rho\sig \tor\ome\rho\sig +\dlagdnanaffiA\mu\nu \DbrfAB\lam\kap \ffiB \rp \metu\kap\pi \Dd\alp{[\bet}{\gam]}\pi\lam\nu \met\alp\veps\nonumber\\
 & + \DIDnator\mu\veps\bet\gam;\label{sec_02_01-67} \\
\Lt\alp\mu\veps\bet\gam & =
\dlagdnanator\mu\alp\veps\bet\gam;\label{sec_02_01-41}
\end{empheq}
\ew
\begin{empheq}[left=\empheqlbrace]{align}
\KfA\mu & = \DIDnaffiA\mu;\label{sec_02_01-68}\\
\LfA\alp\mu & = \dlagdnanaffiA\mu\alp. \label{sec_02_01-27}
\end{empheq}
Here,
\be\label{sec_02_04-60} \ba{l} \DsIDtor\veps\bet\gam  \Def
\dslagdtor\veps\bet\gam - \sna\mu \lp
\dlagdnator\mu\veps\bet\gam \rp\\
\quad + \sna\nu \sna\mu \lp \dlagdnanator\mu\nu\veps\bet\gam \rp,
\ea \ee
and $\Del I/\Del \con\kap\lam\tau$, $\Del I/\Del (\na\mu
\tor\veps\bet\gam)$ and $\Del I/\Del (\na\mu \ffiA)$ are defined by
the formulae \eqref{sec_02_01-17}, \eqref{sec_02_01-18} and
\eqref{sec_02_01-19}, respectively. Thus, calculating the tensors
$\bfK$ and $\bfL$ is finalized.

To simplify calculations remark the following. Comparing the
formulae \eqref{sec_02_01-41} and \eqref{sec_02_01-27} with the
formula \eqref{sec_02_01-40}, it is easily to find that there is a
connection between the quantities $\{ \Lm\alp\mu\bet\gam \}$ and $\{
\Lt\alp\mu\veps\bet\gam \}$, $\{ \LfA\alp\mu \}$
\be\label{sec_02_01-30} \ba{l}
\Lm\alp\mu\bet\gam = \lp \G\kap\lam\mu\nu + \Lt\nu\mu\tau\tet\vphi \Dbrt\lam\kap\tau\tet\vphi\ome\rho\sig \tor\ome\rho\sig\rd\\
\quad \ld + \LfA\nu\mu \DbrfAB\lam\kap \ffiB \rp \metu\kap\pi \Dd\alp{(\bet}{\gam)}\pi\lam\nu\\
\Def \lp \G\kap\lam\mu\nu + \LfA\nu\mu \DbrjAB\lam\kap \jfiB \rp
\metu\kap\pi \Dd\alp{(\bet}{\gam)}\pi\lam\nu.
\ea \ee

 Analogously, comparing the formulae \eqref{sec_02_01-41} and
\eqref{sec_02_01-27} with the formula \eqref{sec_02_01-67}, one
finds the connection between the quantities $\{ \Kt\mu\veps\bet\gam
\}$ and $\{ \Lt\alp\mu\veps\bet\gam \}$, $\{ \LfA\alp\mu \}$:

\be\label{sec_02_01-37} \ba{l}
\Kt\mu\veps\bet\gam = \DIDnator\mu\veps\bet\gam\\
\quad + \lp \G\kap\lam\mu\nu + \Lt\nu\mu\tau\tet\vphi \Dbrt\lam\kap\tau\tet\vphi\ome\rho\sig \tor\ome\rho\sig \rd\\
\quad \ld + \LfA\nu\mu \DbrfAB\lam\kap \ffiB \rp \metu\kap\pi \Dd\alp{[\bet}{\gam]}\pi\lam\nu \met\alp\veps\\
\Def \sKt\mu\veps\bet\gam + \lp \G\kap\lam\mu\nu + \LfA\nu\mu
\DbrjAB\lam\kap \jfiB \rp \metu\kap\pi
\Dd\alp{[\bet}{\gam]}\pi\lam\nu \met\alp\veps, \ea \ee
where
\be\label{sec_02_01-45} \sKt\mu\veps\bet\gam \Def
\DIDnator\mu\veps\bet\gam, \ee
compare with \eqref{sec_02_01-68}. Using the formulae
\eqref{sec_02_01-41} -- \eqref{sec_02_01-27}, \eqref{sec_02_01-45}
in the expression \eqref{sec_02_01-17}, one can present it in a more compact
form:
\bw
\be\label{sec_02_01-70} \ba{l} \DIDcon\kap\lam\tau =
\lp \sna\nu \G\kap\lam\tau\nu + \bfrac12 \G\kap\lam\rho\sig \tor\tau\rho\sig \rp\\
\quad + \sKt\tau\veps\bet\gam \Dbrt\lam\kap\veps\bet\gam\ome\rho\sig \tor\ome\rho\sig + \Lt\nu\tau\veps\bet\gam \lb \Dbrt\lam\kap\veps\bet\gam\ome\rho\sig \na\nu \tor\ome\rho\sig - \kro\lam\nu \na\kap \tor\veps\bet\gam \rb\\
\quad + \KfA\tau \DbrfAB\lam\kap \ffiB + \LfA\nu\tau \lb \DbrfAB\lam\kap \na\nu \ffiB - \kro\lam\nu \na\kap \ffiA \rb\\
\Def \lp \sna\nu \G\kap\lam\tau\nu + \bfrac12 \G\kap\lam\rho\sig
\tor\tau\rho\sig \rp + \sKjA\tau \DbrjAB\lam\kap \jfiB + \LjA\nu\tau
\lb \DbrjAB\lam\kap \na\nu \jfiB - \kro\lam\nu \na\kap \jfiA \rb.
\ea \ee
\ew
%

%
\section{The calculation of the tensors $\bfU$, $\bfM$ and $\bfN$}\label{sec_02_02-00}

Next, we calculate the tensors $\bfU$, $\bfM$ and $\bfN$ following
the formulae \eqref{sec_02_00-08}, \eqref{sec_02_00-09} and
\eqref{sec_02_00-10} in the manifestly covariant theories with the
Lagrangians of the type \eqref{sec_02_01-06}. As we think, we have
found the most economical scheme of calculations. Now, we follow it in
details. For this we need the formulae
\be\label{sec_02_01-44} \meta\alp\bet\gam = 2
\tord{(\bet}{\gam)}\alp; \ee
\be\label{sec_02_01-43} \jfiaA\alp = -\lf \na\alp \jfiA +
\tor\gam\bet\alp \DbrjAB\bet\gam \jfiB\rf; \ee
\be\label{sec_02_01-62} \ba{l} \tora\alp{\veps}\bet\gam = -\na\alp
\tor\veps\bet\gam\\
\quad -(\tor\veps\kap\alp \tor\kap\bet\gam + \tor\veps\kap\bet
\tor\kap\gam\alp + \tor\veps\kap\gam \tor\kap\alp\bet); \ea \ee
\be\label{sec_02_01-31} \metb\alp\bet\kap\lam = -2 \met\alp{(\kap}
\kro\bet{\lam)}; \ee
\be\label{sec_02_01-32} \jfibA\alp\bet = \DbrjAB\bet\alp \jfiB; \ee
\be\label{sec_02_01-69} \torb\alp\lam\veps\bet\gam = \kro\veps\alp
\tor\lam\bet\gam + 2\tor\veps\alp{[\bet} \kro\lam{\gam]}, \ee
proved in Appendix \ref{app_02_c-03}.

\subsection{The calculation of the tensor $\bfN$}

For the Lagrangian \eqref{sec_02_01-06} the formula
\eqref{sec_02_00-10} transforms to
\bse \ba{l}
\N\alp\kap\lam\mu \Def \LgA{(\lam|}{\mu} \gfibA{\alp}{|\kap)}\\
 =\Lm{(\lam|}\mu\bet\gam \metb\alp{|\kap)}\bet\gam +
\Lt{(\lam|}\mu\veps\bet\gam \torb\alp{|\kap)}\veps\bet\gam +
\LfA{(\lam|}\mu \ffibA\alp{|\kap)}. \ea \ese
Let us present the calculation in the next steps.
\bn
\item Return to the united field $\bfjfi = \lf \bftor, \bfffi \rf$ \eqref{sec_02_01-16}.
Then
\bse \N\alp\kap\lam\mu = \Lm{(\lam|}\mu\bet\gam
\metb\alp{|\kap)}\bet\gam + \LjA{(\lam|}\mu \jfibA\alp{|\kap)}. \ese
\item Take into account the connection \eqref{sec_02_01-30} and obtain

\bse \ba{l} \N\alp\kap\lam\mu = \lp \G\veps\eta\mu\nu + \LjA\nu\mu
\DbrjAB\eta\veps \jfiB \rp \metu\veps\pi
\Dd{(\lam|}{(\bet}{\gam)}\pi\eta\nu \metb\alp{|\kap)}\bet\gam\\
\quad + \LjA{(\lam|}\mu \jfibA\alp{|\kap)}.\ea \ese
\item Substitute here the expressions \eqref{sec_02_01-31} and \eqref{sec_02_01-32}.
Then, with using the identity \eqref{sec_02_01-33}, the tensor
$\bfN$ is transformed to the form:
\bse \ba{l} \N\alp\kap\lam\mu = - \lp \G\veps\eta\mu\nu + \LjA\nu\mu
\DbrjAB\eta\veps \jfiB \rp \kro\veps\alp \kro{(\kap}\eta
\kro{\lam)}\nu\\
\quad + \LjA{(\lam|}\mu \DbrjAB{|\kap)}\alp \jfiB =
\G\alp{(\kap}{\lam)}\mu. \ea \ese
\en
Thus,  the expression for the tensor $\bfN$ gets the form:
\be\label{sec_02_01-34} \boxed{ \N\alp\bet\gam\mu =
\G\alp{(\bet}{\gam)}\mu. } \ee
It is important to note that the \emph{tensor $\bfN$ is not equal to
zero only if the Lagrangian contains explicitly the curvature tensor
$\bfcur$} (see the definition \eqref{sec_02_01-35}).

\subsection{The calculation of the tensor $\bfM$}
For the Lagrangian \eqref{sec_02_01-06} the formula
\eqref{sec_02_00-09} transforms to
\bw
\be\label{sec_02_01-36} \ba{rl}
\M\alp\lam\mu & \Def \KgA\mu \gfibA\alp\lam + \LgA\lam\mu \gfiaA\alp +\LgA\kap\mu \lp \na\kap\gfibA\alp\lam - \bfrac{1}{2} \tor\lam\kap\eta \gfibA\alp\eta \rp\\
 & = \lf \Km\mu\bet\gam \metb\alp\lam\bet\gam + \Kt\mu\veps\bet\gam \torb\alp\lam\veps\bet\gam + \KfA\mu \ffibA\alp\lam \rf\\
 & \quad + \lf \Lm\lam\mu\bet\gam \meta\alp\bet\gam + \Lt\lam\mu\veps\bet\gam \tora\alp\veps\bet\gam + \LfA\lam\mu \ffiaA\alp \rf + \lf \Lm\rho\mu\bet\gam \lp \na\rho \metb\alp\lam\bet\gam - \bfrac12 \tor\lam\rho\sig \metb\alp\sig\bet\gam \rp \rd\\
 & \quad \ld + \Lt\rho\mu\veps\bet\gam \lp \na\rho
\torb\alp\lam\veps\bet\gam - \bfrac12 \tor\lam\rho\sig
\torb\alp\sig\veps\bet\gam \rp + \LfA\rho\mu \lp \na\rho
\ffibA\alp\lam - \bfrac12 \tor\lam\rho\sig \ffibA\alp\sig \rp  \rf.
\ea \ee
\ew
Provide the calculation of  \eqref{sec_02_01-36} step by step also.
\bn
\item Denote the first, second and third braces on the
right hand side of \eqref{sec_02_01-36} as $\{\dots\}_1$,
$\{\dots\}_2$ and $\{\dots\}_3$, respectively. Then
\bse \M\alp\lam\mu \Def \{\dots\}_1 + \{\dots\}_2 + \{\dots\}_3.
\ese
\item\label{sec_02_01-38} In $\{\dots\}_1$, take into account the expression
\eqref{sec_02_01-37} and return to the united field $\bfjfi = \lf
\bftor, \bfffi \rf$. Then
\bse \ba{l} \{\dots\}_1 = \Km\mu\bet\gam \metb\alp\lam\bet\gam +
\sKjA\mu
\jfibA\alp\lam\\
\quad + \lp \G\kap\eta\mu\nu + \LjA\nu\mu \DbrjAB\eta\kap \jfiB \rp
\metu\kap\pi \Dd\rho{[\bet}{\gam]}\pi\eta\nu \met\rho\veps \;
\torb\alp\lam\veps\bet\gam, \ea \ese
where
\be\label{sec_02_01-63} \sKjA\mu \jfibA\alp\lam \Def
\sKt\mu\veps\bet\gam \torb\alp\lam\veps\bet\gam + \KjA\mu
\ffibA\alp\lam. \ee
\item In $\{\dots\}_2$, take into account the expression \eqref{sec_02_01-30}
and return to the united field $\bfjfi$ again. Then
\bse \ba{rl} \{\dots\}_2 & = \lp \G\kap\eta\mu\nu + \LjA\nu\mu
\DbrjAB\eta\kap \jfiB \rp \metu\kap\pi
\Dd\lam{(\bet}{\gam)}\pi\eta\nu \;
\meta\alp\bet\gam\\
 & + \LjA\lam\mu \jfiaA\alp.
\ea \ese
\item\label{sec_02_01-39} To provide $\{\dots\}_3$ apply the similar steps and
obtain
\bse \ba{rl}
\{\dots\}_3 & = \lp \G\kap\eta\mu\nu + \LjA\nu\mu \DbrjAB \eta\kap \jfiB \rp \metu\kap\pi\\
& \times\Dd\rho{(\bet}{\gam)}\pi\eta\nu \lp \na\rho \metb\alp\lam\bet\gam - \bfrac{1}{2} \tor\lam\rho\sig \metb\alp\sig\bet\gam \rp\\
 & + \LjA\rho\mu \lp \na\rho \jfibA\alp\lam - \bfrac{1}{2} \tor\lam\rho\sig \jfibA\alp\sig \rp.
\ea \ese
\item Combining the results of the points  \ref{sec_02_01-38} -- \ref{sec_02_01-39}
and collecting the similar terms, find
\bw
\be\label{sec_02_01-42} \ba{l}
\M\alp\lam\mu  = \Km\mu\bet\gam \metb\alp\lam\bet\gam + \sKjA\mu \jfibA\alp\lam + \LjA\nu\mu \lb \kro\lam\nu \jfiaA\alp + \na\nu \jfibA\alp\lam - \bfrac{1}{2} \tor\lam\nu\bet \jfibA\alp\bet \rb  \\
+ \lp \G\kap\eta\mu\nu + \LjA\nu\mu \DbrjAB\eta\kap \jfiB \rp
\metu\kap\pi \Dd\rho\bet\gam\pi\eta\nu \lb \met\rho\veps
\torb\alp\lam\veps\bet\gam + \kro\lam\rho \meta\alp\bet\gam +
\na\rho \metb\alp\lam\bet\gam - \bfrac{1}{2} \tor\lam\rho\sig
\metb\alp\sig\bet\gam \rb. \ea \ee
\ew
\item\label{sec_02_01-48} Corresponding to the formulae \eqref{sec_02_01-26}, \eqref{sec_02_01-70} and
\eqref{sec_02_01-31} one has
\be\label{sec_02_01-64} \ba{l} \Km\mu\bet\gam \metb\alp\lam\bet\gam
= -2
\lb -\lp \sna\nu \Gu\eta\pi\tau\nu + \bfrac{1}{2} \Gu\eta\pi\rho\sig \tor\tau\rho\sig \rp \rd\\
\quad + \sKjA\tau \DbrjuAB\eta\pi \jfiB \\
\quad \ld + \LjA\nu\tau \lp \DbrjuAB\eta\pi \na\nu \jfiB -
\kro\eta\nu \nau\pi \jfiA \rp       \rb
\Dd\mu{(\lam}{\bet)}\pi\eta\tau \met\bet\alp. \ea \ee
\item Taking into account \eqref{sec_02_01-43}
and \eqref{sec_02_01-32}, and adding with subtracting the expression
$\lp \sna\nu \G\alp\lam\mu\nu + \frac{1}{2} \G\alp\lam\rho\sig
\tor\mu\rho\sig \rp$, one obtains for the sum of $\sKjA\mu
\jfibA\alp\lam$ and the next item on the right hand side of
\eqref{sec_02_01-42}:
\bw
\bse \ba{l}
\lb -\lp \sna\nu \Gu\lam\bet\mu\nu + \bfrac{1}{2} \Gu\lam\bet\rho\sig \tor\mu\rho\sig\rp + \sKjA\mu \DbrjuAB\lam\bet \jfiB + \LjA\nu\mu \lp \DbrjuAB\lam\bet \na\nu \jfiB - \kro\lam\nu \nau\bet \jfiA \rp \rb  \met\bet\alp\\
\quad - \lb \sna\nu \G\alp\lam\mu\nu + \bfrac{1}{2}
\G\alp\lam\rho\sig \tor\mu\rho\sig \rb - \bfrac{1}{2} \LjA\nu\mu \lb
2\kro\lam\nu \tor\gam\bet\alp + \tor\lam\nu\bet \kro\gam\alp \rb
\DbrjAB\bet\gam \jfiB. \ea \ese
\ew
\item\label{sec_02_01-49} For the expression inside brackets
of the last item in \eqref{sec_02_01-42}, taking into account the
formulae \eqref{sec_02_01-44}, \eqref{sec_02_01-31} and
\eqref{sec_02_01-69}, and the identities \eqref{sec_02_01-46} and
\eqref{sec_02_01-47}, provide simple identical transformations and
obtain
\bw
\bse \ba{l}
\Dd\rho\bet\gam\pi\eta\nu \lb \met\rho\veps \lp \kro\veps\alp \tor\lam\bet\gam + 2\tor\veps\alp{[\bet} \kro\lam{\gam]} \rp + 2\kro\lam\rho \tord{(\bet}{\gam)}\alp + \tor\lam\rho\sig \met\alp{(\bet} \kro\sig{\gam)} \rb\\
\quad = \Dd\rho\bet\gam\pi\eta\nu \lb \lp \met\alp{(\rho} \tor\lam{\bet)}\gam + 2\kro\lam{(\rho} \tord{\bet)}\gam\alp \rp + \lp \met\alp{[\rho|} \tor\lam\bet{|\gam]} + 2\kro\lam{[\rho} \tord{\gam]}\bet\alp \rp \rb\\
\quad = \Dd{(\rho}{\bet)}\gam\pi\eta\nu \lp \met\alp\rho \tor\lam\bet\gam + 2\kro\lam\rho \tord\bet\gam\alp \rp + \Dd{[\rho|}\bet{|\gam]}\pi\eta\nu \lp \met\alp\rho \tor\lam\bet\gam + 2\kro\lam\rho \tord\gam\bet\alp \rp\\
\quad = \bfrac{1}{2} \kro\rho{(\pi} \kro\bet{\nu)} \kro\gam\eta \lp \met\alp\rho \tor\lam\bet\gam + 2\kro\lam\rho \tord\bet\gam\alp \rp - \bfrac{1}{2} \kro\bet\eta \kro\rho{[\pi} \kro\gam{\nu]} \lp \met\alp\rho \tor\lam\bet\gam + 2\kro\lam\rho \tord\gam\bet\alp \rp\\
\quad =\bfrac{1}{2} \lb 2\lp \kro\lam{(\nu} \tord{\pi)}\eta\alp + \kro\lam{[\nu} \tord{\pi]}\eta\alp \rp - \lp \met\alp{(\pi|} \tor\lam\eta{|\nu)} + \met\alp{[\pi|} \tor\lam\eta{|\nu]} \rp \rb\\
= \bfrac{1}{2} \lp 2\kro\lam\nu \tord\pi\eta\alp - \met\alp\pi
\tor\lam\eta\nu \rp.
\ea \ese
Thus, the last item in \eqref{sec_02_01-42} is equal to
\bse \frac{1}{2} \G\alp\bet\gam\mu \tor\lam\bet\gam - \lp
\G\gam\bet\lam\mu \tor\gam\bet\alp \rp + \frac{1}{2} \LjA\nu\mu \lb
2\kro\lam\nu \tor\gam\bet\alp + \tor\lam\nu\bet \kro\gam\alp \rb
\DbrjAB\bet\gam \jfiB. \ese
\item Combining the results of the points \ref{sec_02_01-48} -- \ref{sec_02_01-49}, one finds
\bse \ba{l}
\M\alp\lam\mu = -2 \lb -\lp \sna\nu \Gu\eta\pi\tau\nu + \bfrac{1}{2} \Gu\eta\pi\rho\sig \tor\tau\rho\sig \rp + \sKjA\tau \DbrjuAB\eta\pi \jfiB + \LjA\nu\tau \lb \DbrjuAB\eta\pi \na\nu \jfiB - \kro\eta\nu \nau\pi \jfiA\rb \rb\\
\quad \times  \lp \Dd\mu{(\lam}{\bet)}\pi\eta\tau - \bfrac{1}{2}
\kro\mu\tau \kro\lam\eta \kro\bet\pi \rp \met\bet\alp - \lb \sna\nu
\G\alp\lam\mu\nu + \bfrac{1}{2} \G\alp\lam\rho\sig \tor\mu\rho\sig
\rb + \bfrac{1}{2} \G\alp\bet\gam\mu \tor\lam\bet\gam - \lp
\G\gam\bet\lam\mu \tor\gam\bet\alp \rp.
\ea \ese
\item At last, using the identity \eqref{sec_02_01-50} and formulae
\eqref{sec_02_01-41}--\eqref{sec_02_01-27}, \eqref{sec_02_01-45},
and denoting
\be\label{sec_02_01-51} \boxed{ \ba{rl}
\spi\pi\rho\sig & \Def -2 \lp \sna\eta G_{\rho\sig}{}^{\pi\eta} + \bfrac{1}{2} G_{\rho\sig}{}^{\alp\bet} \tor\pi\alp\bet \rp\\
 & + 2\lp \DIDnajfiA\pi \DbrjdAB{[\rho}{\sig]} \jfiB + \dlagdnanajfiA\pi\alp \lb \DbrjdAB{[\rho}{\sig]} \na\alp \jfiB - \met\alp{[\rho} \na{\sig]} \jfiA \rb \rp,
\ea } \ee
one obtains the finalized expression for the tensor $\bfM$:

\be\label{sec_02_04-28}\boxed{ \M\kap\lam\mu = - \lp
\Du\mu\lam\alp\pi\rho\sig \spiu\pi\rho\sig \rp \met\alp\kap - \lp
\sna\nu \G\kap\lam\mu\nu + \bfrac{1}{2} \G\kap\lam\rho\sig
\tor\mu\rho\sig \rp + \bfrac{1}{2} \G\kap\alp\bet\mu
\tor\lam\alp\bet - \lp \G\alp\bet\lam\mu \tor\alp\bet\kap \rp. } \ee
\ew
\en
Remark that the tensor $\bfspi \Def \lf \spi\pi\rho\sig \rf$
\eqref{sec_02_01-51} is just the \emph{generalized canonical spin tensor},
corresponding to the Lagrangian \eqref{sec_02_01-06}. This statement
follows from the results of Sec. \ref{sec_02_05-00}. Namely, basing
on the above definition of the ST, one obtains the standard
equations of balance for the EMT. Besides, the gravitational field
equations acquire the form, naturally generalizing the ECT
equations. Remark that the items in the first parentheses on the
right hand side of \eqref{sec_02_01-51} are induced by the
non-minimal coupling with the metric field. These items \emph{in
principal} cannot be obtained with the use of the 1-st Noether theorem in the
Minkowski space and covariantization of the expressions.

\subsection{The calculation of the tensor $\bfU$}
For the Lagrangian \eqref{sec_02_01-06} the formula
\eqref{sec_02_00-08} has the form:
\bw
\be\label{sec_02_01-22} \ba{rl} \U\alp\mu & \Def \lag\kro\mu\alp +
\KgA\mu \gfiaA\alp + \LgA\kap\mu \lp \na\kap\gfiaA\alp
+ \bfrac{1}{2} \cur\veps\alp\kap\lam \gfibA\veps\lam \rp\\
 & = \lag\kro\mu\alp + \lf \Km\mu\bet\gam \meta\alp\bet\gam +
\Kt\mu\veps\bet\gam \tora\alp\veps\bet\gam + \KfA\mu \ffiaA\alp \rf
+ \lf \Lm\kap\mu\bet\gam \lp \na\kap \meta\alp\bet\gam + \bfrac12 \cur\sig\alp\kap\lam \metb\sig\lam\bet\gam \rp \rd\\
 & \quad \ld + \Lt\kap\mu\veps\bet\gam \lp \na\kap
\tora\alp\veps\bet\gam + \bfrac12 \cur\sig\alp\kap\lam
\torb\sig\lam\veps\bet\gam \rp + \LfA\kap\mu \lp \na\kap \ffiaA\alp
+ \bfrac12 \cur\sig\alp\kap\lam \ffibA\sig\lam \rp \rf. \ea \ee
\ew
Transform it. The main steps are as follows.
\bn
\item The first and second braces on the right hand side of
\eqref{sec_02_01-22} denote as $\{\dots\}_4$ and $\{\dots\}_5$,
respectively. Thus,
\bse \U\alp\mu \Def \lag \kro\mu\alp + \{\dots\}_4 + \{\dots\}_5.
\ese
\item\label{sec_02_01-52} In $\{\dots\}_4$, take into account the expression \eqref{sec_02_01-37}
and return to the united field $\bfjfi = \lf \bftor, \bfffi \rf$.
Then
\bse \ba{l}
\{\dots\}_4 = \Km\mu\bet\gam \meta\alp\bet\gam + \sKjA\mu \jfiaA\alp\\
\quad + \lp \Gu\pi\eta\mu\nu + \LjA\nu\mu \DbrjuAB\eta\pi \jfiB \rp
\Dd\rho{[\bet}{\gam]}\pi\eta\nu \met\rho\veps \;
\tora\alp\veps\bet\gam, \ea \ese
where
\be \sKjA\mu \jfiaA\alp \Def \sKt\mu\veps\bet\gam
\tora\alp\veps\bet\gam + \KfA\mu \ffiaA\alp. \ee
\item\label{sec_02_01-53} In $\{\dots\}_5$, take into account the expression
\eqref{sec_02_01-30} and return to the united field $\bfjfi = \lf
\bftor, \bfffi \rf$. Then
\bse \ba{l}
\{\dots\}_5 = \lp \Gu\pi\eta\mu\nu + \LjA\nu\mu \DbrjuAB\eta\pi \jfiB \rp \\
\quad \times \Dd\kap{(\bet}{\gam)}\pi\eta\nu \lp \na\kap \meta\alp\bet\gam + \bfrac{1}{2} \cur\veps\alp\kap\lam \metb\veps\lam\bet\gam \rp\\
\quad + \LjA\kap\mu \lp \na\kap \jfiaA\alp + \bfrac{1}{2}
\cur\veps\alp\kap\lam \jfibA\veps\lam \rp. \ea \ese
\item Combining the results of the points \ref{sec_02_01-52} --
\ref{sec_02_01-53} and collecting the similar terms, find
\bw
\be\label{sec_02_01-54} \ba{rl}
\U\alp\mu & = \lag \kro\mu\alp + \Km\mu\bet\gam \meta\alp\bet\gam + \sKjA\mu \jfiaA\alp + \lp \Gu\pi\eta\mu\nu + \LjA\nu\mu \DbrjuAB\eta\pi \jfiB \rp \Dd\kap\bet\gam\pi\eta\nu\\
 & \times  \lb \met\kap\veps \tora\alp\veps\bet\gam + \na\kap \meta\alp\bet\gam + \bfrac{1}{2} \cur\veps\alp\kap\lam \metb\veps\lam\bet\gam \rb + \LjA\kap\mu \lp \na\kap \jfiaA\alp + \bfrac{1}{2} \cur\veps\alp\kap\lam \jfibA\veps\lam \rp.
\ea \ee
\ew
\item\label{sec_02_01-55} Remark that, corresponding to the formulae \eqref{sec_02_01-26}, \eqref{sec_02_01-70} and
\eqref{sec_02_01-44},

\bse \ba{l}
\Km\mu\bet\gam \meta\alp\bet\gam\\
 \\
\quad = 2\lb -\lp \sna\nu \Gu\eta\pi\tau\nu + \bfrac{1}{2} \Gu\eta\pi\rho\sig \tor\tau\rho\sig \rp + \sKjA\tau \DbrjuAB\eta\pi \jfiB \rd\\
\qquad \ld + \LjA\nu\tau \lp \DbrjuAB\eta\pi \na\nu \jfiB -
\kro\eta\nu \nau\pi \jfiA \rp \rb
 \Dd\mu{(\bet}{\gam)}\pi\eta\tau \;  \tord\gam\bet\alp.
\ea \ese

\item\label{sec_02_01-56} Turn to the right hand side of \eqref{sec_02_01-54}.
Substitute the expression \eqref{sec_02_01-43} into the third item,
and substitute the expressions \eqref{sec_02_01-43} and
\eqref{sec_02_01-32} into the last one. After that commutate the
second covariant derivatives $\na\kap \na\alp \jfiA$ by the rule
\eqref{sec_02_02-01}. In the result one obtains

\bse \ba{l}
\LjA\nu\mu \lp \na\nu \jfiaA\alp + \bfrac{1}{2} \cur\gam\alp\nu\bet \jfibA\gam\bet \rp\\
\quad = - \LjA\nu\mu \na\alp\na\nu \jfiA\\
\qquad - \LjA\nu\mu \lb \DbrjuAB\bet\gam \na\nu \jfiB - \kro\bet\nu \nau\gam \jfiA \rb \tord\gam\bet\alp\\
\qquad + \LjA\nu\mu \DbrjAB\bet\gam \jfiB \lp \cur\gam\bet\alp\nu -
\na\nu \tor\gam\bet\alp + \bfrac{1}{2} \cur\gam\alp\nu\bet \rp. \ea
\ese
\item Combining the results of the points  \ref{sec_02_01-55} and
\ref{sec_02_01-56}, one can see that the sum of the first, second, third
and fifth items, after adding and subtracting the combination $\lp
\sna\nu \Gu\bet\gam\mu\nu + \frac{1}{2} \Gu\bet\gam\rho\sig
\tor\mu\rho\sig \rp \tord\gam\bet\alp$, becomes
\bw
\be\label{sec_02_01-57} \ba{l}
\lb \lag \kro\mu\alp - \sKjA\mu \na\alp \jfiA - \LjA\nu\mu \na\alp\na\nu \jfiA \rb + 2\lb -\lp \sna\nu \Gu\eta\pi\tau\nu + \bfrac{1}{2} \Gu\eta\pi\rho\sig \tor\tau\rho\sig \rp \rd\\
\quad \ld + \sKjA\tau \DbrjuAB\eta\pi \jfiB + \LjA\nu\tau \lp \DbrjuAB\eta\pi \na\nu \jfiB - \kro\eta\nu \nau\pi \jfiA \rp \rb \times \lp \Dd\mu{(\bet}{\gam)}\pi\eta\tau - \bfrac{1}{2} \kro\mu\tau \kro\bet\eta \kro\gam\pi \rp \tord\gam\bet\alp\\
\quad -\lp \sna\nu \Gu\bet\gam\mu\nu + \bfrac{1}{2} \Gu\bet\gam\rho\sig \tor\mu\rho\sig \rp \tord\gam\bet\alp + \LjA\nu\mu \DbrjAB\bet\gam \jfiB \lp \cur\gam\bet\alp\nu - \na\nu \tor\gam\bet\alp + \bfrac{1}{2} \cur\gam\alp\nu\bet \rp\\
\\
= \lb \lag \kro\mu\alp - \sKjA\mu \na\alp \jfiA - \LjA\nu\mu \na\alp\na\nu \jfiA \rb + \lp \Du\mu\bet\gam\tau\eta\pi \spiu\tau\eta\pi \rp \tord\gam\bet\alp\\
\quad -\lp \sna\nu \Gu\bet\gam\mu\nu + \bfrac{1}{2}
\Gu\bet\gam\rho\sig \tor\mu\rho\sig \rp \tord\gam\bet\alp +
\LjA\nu\mu \DbrjAB\bet\gam \jfiB \lp \cur\gam\bet\alp\nu - \na\nu
\tor\gam\bet\alp + \bfrac{1}{2} \cur\gam\alp\nu\bet \rp, \ea \ee
\ew
where the identity \eqref{sec_02_01-50} and the definition
\eqref{sec_02_01-51} have been taken into account.
\item\label{sec_02_01-58} After substituting the equality
\bse -\sna\nu \Gu\bet\gam\mu\nu \tord\gam\bet\alp= \sna\nu \lp
\Gu\bet\gam\mu\nu \tord\bet\gam\alp \rp - \G\gam\bet\mu\nu \na\nu
\tor\gam\bet\alp, \ese

and adding with subtracting the expression
\bse \ba{l}
\Gu\gam\bet\nu\mu \curd\gam\bet\nu\alp - \frac{1}{2} \Gu\gam\bet\nu\mu \curd\alp\gam\bet\nu\\
\quad = \G\gam\bet\mu\nu \lp \cur\gam\bet\alp\nu + \frac{1}{2}
\cur\gam\alp\nu\bet \rp, \ea \ese in the right hand side of
\eqref{sec_02_01-57}, it acquires the form:
\bw
\be\label{sec_02_01-59} \ba{l}
 \lb \lag \kro\mu\alp - \sKjA\mu \na\alp \jfiA - \LjA\nu\mu \na\alp\na\nu \jfiA -\Gu\gam\bet\nu\mu \curd\gam\bet\nu\alp \rb + \lp \Du\mu\bet\gam\tau\eta\pi \spiu\tau\eta\pi \rp \tord\gam\bet\alp\\
\quad + \lb \sna\nu \lp \Gu\bet\gam\mu\nu \tord\bet\gam\alp \rp + \bfrac{1}{2} \lp \Gu\bet\gam\rho\sig \tord\bet\gam\alp \rp \tor\mu\rho\sig \rb\\
\quad + \lp \G\gam\bet\mu\nu + \LjA\nu\mu \DbrjAB\bet\gam \jfiB \rp
\lb \cur\gam\bet\alp\nu - \na\nu \tor\gam\bet\alp + \bfrac{1}{2}
\cur\gam\alp\nu\bet \rb. \ea \ee
\ew
\item At last, turn to the fourth item in the formula \eqref{sec_02_01-54}.
Substituting the expressions \eqref{sec_02_01-44},
\eqref{sec_02_01-31} and \eqref{sec_02_01-62} into the brackets
\bse [\dots] \Def \lb \met\kap\veps \tora\alp\veps\bet\gam + \na\kap
\meta\alp\bet\gam + \frac{1}{2} \cur\veps\alp\kap\lam
\metb\veps\lam\bet\gam \rb, \ese
one presents it as
\bw
\bse \ba{rl}
[\dots] & = -\met\kap\veps \lf \na\alp \tor\veps\bet\gam + \lp \tor\veps\lam\alp \tor\lam\bet\gam + \tor\veps\lam\bet \tor\lam\gam\alp + \tor\veps\lam\gam \tor\lam\alp\bet \rp \rf + 2\na\kap \tord{(\bet}{\gam)}\alp - \cur\veps\alp\kap\lam \met\veps{(\bet} \kro\lam{\gam)}\\
  & = -\met\kap\veps \lf -\na\bet \tor\veps\gam\alp - \na\gam \tor\veps\alp\bet + \cur\veps\alp\bet\gam + \cur\veps\bet\gam\alp + \cur\veps\gam\alp\bet \rf + \na\kap \tord\bet\gam\alp + \na\kap \tord\gam\bet\alp - \bfrac{1}{2} \curd\bet\alp\kap\gam - \bfrac{1}{2}
\curd\gam\alp\kap\bet, \ea \ese
where the Ricci identity $\cur\veps{[\alp}\bet{\gam]} \eq
\na{[\alp}\tor\veps\bet{\gam]} + \tor\veps\lam{[\alp}
\tor\lam\bet{\gam]}$ has been used. Regrouping  terms one obtains
%
%
\bse \ba{rl}
[\dots] & = \na\bet \tord\kap\gam\alp - \na\gam \tord\kap\bet\alp + \curd\alp\kap\bet\gam + \curd\bet\kap\gam\alp + \curd\kap\gam\bet\alp + \na\kap \tord\bet\gam\alp + \na\kap \tord\gam\bet\alp + \bfrac{1}{2} \curd\alp\bet\kap\gam + \bfrac{1}{2} \curd\alp\gam\kap\bet\\
 & = \lp 2\na{(\kap} \tord{\bet)}\gam\alp + \curd\alp{(\kap}{\bet)}\gam
\rp + \lp 2\na{[\kap} \tord{\gam]}\bet\alp -
\curd\alp{[\kap}{\gam]}\bet + \curd{[\kap}{\gam]}\bet\alp \rp -
\curd\kap\bet\gam\alp. \ea \ese
%
%
\item From the last, using the identities \eqref{sec_02_01-46} and \eqref{sec_02_01-47}, and the definition \eqref{app_02_a-01}, one gets
%
%
\bse \ba{rl}
\Dd\kap\bet\gam\pi\eta\nu [\dots] & = \Dd{(\kap}{\bet)}\gam\pi\eta\nu \lp 2\na\kap \tord\bet\gam\alp + \curd\alp\kap\bet\gam \rp + \Dd{[\kap|}\bet{|\gam]}\pi\eta\nu \lp 2\na\kap \tord\gam\bet\alp - \curd\alp\kap\gam\bet + \curd\kap\gam\bet\alp \rp - \Dd\kap\bet\gam\pi\eta\nu \curd\kap\bet\gam\alp\\
  & = \bfrac{1}{2} \kro\kap{(\pi} \kro\bet{\nu)} \kro\gam\eta \lp 2\na\kap \tord\bet\gam\alp + \curd\alp\kap\bet\gam \rp -\bfrac{1}{2} \kro\bet\eta \kro\kap{[\pi} \kro\gam{\nu]} \lp 2\na\kap \tord\gam\bet\alp - \curd\alp\kap\gam\bet + \curd\kap\gam\bet\alp \rp\\
 & \quad - \bfrac{1}{2} \lp \curd\eta\pi\nu\alp + \curd\nu\pi\eta\alp -
\curd\pi\eta\nu\alp \rp = -\curd\pi\eta\alp\nu + \na\nu
\tord\pi\eta\alp - \bfrac{1}{2} \curd\pi\alp\nu\eta. \ea \ese
\ew
\item Taking into account the result of the previous point,
one concludes that the fourth item in the formula \eqref{sec_02_01-54}
is equal to
\be\label{sec_02_01-60} \ba{l}
 -\lp \G\pi\eta\mu\nu + \LjA\nu\mu
\DbrjAB\eta\pi \jfiB \rp\\
\quad\times \lb \cur\pi\eta\alp\nu - \na\nu \tor\pi\eta\alp +
\bfrac{1}{2} \cur\pi\alp\nu\eta \rb. \ea \ee

Notice that this expression exactly equal (up to a sign) to the last
term in the formula \eqref{sec_02_01-59}.
\item Summing \eqref{sec_02_01-59} and \eqref{sec_02_01-60},
keeping in mind \eqref{sec_02_01-41} -- \eqref{sec_02_01-27},
\eqref{sec_02_01-45} and denoting
\en
\bw
\be\label{sec_02_01-61} \boxed{ \sem\mu\alp \Def \lag \kro\mu\alp -
\DIDnajfiA\mu \na\alp \jfiA - \dlagdnanajfiA\mu\nu \na\alp \na\nu
\jfiA - \Gu\bet\gam\veps\mu \curd\bet\gam\veps\alp, } \ee
one obtains the finalized expression for the tensor $\bfU$:
\be\label{sec_02_04-34} \boxed{ \U\alp\mu = \sem\mu\alp + \lp
\Du\mu\bet\gam\pi\rho\sig \spiu\pi\rho\sig \rp \tord\gam\bet\alp +
\frac{1}{2} \Gu\bet\gam\veps\mu \curd\alp\bet\gam\veps + \lb \sna\nu
\lp \Gu\bet\gam\mu\nu \tord\bet\gam\alp \rp + \frac{1}{2} \lp
\Gu\bet\gam\rho\sig \tord\bet\gam\alp \rp \tor\mu\rho\sig \rb. } \ee
\ew
Notice also that the tensor $\bfsem \Def \lf \sem\mu\nu \rf$
\eqref{sec_02_01-61} is just the \emph{generalized canonical
energy-momentum tensor}, corresponding to the Lagrangian
\eqref{sec_02_01-06}. This statement follows also from the results
of Sec. \ref{sec_02_05-00}. Namely, basing on the above definition
of the EMT, one obtains the standard equations of balance for
itself. Besides, the gravitational field equations acquire
the form, naturally generalizing the ECT equations. It is worse to
note that the sequence of the second derivative in the multiplier $\{
\na\alp \na\nu \jfiA\}$ in \eqref{sec_02_01-61} is \emph{reverse} to
the sequence that follows from the construction of the canonical
EMT by the direct application of the 1-st Noether theorem. The last
term in \eqref{sec_02_01-61} has appeared due to the non-minimal
coupling with the metric field, and also cannot be obtained \emph{in
principal} by the application of the 1-st Noether theorem in the
Minkowski space and covariantization of the expressions.
%

%
\section{A physical sense of the Klein and Noether identities}\label{sec_02_03-00}

\subsection{Structure of the variational derivatives}
In the following subsections of the present section, we discuss the
physical sense of the Klein and Noether identities in the manifestly
generally covariant theories. The identities include various
combinations of the variational derivatives of the action functional
$I$ with respect to fields $\bfmet$, $\bftor$ and $\bfffi$ as
ingredients. By this, it is useful to analyze in more details the
structure of such derivatives. At first, define the tensors:
\be\label{sec_02_04-07} \mspi\pi\rho\sig \Def (-2)\lp \sna\eta
\Gd\rho\sig\pi\eta + \frac{1}{2} \Gd\rho\sig\alp\bet \tor\pi\alp\bet
\rp; \ee

\be\label{sec_02_04-08} \ba{l}
 \fj\pi\rho\sig \Def 2 \DIDnajfiA\pi
\DbrjdAB\rho\sig \jfiB\\
\quad + 2\dlagdnanajfiA\pi\eta \lb \DbrjdAB\rho\sig \na\eta \jfiB -
\met\eta\rho \na\sig \jfiA \rb, \ea \ee
which are determined by the dependence of the Lagrangian $\lag$ on
the curvature tensor $\bfcur$ and on the fields $\bfjfi$,
respectively. It is worse to note that the sum of the tensor
\eqref{sec_02_04-07} and the antisymmetrical part of the tensor
\eqref{sec_02_04-08}: $\jspi\pi\rho\sig \Def \fj\pi{[\rho}{\sig]}$
presents the canonical ST \eqref{sec_02_01-51}:
\be\label{sec_02_04-18} \spi\pi\rho\sig = \mspi\pi\rho\sig +
\jspi\pi\rho\sig. \ee

Now, let us discuss the structure of the variational derivative with
respect to the torsion tensor. In the terms of the quantities
\eqref{sec_02_04-07} and \eqref{sec_02_04-08} the derivative
\eqref{sec_02_02-02} can be rewritten as
\be\label{sec_02_04-16} \DIDcon\kap\lam\tau = \frac{1}{2} \lp
\mspiud\tau\lam\kap + \fjud\tau\lam\kap \rp. \ee
Then, due to the identity \eqref{sec_02_04-09}, the variational
derivative \eqref{sec_02_04-10} is equal to
\be \DIDtor\veps\bet\gam = \DsIDtor\veps\bet\gam + \frac{1}{2} \lp
\Du\gam\bet\alp\pi\rho\sig \spiu\pi\rho\sig \rp \met\alp\veps, \ee
where the first term on the right hand side is defined by the
formula \eqref{sec_02_04-60}. Denote the quantity
\be\label{sec_02_04-29} \belu\gam\bet\alp \Def
\Du\gam\bet\alp\pi\rho\sig \spiu\pi\rho\sig;\qquad
\belu{[\gam}{\bet]}\alp = \belu\gam\bet\alp \ee
and call it as the \emph{Belinfante tensor: $\bfbel \Def \lf
\belu\gam\bet\alp \rf$, induced by the ST $\bfspi$}. Then
\be\label{sec_02_04-11} \DIDtor\veps\bet\gam = \DsIDtor\veps\bet\gam
+ \frac{1}{2} \belud\gam\bet\veps. \ee
This means \emph{that in the case of only minimal $\tor{}{}{}$-coupling
(when the Lagrangian $\lag$ does not contain the torsion tensor
$\bftor$ explicitly) one has}
\be\label{sec_02_04-11a} \DIDtor\veps\bet\gam = \frac{1}{2}
\belud\gam\bet\veps. \ee
Earlier, the same result \eqref{sec_02_00-12} has been proved only
for the Lagrangians of the type $\lag = \lag (\bfmet; \bfffi,
\bfna\bfffi)$ \eqref{sec_02_00-11} with a more simple presentations
both of the ST and of the Belinfante tensor (see Refs.~\cite{Trautman_1972_a, Trautman_1972_b, Trautman_1973_b,
Trautman_1980, Hehl_1973, Hehl_1974, Hehl_Heyde_Kerlick_Nester_1976}). We have proved a
more general claim: the formula \eqref{sec_02_00-12} is left valid
for the Lagrangians of a more general type $\lag = \lag (\bfmet,
\bfcur; \bfffi, \bfna\bfffi, \bfna\bfna\bfffi)$, as this follows
from \eqref{sec_02_04-11a} .

\emph{The formula \eqref{sec_02_04-11} shows that the presence of a
non-minimal coupling with torsion changes \eqref{sec_02_04-11a}}. The
requirement (the desire) to conserve a sense of the variational
derivative \eqref{sec_02_04-11a} even at the presence of a
non-minimal $\tor{}{}{}$-coupling leads to a necessity to modify
both the initial Belinfante tensor and the initial ST. Let us
demonstrate the modification step by step. Rewrite the formula
\eqref{sec_02_04-11} in form of \eqref{sec_02_04-11a}:
\be\label{sec_02_04-21} \boxed{ \DIDtor\veps\bet\gam = \frac12
\beliud\gam\bet\veps, } \ee
where the modified Belinfante tensor $\bfbeli = \{
\beliu\gam\bet\alp \}$ is defined analogously to the initial one
(that is with the use of \emph{any} ST):
\be\label{sec_02_04-41} \beliu\gam\bet\alp \Def
\Du\gam\bet\alp\pi\rho\sig \spiiu\pi\rho\sig. \ee
The \emph{modified } Belinfante tensor and canonical ST can be
represented as initial ones and correspondent additions:
\be\label{sec_02_04-41+} \beliu\gam\bet\alp \Def
 \belu\gam\bet\alp +
\belau\gam\bet\alp  \ee
and
\be\label{sec_02_04-53} { \spii\pi\rho\sig \Def \spi\pi\rho\sig +
\spia\pi\rho\sig. } \ee
Finally, combining \eqref{sec_02_04-11} - \eqref{sec_02_04-41}, one
obtains the definitions for the additional Belinfante tensor and ST:
\be\label{sec_02_04-13} \belaud\gam\bet\veps =
2\DsIDtor\veps\bet\gam; \ee
\be\label{sec_02_04-52} { \spiau\pi\rho\sig = -4 \metu{[\sig|}\veps
\DsIDtor\veps{|\rho]}\pi. } \ee

Now, let us turn to the variational derivative with respect to the
metric tensor $\bfmet$. By the formulae \eqref{sec_02_04-15} and
\eqref{sec_02_04-16},
\be \ba{l} \DIDmet\bet\gam = \frac12 \lag \metu\bet\gam +
\dslagdmet\bet\gam\\
\quad - \bfrac12 \sna\mu \lp \mspiu\tau\lam\pi + \fju\tau\lam\pi \rp
\Dd\mu{(\bet}{\gam)}\pi\lam\tau. \ea \ee
Using the identity \eqref{sec_02_01-50} and the formula
\eqref{sec_02_04-18}, rewrite the last as
\be\label{sec_02_04-19} \ba{l} \DIDmet\bet\gam = \frac12 \lag
\metu\bet\gam
+ \dslagdmet\bet\gam\\
\quad - \bfrac14 \sna\mu \lp \mspiu\mu\bet\gam + \fju\mu\bet\gam \rp
- \frac12 \sna\mu \belu\mu\bet\gam. \ea \ee
Substituting this expression into the standard definition of the
\emph{metric EMT} $\bfsemm \Def \{ \semmu\bet\gam \}$:
\be\label{sec_02_04-20} \boxed{ \bfrac{1}{2} \semmu\bet\gam \Def
\DIDmet\bet\gam }, \ee
one obtains
\bw
\be \boxed{ \semmu\bet\gam = \lag \metu\bet\gam +
2\dslagdmet\bet\gam - \bfrac12 \sna\mu \lp \mspiu\mu\bet\gam +
\fju\mu\bet\gam \rp - \sna\mu \belu\mu\bet\gam.} \ee
\ew

\subsection{The physical sense of the Noether identity}
Turn to the Noether identity
\eqref{sec_02_00-01}

\be\label{sec_02_04-24} \sna\mu \Ib\nu\mu \eq \Ia\nu. \ee
Here, corresponding to the identities \eqref{sec_02_04-62} and
\eqref{sec_02_04-61},
\bw
\be \Ib\nu\mu \Def \DIDgfiA \gfibA\nu\mu = \DIDmet\bet\gam
\metb\nu\mu\bet\gam + \DIDtor\veps\bet\gam \torb\nu\mu\veps\bet\gam
+ \DIDffiA \ffibA\nu\mu \Def \DIDmet\bet\gam \metb\nu\mu\bet\gam +
\DIDjfiA \jfibA\nu\mu; \ee
\be \Ia\nu \Def \DIDgfiA \gfiaA\nu = \DIDmet\bet\gam
\meta\nu\bet\gam + \DIDtor\veps\bet\gam \tora\nu\veps\bet\gam +
\DIDffiA \ffiaA\nu \Def \DIDmet\bet\gam \meta\nu\bet\gam + \DIDjfiA
\jfiaA\nu. \ee
\ew
Taking into account the formulae \eqref{sec_02_04-20},
\eqref{sec_02_01-44}, \eqref{sec_02_01-43}, \eqref{sec_02_01-31} and
\eqref{sec_02_01-32}, we rewrite these expressions as
\be\label{sec_02_04-22} \Ib\nu\mu = -\semm\mu\nu + \DIDjfiA
\DbrjAB\mu\nu \jfiB; \ee

\be\label{sec_02_04-23} \Ia\nu = \semm\mu\lam \tor\lam\mu\nu -
\DIDjfiA \lb \na\nu \jfiA + \DbrjAB\mu\lam \jfiB \; \tor\lam\mu\nu
\rb. \ee
Using the formulae \eqref{sec_02_04-21} and \eqref{app_02_c-11}, we
present the above tensors in the expanded form:
\bw
\be\label{sec_02_04-26} \Ib\nu\mu = -\lb \semm\mu\nu + \lp \frac12
\beliud\bet\gam\nu \tor\mu\bet\gam + \beliud\mu\bet\gam
\tor\gam\bet\nu \rp \rb + \DIDffiA \DbrfAB\mu\nu \ffiB; \ee
\be\label{sec_02_04-27} \ba{rl} \Ia\nu & = \lb \semm\mu\lam + \lp
\bfrac12 \beliud\bet\gam\lam \tor\mu\bet\gam +
\beliud\mu\bet\gam \tor\gam\bet\lam \rp \rb \tor\lam\mu\nu - \bfrac12 \beliud\gam\bet\veps \na\nu \tor\veps\bet\gam\\
 & \quad - \DIDffiA \lb \na\nu \ffiA + \DbrfAB\mu\lam \ffiB \;  \tor\lam\mu\nu \rb.
\ea \ee
Substituting the expressions \eqref{sec_02_04-22} and
\eqref{sec_02_04-23} into \eqref{sec_02_04-24}, we obtain the
explicit form of the Noether identity:
\be\label{sec_02_04-25} \boxed{ \sna\mu \semm\mu\nu \eq
-\semm\mu\lam \tor\lam\mu\nu + \lf \sna\mu \lb \DIDjfiA
\DbrjAB\mu\nu \jfiB \rb + \DIDjfiA \lb \na\nu \jfiA + \DbrjAB\mu\lam
\jfiB \;  \tor\lam\mu\nu \rb \rf. } \ee
\ew
After simplifying the Lagrangian $\lag = \lag (\bfmet, \bfcur; \;
\bftor, \bfna\bftor, \bfna\bfna\bftor; \; \bfffi, \bfna\bfffi,
\bfna\bfna\bfffi)$ to the form $\lag = \lag (\bfmet; \; \bfffi,
\bfna\bfffi)$ the identity \eqref{sec_02_04-25} degenerates to the
identity obtained in Refs.~\cite{Trautman_1972_a, Trautman_1972_b,
Trautman_1973_b, Trautman_1980, Hehl_1973, Hehl_1974, Hehl_Heyde_Kerlick_Nester_1976} and,
thus, generalizes the result of these works.

The formula \eqref{sec_02_04-25} shows that, when the equations of
motion for fields $\lf \bftor, \bfffi \rf = \bfjfi$ hold (on
$\bfjfi$-equations), the Noether identity transforms to the
\emph{equations of balance} for the metric EMT $\bfsemm$:
\be\label{sec_02_04-25a}  \boxed{ \sna\mu \semm\mu\nu =
-\semm\mu\lam \tor\lam\mu\nu \qquad\mbox{(on the
$\bfjfi$-equations)}. } \ee

\bw \noindent If one substitutes the expressions
\eqref{sec_02_04-26} and  \eqref{sec_02_04-27} into
\eqref{sec_02_04-24}, then, after taking into account the identity
\eqref{sec_02_03-07}, one obtains the Noether identity in the
expanded form:
%
%
\be \boxed{ \ba{l}
\sna\mu \lp \semm\mu\nu + \sna\eta \beliud\eta\mu\nu \rp \eq -\lp \semm\mu\lam + \sna\eta \beliud\eta\mu\lam \rp \tor\lam\mu\nu + \bfrac12 \spii\pi\rho\sig \curud\rho\sig\pi\nu\\
\qquad +\lf \sna\mu \lb \DIDffiA \DbrfAB\mu\nu \ffiB \rb + \DIDffiA
\lb \na\nu \ffiA + \DbrfAB\mu\lam \ffiB \;  \tor\lam\mu\nu \rb \rf.
\ea } \ee
\ew
It is clearly that on the equations of motion of $\lf \bftor, \bfffi
\rf = \bfjfi$-fields this equation turns again to
\eqref{sec_02_04-25a}. However, on the equations of motion of
\emph{only $\bfffi$-fields} the expanded equations of balance
acquire the form:

\bw
\be \boxed{ \sna\mu \lp \semm\mu\nu + \sna\eta \beliud\eta\mu\nu \rp
= -\lp \semm\mu\lam + \sna\eta \beliud\eta\mu\lam \rp \tor\lam\mu\nu
+ \bfrac12 \spii\pi\rho\sig \curud\rho\sig\pi\nu \qquad\mbox{(on the
$\bfffi$-equations)}. } \ee
\ew
\emph{Thus, the Noether identity is the basis for defining the
equations of balance for the metric EMT.}

\subsection{The 4-th and 3-rd Klein identities}
Notice that, by the definition \eqref{sec_02_01-35}, the tensor
$\bfG \Def \{ \G\kap\lam\mu\nu \}$ has the same symmetries, like the
curvature tensor. Using the definition of the
tensor $\bfN$ \eqref{sec_02_01-34} and the antisymmetry of $\bfG$ in the second pare of
indexes, $\G\alp\bet\del\gam = -\G\alp\bet\gam\del$, we are
convinced that \emph{the 4-th Klein identity \eqref{sec_02_00-04}:
\be \label{sec_02_04-25b}\N\alp{(\bet}\gam{\del)} = \frac13 \lp
\N\alp\bet\gam\del + \N\alp\gam\del\bet + \N\alp\del\bet\gam \rp \eq
0 \ee
is satisfied automatically}. By the antisymmetry of $\bfG$ in
the first pare of indexes, $\Gu\bet\alp\gam\del =
-\Gu\alp\bet\gam\del$, \emph{the tensor $\bfN$ satisfies also the
new identity}:
\be \label{sec_02_04-25c}{\Nud{(\alp}\bet{\gam)}\del = \frac13 \lp
\Nud\alp\bet\gam\del +  \Nud\bet\gam\alp\del + \Nud\gam\alp\bet\del
\rp \eq 0. } \ee
In the case of a pure metric theory another logic leads also to this
conclusion. In the Riemannian geometry $\rsR (1,D)$ (but not in the
Riemann-Cartan geometry $\rsC (1,D)$!) the tensor $\{
\Gu\alp\bet\gam\del \}$ is symmetrical with respect to the
permutation of the first and the second pairs of indexes: $\{
\Gu\gam\del\alp\bet = \Gu\alp\bet\gam\del \}$, like the curvature
tensor  $\{ \curd\alp\bet\gam\del \}$. Then, the tensor $\bfN$
becomes also symmetrical in external indexes, $\Nu\alp\bet\gam\del =
\Nu\del\bet\gam\alp$. Namely this property together with
\eqref{sec_02_04-25b} gives \eqref{sec_02_04-25c}.

In arbitrary generally covariant theories with the Lagrangians
$\lag$, containing derivatives of the metric up to a second order, see Refs.~\cite{Rund_1966, Lovelock_1971, Petrov_2004_b_en, Petrov_2008,
Petrov_2009_a, Petrov_2010_a, Petrov_2011, Petrov_Lompay_2013}, the
quantity $\bf n \Def \{n^{\alp\mu\nu\bet}\}$ is defined as
$n^{\alp\mu\nu\bet}\Def \pa{} {\lag}/ \pa{} g_{\mu\nu,\alp\bet}$.
The same as the tensor $\bfN$, it satisfies the identity of the type
\eqref{sec_02_04-25b}. Then, because $\bf n$ is symmetrical both in
inner and in external indexes, it satisfies also the identity of the type
\eqref{sec_02_04-25c}. Thus, our conclusions related to the
properties of $\bfN$ in the manifestly generally covariant theories
generalize the results of the aforementioned works.

Now, let us turn to the 3-rd Klein identity \eqref{sec_02_00-03}
\be \M\kap{(\lam}{\mu)} + \sna\nu \N\kap\lam\mu\nu +
\N\kap{(\lam|}\rho\sig \tor{|\mu)}\rho\sig \eq 0. \ee
Taking into account \eqref{sec_02_04-29}, and \eqref{sec_02_01-34},
calculate the symmetrical in the upper indexes part of the tensor
$\bfM$ \eqref{sec_02_04-28}: \be \M\kap{(\lam}{\mu)} = -\sna\nu
\N\kap\lam\mu\nu -\N\kap{(\lam|}\rho\sig \tor{|\mu)}\rho\sig. \ee
From here it follows that \emph{the 3-rd Klein identity is satisfied
automatically also}. Thus, the concrete form of the Lagrangian
\eqref{sec_02_01-06} is enough to be convinced in the identities
\eqref{sec_02_00-03} - \eqref{sec_02_00-04}.

\subsection{The physical sense of the 2-nd Klein identity}
It is convenient to represent the 2-nd Klein identity
\eqref{sec_02_00-02} in the form:
\be\label{sec_02_04-38} \Us\nu\mu \eq -\Ib\nu\mu \ee
(see the Paper~I, Sec. V, formulae  (83) and (84)), where
\be\label{sec_02_04-36} \ba{rl} \Us\nu\mu & = \lp \U\nu\mu - \bfrac13
\N\alp\mu\bet\gam \cur\alp\nu\bet\gam \rp\\
 & \quad -  \lp \sna\lam
\potb\nu\mu\lam + \bfrac12 \potb\nu\alp\bet \tor\mu\alp\bet \rp; \ea
\ee
\be\label{sec_02_04-64} \potb\nu\mu\lam \Def -\M\nu{[\mu}{\lam]} +
\frac23 \lp \sna\eta \N\nu\eta{[\mu}{\lam]} + \frac12
\tor{[\mu}\rho\sig \N\nu{\lam]}\rho\sig \rp. \ee
Let us calculate the expression \eqref{sec_02_04-36}.
\bn
\item At first, notice that, by \eqref{sec_02_01-34}, the relation
\be\label{sec_02_04-59} \N\nu\eta{[\mu}{\lam]} = \frac12 \lp
\G\nu\eta\mu\lam - \G\nu{[\mu}{\lam]}\eta \rp \ee
takes a place. From the last the formulae
\be\label{sec_02_04-31} \ba{l} \G\nu{[\mu}{\lam]}\eta + \bfrac23
\N\nu\eta{[\mu}{\lam]}\\
\quad = \bfrac13 \lp \G\nu\mu\lam\eta + \G\nu\lam\eta\mu +
\G\nu\eta\mu\lam \rp = \G\nu{[\mu}\lam{\eta]}; \ea \ee
\be\label{sec_02_04-32} \frac12 \lp \G\nu\mu\rho\sig +
\G\nu{[\rho}{\sig]}\mu \rp - \frac13 \N\nu\mu{[\rho}{\sig]} =
\G\nu{[\mu}\rho{\sig]} \ee
follow. It is useful also the formula, which follows after
contracting \eqref{sec_02_04-31} with the Riemannian tensor:
\be\label{sec_02_04-31a} \frac12 \Gu\alp\bet\gam\mu
\curd\nu\alp\bet\gam - \bfrac13 \N\alp\mu\bet\gam
\cur\alp\nu\bet\gam =  -\bfrac12 \G\alp{[\mu}\rho{\sig]}
\cur\alp\nu\rho\sig. \ee
\item Substituting into the right hand side of \eqref{sec_02_04-64}
the expressions of the tensors $\bfM$ and $\bfN$,
\eqref{sec_02_04-28} and  \eqref{sec_02_01-34}, and using the
definition \eqref{sec_02_04-29} and the relations
\eqref{sec_02_04-31} and \eqref{sec_02_04-32}, one finds
\be\label{sec_02_05-05} \ba{l} \potb\nu\mu\lam = \lb
-\belud\mu\lam\nu +
\G\alp\bet\mu\lam \tor\alp\bet\nu \rb\\
\quad + \lb \sna\eta \G\nu{[\mu}\lam{\eta]} + \frac12 \lp
\G\nu{[\mu}\rho{\sig]} \tor\lam\rho\sig - \G\nu{[\lam}\rho{\sig]}
\tor\mu\rho\sig \rp \rb. \ea \ee
\item Substituting \eqref{sec_02_05-05} into the expression
in the second parentheses in \eqref{sec_02_04-36}, one gets
\bw
\be\label{sec_02_04-33} \ba{l}
\sna\lam \potb\nu\mu\lam + \bfrac12 \potb\nu\alp\bet \tor\mu\alp\bet\\
\quad = -\lb \sna\lam \belud\mu\lam\nu + \bfrac12 \belud\rho\sig\nu \tor\mu\rho\sig \rb + \lb \sna\lam \lp \Gu\alp\bet\mu\lam \tord\alp\bet\nu \rp + \bfrac12 \lp \Gu\alp\bet\rho\sig \tord\alp\bet\nu \rp \tor\mu\rho\sig \rb\\
\qquad -\bfrac12 \G\alp{[\mu}\rho{\sig]} \cur\alp\nu\rho\sig +\lb
\G\nu\pi\rho\sig \lp \cur\mu{[\pi}\rho{\sig]} - \na{[\pi}
\tor\mu\rho{\sig]} - \tor\mu\veps{[\pi} \tor\veps\rho{\sig]} \rp
\rb. \ea \ee
Here, the use of the identity
\be\label{sec_02_05-18} \sna\lam \lb \sna\eta \tpotc\nu\mu\lam\eta +
\bfrac{1}{2} \tpotc\nu\mu\rho\sig \tor\lam\rho\sig \rb \eq
\bfrac{1}{2} \lp -\cur\lam\nu\rho\sig \tpotc\lam\mu\rho\sig +
\cur\mu\lam\rho\sig \tpotc\nu\lam\rho\sig \rp \ee
\ew
has been essential (see the Paper~I, Appendix C.1, formula (C3)),
with the exchange:
\bse\tpotc\nu\mu\lam\eta = \G\nu{[\mu}\lam{\eta]}. \ese
\item By the Ricci identity:
$ \cur\mu{[\pi}\rho{\sig]} \eq \na{[\pi} \tor\mu\rho{\sig]} -
\tor\mu\veps{[\pi} \tor\veps\rho{\sig]} $, the last item on the
right hand side of the formula \eqref{sec_02_04-33} disappears and
it acquires the form:
\bw
\be\label{sec_02_04-33a} \ba{l}
\sna\lam \potb\nu\mu\lam + \bfrac12 \potb\nu\alp\bet \tor\mu\alp\bet\\
\quad = -\lb \sna\lam \belud\mu\lam\nu + \bfrac12 \belud\rho\sig\nu
\tor\mu\rho\sig \rb + \lb \sna\lam \lp \Gu\alp\bet\mu\lam
\tord\alp\bet\nu \rp + \bfrac12 \lp \Gu\alp\bet\rho\sig
\tord\alp\bet\nu \rp \tor\mu\rho\sig \rb -\bfrac12
\G\alp{[\mu}\rho{\sig]} \cur\alp\nu\rho\sig . \ea \ee
\ew
\item Substituting into the first parentheses in \eqref{sec_02_04-36}
the expression for the tensor $\bfU$ \eqref{sec_02_04-34} and using
\eqref{sec_02_04-31a}, one obtains
\be\label{sec_02_04-35} \ba{l}
 \U\nu\mu - \bfrac13 \N\alp\mu\bet\gam \cur\alp\nu\bet\gam\\
\quad = \sem\mu\nu + \belu\mu\bet\gam \tord\gam\bet\nu - \bfrac12 \G\alp{[\mu}\rho{\sig]} \cur\alp\nu\rho\sig\\
\qquad +\lb \sna\eta \lp \Gu\bet\gam\mu\eta \tord\bet\gam\nu \rp +
\bfrac12 \lp \Gu\bet\gam\rho\sig \tord\bet\gam\nu \rp
\tor\mu\rho\sig \rb, \ea \ee

\item Finally, substituting \eqref{sec_02_04-33a} and \eqref{sec_02_04-35}
into the right hand side of \eqref{sec_02_04-36}, one finds the
search expression:
\en
\be\label{sec_02_04-35a} \Us\nu\mu = \sem\mu\nu + \lb \sna\lam
\belud\mu\lam\nu + \frac12 \belud\rho\sig\nu \tor\mu\rho\sig +
\belu\mu\bet\gam \tord\gam\bet\nu \rb. \ee
At the absence of the torsion, the right hand side of this
expression presents the known expression for the Belinfante
symmetrized EMT  \cite{Belinfante_1939, Belinfante_1940}. Therefore,
when the torsion presents it is naturally to define the right hand
side of \eqref{sec_02_04-35a} as \emph{symmetrized
EMT} $\bfsems \Def \{ \sems\mu\nu \}$:
\be\label{sec_02_04-40} \boxed{ \ba{rl}
\sems\mu\nu & \Def \sem\mu\nu + \lb \sna\lam \belud\mu\lam\nu + \bfrac12 \belud\rho\sig\nu \tor\mu\rho\sig + \belud\mu\bet\alp \tor\alp\bet\nu \rb\\
 & = \sem\mu\nu - \lb \sna\lam \belud\lam\mu\nu + \bfrac12 \belud\gam\bet\alp \Dbrt\mu\nu\alp\bet\gam\tet\vphi\xi \tor\tet\vphi\xi
 \rb,
\ea } \ee
and, in the result one obtains
\be\label{sec_02_04-37} \Us\nu\mu = \sems\mu\nu. \ee
Substituting \eqref{sec_02_04-37} and \eqref{sec_02_04-22} into the
2-nd Klein identity \eqref{sec_02_04-38}, one has

\be\label{sec_02_04-39} \boxed{ \sems\mu\nu \eq \semm\mu\nu -
\DIDjfiA \DbrjAB\mu\nu \jfiB. } \ee
From the identity \eqref{sec_02_04-39} it follows that \emph{on the
$\bfjfi$-equations symmetrized EMT $\bfsems$ \eqref{sec_02_04-40}
is equal to the metric EMT $\bfsemm$ \eqref{sec_02_04-20}:}
\be\label{sec_02_04-66} \boxed{ \sems\mu\nu =
\semm\mu\nu\qquad\mbox{(on the $\bfjfi$-equations)}. } \ee
On the other hand, if on the right hand side of \eqref{sec_02_04-38}
instead of \eqref{sec_02_04-22} one uses \eqref{sec_02_04-26} then,
keeping in mind \eqref{sec_02_04-37}, \eqref{sec_02_04-40} and
\eqref{sec_02_04-41}, the 2-nd Klein identity can be presented in
the other form:
\bw
\be \boxed{ \sem\mu\nu - \sna\lam \belud\lam\mu\nu \eq \semm\mu\nu +
\lp \frac12 \belaud\rho\sig\nu \tor\mu\rho\sig + \belaud\mu\bet\alp
\tor\alp\bet\nu \rp - \DIDffiA \DbrfAB\mu\nu \ffiB. } \ee
In particular, on $\bfffi$-equations one has
\be \boxed{ \sem\mu\nu - \sna\lam \belud\lam\mu\nu = \semm\mu\nu +
\lp \frac12 \belaud\rho\sig\nu \tor\mu\rho\sig + \belaud\mu\bet\alp
\tor\alp\bet\nu \rp \qquad\mbox{(on the $\bfffi$-equations)}. } \ee
\ew
Finalizing, one can conclude that \emph{the 2-nd Klein identity permits
to define the Belinfante symmetrized EMT $\bfsems$} (see
\eqref{sec_02_04-40}), \emph{and to prove an equivalence of the
symmetrized $\bfsems$ and metric $\bfsemm$ EMTs on $\bfjfi$-fields
equations} (see \eqref{sec_02_04-66}).
\subsection{The physical sense of the 1-st Klein identity}
We call the identity
\bw
\be\label{sec_02_04-45} \sna\mu \U\nu\mu - \frac12 \M\mu\rho\sig
\cur\mu\nu\rho\sig -  \frac13 \N\kap\lam\rho\sig \lp \na\lam
\cur\kap\nu\rho\sig + \frac12 \cur\kap\nu\lam\veps \tor\veps\rho\sig
\rp \eq - \Ia\nu \ee
\ew
as the 1-st Klein identity (see the Paper~I, Sec. III, formula
(38)). Calculate the left hand side of this for the manifestly
generally covariant theories.
\bn
\item\label{sec_02_04-42} Using the expression for the tensor
$\bfU$ \eqref{sec_02_04-34}, the definition \eqref{sec_02_04-29},
the identity
\be\label{sec_02_05-17} \sna\mu \lb \sna\eta \tpotb\nu\mu\eta +
\bfrac{1}{2}\tpotb\nu\rho\sig \tor\mu\rho\sig \rb \eq -\bfrac{1}{2}
\cur\lam\nu\rho\sig \tpotb\lam\rho\sig, \ee
(see the Paper~I, Appendix C.1, formula (C2)) and $\tpotb\nu\mu\eta
= \Gu\alp\bet\mu\eta \tord\alp\bet\nu$, one obtains the expression
for the first item in
\eqref{sec_02_04-45}:
\bw
\bse \sna\mu \U\nu\mu = \sna\mu \sem\mu\nu + \sna\mu \lp
\belud\mu\bet\alp \tor\alp\bet\nu \rp - \bfrac12 \lp \sna\mu
\G\alp\rho\sig\mu \rp \cur\alp\nu\rho\sig - \bfrac12
\G\alp\bet\gam\mu \lp \na\mu \cur\alp\nu\bet\gam \rp - \bfrac12 \lp
\Gu\alp\bet\rho\sig \tord\alp\bet\lam \rp \cur \lam\nu\rho\sig. \ese
\ew
\item\label{sec_02_04-43} The expression
for the second item in \eqref{sec_02_04-45} can be obtained after
using the formulae for $\bfM$ \eqref{sec_02_04-28} and the
definition  \eqref{sec_02_04-29}:
\bse
\ba{l}
-\bfrac12 \M\mu\rho\sig \cur\mu\nu\rho\sig\\
= -\bfrac12 \belu\sig\rho\pi \curd\nu\pi\rho\sig + \bfrac12 \lp \sna\eta \G\mu\rho\sig\eta \rp \cur\mu\nu\rho\sig\\
 \quad + \bfrac14 \lp \G\mu\rho{[\alp}{\bet]} + \G\mu{[\alp}{\bet]}\rho \rp \lb \cur\mu\nu\rho\sig \tor\sig\alp\bet \rb\\
 \quad + \bfrac12 \lp \Gu\alp\bet\rho\sig \tord\alp\bet\mu \rp \cur\mu\nu\rho\sig.
\ea
\ese
\item Combining the results of the points \ref{sec_02_04-42}
and \ref{sec_02_04-43}, taking into account \eqref{sec_02_04-31} and
\eqref{sec_02_04-32}, and using the Bianchi identity $ \na{[\pi|}
\cur\mu\nu{|\rho}{\sig]} \eq - \cur\mu\nu\veps{[\pi}
\tor\veps\rho{\sig]}$, one gets
\be\label{sec_02_04-44} \ba{l}
\sna\mu \U\nu\mu - \bfrac12 \M\mu\rho\sig \cur\mu\nu\rho\sig\\
\quad - \bfrac13 \N\kap\lam\rho\sig \lp \na\lam \cur\kap\nu\rho\sig + \bfrac12 \cur\kap\nu\lam\veps \tor\veps\rho\sig \rp\\
=\sna\mu \sem\mu\nu - \sna\mu \lp \belud\mu\bet\alp \tor\alp\bet\nu
\rp -\bfrac12 \belu\sig\rho\pi \curd\nu\pi\rho\sig. \ea \ee
Substituting the expressions \eqref{sec_02_04-44} and
\eqref{sec_02_04-23} into \eqref{sec_02_04-45}, one finds the
explicit form of the 1-st Klein identity:
\be\label{sec_02_04-46} \ba{rl}
\sna\mu \sem\mu\nu & \eq -\lp \semm\mu\lam + \sna\eta \belud\eta\mu\lam \rp \tor\lam\mu\nu\\
 & \quad + \bfrac12 \belu\sig\rho\pi \curd\nu\pi\rho\sig - \belud\eta\mu\lam \na\eta \tor\lam\mu\nu\\
 & \quad + \DIDjfiA \lb \na\nu \jfiA + \DbrjAB\mu\lam \jfiB \;  \tor\lam\mu\nu \rb.
\ea \ee
\en Transform this identity as follows.
\bn
\item Using the 2-nd Klein identity \eqref{sec_02_04-39}
and the formula \eqref{sec_02_04-40}, transform the combination
$(\bfsemm + \bfsna \bfbel )$:
\bse \ba{rl} \semm\mu\lam + \sna\eta \belud\eta\mu\lam & \eq
\sem\mu\lam - \frac12 \belud\gam\bet\alp
\Dbrt\mu\lam\alp\bet\gam\tet\vphi\xi
\tor\tet\vphi\xi\\
 & \quad + \DIDjfiA \DbrjAB\mu\lam \jfiB.
\ea \ese
\item Substitute the above expression into \eqref{sec_02_04-46} and collect
the similar terms, then the terms containing the Belinfante tensor
$\bfbel$ are presented by the combination
\bw
\bse \ba{l}
\bfrac12 \belud\gam\bet\alp \lb \lp \na\bet \tor\alp\gam\nu + \na\gam \tor\alp\nu\bet \rp + \lp \tor\alp\veps\nu \tor\veps\bet\gam + \tor\alp\veps\bet \tor\veps\gam\nu + \tor\alp\veps\gam \tor\veps\nu\bet \rp - \cur\alp\nu\bet\gam \rb\\
\quad = \bfrac12 \belud\gam\bet\alp \lb -\na\nu \tor\alp\bet\gam +
\cur\alp\bet\gam\nu + \cur\alp\gam\nu\bet \rb = \bfrac12
\belud\gam\bet\alp \na\nu \tor\alp\bet\gam + \belu\gam{[\bet}{\alp]}
\curd\alp\bet\gam\nu = \bfrac12 \belud\gam\bet\alp \na\nu
\tor\alp\bet\gam + \bfrac12 \spi\pi\rho\sig \curud\rho\sig\pi\nu, \ea
\ese
\ew
where at the first equality the Ricci identity has been used,
whereas at the third equality the definition \eqref{sec_02_04-29}
and identity \eqref{sec_02_04-14} have been taken into account.
\en
After all the above steps the 1-st Klein identity
\eqref{sec_02_04-46} acquires the form:
\bw
\be\label{sec_02_04-47} \boxed{ \sna\mu \sem\mu\nu \eq - \sem\mu\lam
\tor\lam\mu\nu + \bfrac12 \spi\pi\rho\sig \curud\rho\sig\pi\nu -
\bfrac12 \belud\gam\bet\alp \na\nu \tor\alp\bet\gam + \DIDjfiA
\na\nu \jfiA } \ee
or, in the decomposed form:
\be\label{sec_02_04-48} \boxed{ \sna\mu \sem\mu\nu \eq - \sem\mu\lam
\tor\lam\mu\nu + \bfrac12 \spi\pi\rho\sig \curud\rho\sig\pi\nu +
\bfrac12 \belaud\gam\bet\alp \na\nu \tor\alp\bet\gam + \DIDffiA
\na\nu \ffiA. } \ee
From \eqref{sec_02_04-47} and \eqref{sec_02_04-48} the
\emph{equations of balance for the canonical EMT $\bfsem$} follow
\be \boxed{ \sna\mu \sem\mu\nu = - \sem\mu\lam \tor\lam\mu\nu +
\bfrac12 \spi\pi\rho\sig \curud\rho\sig\pi\nu - \bfrac12
\belud\gam\bet\alp \na\nu \tor\alp\bet\gam \qquad\mbox{(on the
$\bfjfi$-equations)}; } \ee
and
\be\label{sec_02_04-50} \boxed{ \sna\mu \sem\mu\nu = - \sem\mu\lam
\tor\lam\mu\nu + \bfrac12 \spi\pi\rho\sig \curud\rho\sig\pi\nu +
\bfrac12 \belaud\gam\bet\alp \na\nu \tor\alp\bet\gam \qquad\mbox{(on
the $\bfffi$-equations)}. } \ee
\ew

In Refs.~\cite{Trautman_1972_a, Trautman_1972_b, Trautman_1973_b,
Trautman_1980, Hehl_1973, Hehl_1974, Hehl_Heyde_Kerlick_Nester_1976}, for the Lagrangian
of the type $\lag = \lag (\bfmet; \; \bfffi, \bfna\bfffi)$ the
equation of balance for the canonical EMT \eqref{sec_02_00-13} has
been obtained. The result \eqref{sec_02_00-13} is left valid in a
more general case also, when the Lagrangian has a form: $\lag = \lag
(\bfmet, \bfcur; \; \bfffi, \bfna\bfffi, \bfna\bfna\bfffi)$ because
the last term in \eqref{sec_02_04-50} does not appear. In the case
of non-minimal $\bftor$-coupling the right hand side of
\eqref{sec_02_04-50} contains \emph{additional} term $\lp \frac12
\belaud\gam\bet\alp \na\nu \tor\alp\bet\gam \rp$. However, the
\emph{new} equation \eqref{sec_02_04-50} can be also transformed to
the form \eqref{sec_02_00-13}. For this, using the identity \eqref{sec_02_03-07} in the last term in
\eqref{sec_02_04-48}, one obtains
\bw
\be \ba{l}
\sna\mu \lb \sem\mu\nu - \lp \sna\lam \belaud\mu\lam\nu + \bfrac12 \belaud\kap\lam\nu \tor\mu\kap\lam + \belaud\mu\lam\kap \tor\kap\lam\nu \rp \rb\\
\quad \eq - \lb \sem\mu\lam - \lp \sna\eta \belaud\mu\eta\lam +
\bfrac12 \belaud\kap\eta\lam \tor\mu\kap\eta + \belaud\mu\eta\kap
\tor\kap\eta\lam \rp \rb \tor\lam\mu\nu + \bfrac12 \spii\pi\rho\sig
\curud\rho\sig\pi\nu + \DIDffiA \na\nu \ffiA. \ea \ee
\ew
\noindent Here, the expression in the brackets we denote as
the \emph{modified canonical EMT} $\bfsemi \Def \{ \semi\mu\nu \}$:

\be\label{sec_02_04-54} \semi\mu\nu \Def \sem\mu\nu + \sema\mu\nu,
\ee
where the \emph{additional EMT} $\bfsema \Def \{ \sema\mu\nu \}$ is
defined as
\bw
\be\label{sec_02_04-56} \sema\mu\nu \Def - \lp \sna\lam
\belaud\mu\lam\nu + \bfrac12 \belaud\kap\lam\nu \tor\mu\kap\lam +
\belaud\mu\lam\kap \tor\kap\lam\nu \rp = \sna\lam \belaud\lam\mu\nu
+ \bfrac12 \belaud\gam\bet\alp \Dbrt\mu\nu\alp\bet\gam\tet\vphi\xi
\tor\tet\vphi\xi. \ee
\ew
Note that this modification is analogous to the modification of the
canonical ST in \eqref{sec_02_04-53} and \eqref{sec_02_04-52}. It is
evidently that the canonical EMT $\bfsemi$ \eqref{sec_02_04-54} in
the case of minimal $\tor{}{}{}$-coupling only transforms to (usual)
canonical EMT $\bfsem$. By the definition \eqref{sec_02_04-56}, The
Belinfante symmetrization of the type \eqref{sec_02_04-40} applied
to $\bfsema$ leads to zero identically. Therefore, the
\emph{symmetrized EMT $\bfsems$} constructed by the symmetrization
of $\bfsemi$ with the use of $\bfbeli$ by the rule
\bw
\be\label{sec_02_04-65} \ba{rl}
\sems\mu\nu & = \semi\mu\nu + \lb \sna\lam \beliud\mu\lam\nu + \bfrac12 \beliud\kap\lam\nu \tor\mu\kap\lam + \beliud\mu\lam\kap \tor\kap\lam\nu \rb\\
 & = \semi\mu\nu - \lb \sna\lam \beliud\lam\mu\nu + \bfrac12 \beliud\gam\bet\alp \Dbrt\mu\nu\alp\bet\gam\tet\vphi\xi \tor\tet\vphi\xi \rb
\ea \ee
\ew
\emph{exactly coincides} with the (usual) symmetrized EMT $\bfsems$
\eqref{sec_02_04-40}.

In the terms of the modified canonical EMT $\bfsemi$ the identities
\eqref{sec_02_04-47} and \eqref{sec_02_04-48} can be rewritten as
\be \boxed{ \ba{rl} \sna\mu \semi\mu\nu & \eq - \semi\mu\lam
\tor\lam\mu\nu +
\bfrac12 \spii\pi\rho\sig \curud\rho\sig\pi\nu\\
 & \quad - \bfrac12 \beliud\gam\bet\alp \na\nu \tor\alp\bet\gam + \DIDjfiA \na\nu \jfiA
\ea }\ee
and
\be\label{sec_02_05-22} \boxed{ \ba{rl} \sna\mu \semi\mu\nu & \eq -
\semi\mu\lam \tor\lam\mu\nu + \bfrac12 \spii\pi\rho\sig
\curud\rho\sig\pi\nu\\
 & \quad + \DIDffiA \na\nu \ffiA.
\ea } \ee
They are the basis for the \emph{equations of balance for the
modified canonical EMT} $\bfsemi$:

\be \boxed{ \ba{l} \sna\mu \semi\mu\nu = - \semi\mu\lam
\tor\lam\mu\nu +
\bfrac12 \spii\pi\rho\sig \curud\rho\sig\pi\nu\\
\quad - \bfrac12 \beliud\gam\bet\alp \na\nu \tor\alp\bet\gam
\qquad\mbox{(on the $\bfjfi$-equations)} \ea} \ee
and
\be\label{sec_02_04-55} \boxed{ \ba{r} \sna\mu \semi\mu\nu = -
\semi\mu\lam
\tor\lam\mu\nu + \bfrac12 \spii\pi\rho\sig \curud\rho\sig\pi\nu\\
\mbox{(on the $\bfffi$-equations)}. \ea} \ee
Now, the equation \eqref{sec_02_04-55} has the same structure
as the equation \eqref{sec_02_00-13}. Also notice that, when the
equations for the torsion field $\Del I/ \Del\bftor = 0$ hold, then
\be \semi\mu\nu = \sems\mu\nu \qquad\mbox{(on the
$\bftor$-equations)}, \ee
as it follows from \eqref{sec_02_04-65} and \eqref{sec_02_04-21}.

At last, let us find the identities and the equations of balance for
the symmetrized EMT $\bfsems$. Use \eqref{sec_02_04-40} for
rewriting $\bfsem$ as a function of $\bfsems$ and $\bfbel$,
substitute the result into \eqref{sec_02_04-47} and
\eqref{sec_02_04-48} and find, respectively,
\be\label{sec_02_04-57} \boxed{ \sna\mu \sems\mu\nu \eq
-\sems\mu\lam \tor\lam\mu\nu + \DIDjfiA \na\nu \jfiA;} \ee
and
\be\label{sec_02_04-58} \boxed{ \ba{l} \sna\mu \sems\mu\nu \eq
-\sems\mu\lam \tor\lam\mu\nu\\
\quad + \bfrac12 \beliud\gam\bet\alp \na\nu \tor\alp\bet\gam +
\DIDffiA \na\nu \ffiA. \ea} \ee
The same identities can be obtained by another way. Namely, express
the metric EMT $\bfsemm$ through the symmetrized EMT $\bfsems$ from
the 2-nd Klein identity \eqref{sec_02_04-39} and substitute the
result into the Noether identity \eqref{sec_02_04-25}.

Next, the \emph{equations of balance for the symmetrized EMT}
$\bfsems$, which follow from the identities \eqref{sec_02_04-57} and
\eqref{sec_02_04-58} are
\be \boxed{ \sna\mu \sems\mu\nu = -\sems\mu\lam \tor\lam\mu\nu
\qquad\mbox{(on the $\bfjfi$-equations)} } \ee
and
\be\label{sec_02_04-63} \boxed{ \ba{l} \sna\mu \sems\mu\nu =
-\sems\mu\lam
\tor\lam\mu\nu\\
\quad + \bfrac12 \beliud\gam\bet\alp \na\nu \tor\alp\bet\gam \qquad
\mbox{(on the $\bfffi$-equations)}. \ea} \ee

Finalizing subsection, one concludes that \emph{the 1-st Klein
identity is the basis for constructing the equations of balance for
the canonical EMT $\bfsem$.} These relations coincide with the known
(standard) ones, when a non-minimal coupling with torsion is absent.
When a non-minimal coupling with torsion presents the canonical EMT $\bfsem$ is changed by $\bfsemi$ with the use of the modified
Belinfante tensor, and then the equations of balance for the
modified EMT $\bfsemi$ acquire the standard form again. Also, the
\emph{1-st Klein identity, as well as the 2-nd one, are the basis
for constructing the equations of balance for the symmetrized EMT $\bfsems$}.

%
\section{The generalized superpotential and Noether current }\label{sec_02_04-00}

\subsection{The calculation of the superpotential}
At first, let us calculate the generalized superpotential
$\bfpotdpara = \{ \potdpara{[\mu}{\nu]} = \potdpara\mu\nu \}$ in the
explicit form. Corresponding the formula \eqref{sec_02_00-06}, one
has
\be \potdpara\mu\nu = \potb\alp\mu\nu \dpara\alp +
\potc\alp\bet\mu\nu \na\bet \dpara\alp, \ee
where
\be \potb\alp\mu\nu = -\M\alp{[\mu}{\nu]} + \frac23 \lp \sna\lam
\N\alp\lam{[\mu}{\nu]} + \frac12 \tor{[\mu}\rho\sig
\N\alp{\nu]}\rho\sig \rp; \ee
\be \potc\alp\bet\mu\nu = -\frac43 \N\alp\bet{[\mu}{\nu]}. \ee
In fact, we have calculated the tensor $\{ \potb\alp\mu\nu \}$
already. It is defined by the expression \eqref{sec_02_05-05}. For
the tensor $\{ \potc\alp\bet\mu\nu \}$, using \eqref{sec_02_04-59},
one finds
\be \ba{l} \potc\alp\bet\mu\nu = -\frac23 \G\alp\bet\mu\nu + \frac23
\G\alp{[\mu}{\nu]}\bet\\
\quad = -\G\alp\bet\mu\nu + \frac13 \lp
\G\alp\bet\mu\nu + \G\alp\mu\nu\bet + \G\alp\nu\bet\mu \rp\\
\quad = -\G\alp\bet\mu\nu + \G\alp{[\bet}\mu{\nu]}. \ea \ee
Finally one obtains
\bw
\be\label{sec_02_05-15} \boxed{ \ba{rl}
\potdpara\mu\nu & = \lf \lb -\belud\mu\nu\alp + \G\gam\bet\mu\nu \tor\gam\bet\alp \rb + \lb \sna\lam \G\alp{[\mu}\nu{\lam]} + \bfrac12 \lp \G\alp{[\mu}\rho{\sig]} \tor\nu\rho\sig - \G\alp{[\nu}\rho{\sig]} \tor\mu\rho\sig \rp \rb \rf \dpara\alp\\
 & \quad + \lf \G\alp\bet\mu\nu + \G\alp{[\bet}\mu{\nu]} \rf \na\bet \dpara\alp.
\ea } \ee
\ew

\subsection{Dynamical quantities in the structure of the generalized currents}
More useful and interesting, however, to construct the
superpotential starting from the generalized canonical Noether
current $\bfJdpara$ \eqref{sec_02_00-05}
\be\label{sec_02_05-06} \Jdpara\mu = \U\alp\mu \dpara\alp +
\M\alp\bet\mu \na\bet \dpara\alp + \N\alp\bet\gam\mu \na{(\gam}
\na{\bet)} \dpara\alp. \ee
Such a construction lets us understand better the connections of the
generalized currents  $\bfJdpara$, $\bfJsdpara$, on the one hand,
with the dynamical characteristics $\bfsem$, $\bfsems$, $\bfspi$,
\dots, on the other hand.

Substituting \eqref{sec_02_04-34}, \eqref{sec_02_04-28} and
\eqref{sec_02_01-34} into the formula \eqref{sec_02_05-06}, one
finds the explicit presentation for the current $\bfJdpara$:
\bw
\be\label{sec_02_05-16} \ba{rl}
\Jdpara\mu & = \lf \sem\mu\alp + \belud\mu\lam\kap \tor\kap\lam\alp + \bfrac12 \Gu\pi\rho\sig\mu \curd\alp\pi\rho\sig + \lb \sna\nu \lp \G\kap\lam\mu\nu \tor\kap\lam\alp \rp + \bfrac12 \lp \G\kap\lam\rho\sig \tor\kap\lam\alp \rp \tor\mu\rho\sig \rb \rf \dpara\alp\\
 & \quad +\lf -\belud\mu\bet\alp - \lb \sna\nu \G\alp\bet\mu\nu + \bfrac12 \G\alp\bet\rho\sig \tor\mu\rho\sig \rb + \bfrac12 \G\alp\rho\sig\mu \tor\bet\rho\sig - \lp \G\kap\lam\bet\mu \tor\kap\lam\alp \rp \rf \na\bet \dpara\alp\\
 & \quad + \lf \G\alp{(\bet}{\gam)}\mu \rf \na{(\gam} \na{\bet)}
 \dpara\alp.
\ea \ee
\ew
As is seen, the \emph{canonical} current $\bfJdpara$ essentially is
constructed by the \emph{canonical} dynamic quantities $\bfsem$,
$\bfspi$ and the tensor $\bfG$.

Now, apply the identical transformations to the terms at the right
hand side of \eqref{sec_02_05-16} as follows.
\bn
\item\label{sec_02_05-09} For a first part of items in \eqref{sec_02_05-16},
differentiating by parts, adding and subtracting the combination
$\lp \frac12 \belud\kap\lam\alp \tor\mu\kap\lam \dpara\alp \rp$, one
finds
\bw
\be \ba{l}
\lp \sem\mu\alp + \belud\mu\lam\kap \tor\kap\lam\alp \rp \dpara\alp - \belud\mu\bet\alp \na\bet \dpara\alp = \lf \sem\mu\alp + \lp \sna\lam \belud\mu\lam\alp + \bfrac12 \belud\kap\lam\alp \tor\mu\kap\lam + \belud\mu\lam\kap \tor\kap\lam\alp \rp \rf \dpara\alp\\
\quad + \lf \sna\nu \lp -\belud\mu\nu\alp \dpara\alp \rp + \bfrac12
\lp -\belud\kap\lam\alp \dpara\alp \rp \tor\mu\kap\lam \rf = \lf
\sems\mu\alp \rf \dpara\alp + \lf \sna\nu \lb -\belud\mu\nu\alp
\dpara\alp \rb + \bfrac12 \lb -\belud\rho\sig\alp \dpara\alp \rb
\tor\mu\rho\sig \rf, \ea \ee
\ew
where, at the second equality, the definition \eqref{sec_02_04-40}
has been taken into account.
\item For a second part of items in \eqref{sec_02_05-16},
differentiating by parts and collecting the similar terms, one
obtains
\bse \ba{l}
\lb \sna\nu \lp \G\kap\lam\mu\nu \tor\kap\lam\alp \rp + \bfrac12 \lp \G\kap\lam\rho\sig \tor\kap\lam\alp \rp \tor\mu\rho\sig \rb \dpara\alp\\
\quad - \lp \G\kap\lam\bet\mu \tor\kap\lam\alp \rp \na\bet \dpara\alp\\
\ea \ese
\be \ba{l}
 = \sna\nu \lb \G\kap\lam\mu\nu \tor\kap\lam\alp \dpara\alp \rb + \bfrac12 \lb \G\kap\lam\rho\sig \tor\kap\lam\alp \dpara\alp \rb
\tor\mu\rho\sig. \ea \ee
\item For the last part of items in \eqref{sec_02_05-16}, again
differentiating by parts and collecting the similar terms, one
finds
\bw
\be\label{sec_02_05-07} \ba{l}
\lp \bfrac12 \Gu\pi\rho\sig\mu \curd\alp\pi\rho\sig \rp \dpara\alp - \lb \sna\nu \G\alp\bet\mu\nu + \bfrac12 \G\alp\bet\rho\sig \tor\mu\rho\sig \rb \na\bet \dpara\alp + \bfrac12 \G\alp\rho\sig\mu \tor\bet\rho\sig \na\bet \dpara\alp + \G\alp{(\bet}{\gam)}\mu \na\gam \na\bet \dpara\alp\\
= \bfrac12 \lp \Gu\lam{[\rho}{\sig]}\mu \dpara\alp \rp
\curd\alp\lam\rho\sig + \lb \sna\gam \G\alp{[\bet}{\gam]}\mu +
\bfrac12 \G\alp{[\rho}{\sig]}\mu \tor\bet\rho\sig \rb \na\bet
\dpara\alp + \sna\nu \lb \G\alp{(\bet}{\nu)}\mu \na\bet \dpara\alp
\rb + \bfrac12 \lb -\G\alp\bet\rho\sig \na\bet \dpara\alp \rb
\tor\mu\rho\sig.
\ea \ee
\ew
\item Differentiating by parts the second term on the
right hand side of \eqref{sec_02_05-07}, subsequently using the
identities \eqref{sec_02_05-18} and \eqref{sec_02_05-17} with
$\tpotc\alp\mu\nu\kap = \G\alp{[\nu}\kap{\mu]}$ and
$\tpotbu\nu\kap\mu = \G\alp{[\nu}\kap{\mu]} \dpara\alp$,
respectively, and collecting the similar terms, one gets
\be\label{sec_02_05-08} \ba{l}
\lb \sna\gam \G\alp{[\bet}{\gam]}\mu + \bfrac12 \G\alp{[\rho}{\sig]}\mu \tor\bet\rho\sig \rb \na\bet \dpara\alp\\
= \sna\nu \lb \sna\kap \lp \G\alp{[\nu}{\kap]}\mu \dpara\alp \rp + \bfrac12 \lp \G\alp{[\rho}{\sig]}\mu \dpara\alp \rp \tor\nu\rho\sig \rb\\
\quad - \sna\nu \lb \G\alp{[\nu}{\bet]}\mu \na\bet \dpara\alp \rb\\
\quad - \bfrac12 \lp \G\alp{[\rho}{\sig]}\lam \cur\mu\lam\rho\sig - \G\lam{[\rho}{\sig]}\mu \cur\lam\alp\rho\sig \rp \dpara\alp\\
= \sna\nu \lb \G\alp{[\bet}{\nu]}\mu \na\bet \dpara\alp \rb -
\bfrac12 \lp \Gu\lam{[\rho}{\sig]}\mu \dpara\alp \rp
\curd\alp\lam\rho\sig.
\ea \ee
\item\label{sec_02_05-10} Substituting this result into \eqref{sec_02_05-07} one obtains
\be \ba{l}
\mbox{R.H.S. of the eq. \eqref{sec_02_05-07}}\\
= \sna\nu \lb -\G\alp\bet\mu\nu \na\bet \dpara\alp \rb + \frac12 \lb
-\G\alp\bet\rho\sig \na\bet \dpara\alp \rb \tor\mu\rho\sig. \ea \ee
\en
Combining the points \ref{sec_02_05-09} -- \ref{sec_02_05-10}, one
finds that the formula \eqref{sec_02_05-16} is presented
equivalently as
\be\label{sec_02_05-13} \boxed{ \Jdpara\mu = \Jsdpara\mu + \lf
\sna\nu \potpdpara\mu\nu + \bfrac12 \potpdpara\rho\sig
\tor\mu\rho\sig \rf, } \ee
where
\be\label{sec_02_05-11} \boxed{ \Jsdpara\mu \Def \sems\mu\alp
\dpara\alp } \ee
is the generalized symmetrized Noether current (see the Paper~I,
Sec. V), and
\be\label{sec_02_05-12} \boxed{ \ba{rl} \potpdpara\mu\nu & \Def \lb
-\belud\mu\nu\alp + \G\kap\lam\mu\nu \tor\kap\lam\alp \rb \dpara\alp\\
 & \quad + \lb -\G\alp\bet\mu\nu \rb \na\bet \dpara\alp. \ea } \ee

The formula \eqref{sec_02_05-11} shows that the \emph{symmetrized
current } $\bfJsdpara$ is expressed thorough \emph{only the
symmetrized EMT} $\bfsems$ even in the case of the Lagrangian of the
most general type \eqref{sec_02_01-06}. Analogously, the formula
\eqref{sec_02_05-12} shows that the superpotential $\bfpotpdpara$ is
expressed through \emph{only the Belinfante tensor} $\bfbel$ induced
by the \emph{canonical ST} $\bfspi$ and the tensor $\bfG$.

Combining the 2-nd Klein identity \eqref{sec_02_04-39} and
\eqref{sec_02_04-22}, one finds
 \be \sems\mu\nu = -\Ib\mu\nu, \ee
that is the symmetrized EMT $\bfsems$ depends on only the Lagrangian
derivatives (see definition \eqref{sec_02_04-62}), and,
consequently, \emph{does not depend on divergences in Lagrangian
$\lag$.} By \eqref{sec_02_04-39} and \eqref{sec_02_04-22} also,
the formula \eqref{sec_02_05-11}, can be represented as
\be\label{sec_02_05-14} \Jsdpara\mu = \lb \semm\mu\alp - \DIDjfiA
\DbrjAB\mu\alp \jfiB \rb \dpara\alp = -\Ib\alp\mu \dpara\alp. \ee

Comparing \eqref{sec_02_05-13} and \eqref{sec_02_05-14} with the
boundary Klein-Noether theorem \eqref{sec_02_05-19}, one concludes
that the superpotential $\bfpotpdpara$ \eqref{sec_02_05-12} has to
be equivalent to the canonical superpotential \eqref{sec_02_00-06}.
Nevertheless, comparing the right hand sides of \eqref{sec_02_05-12}
and \eqref{sec_02_05-15} directly, we do not see this! However, the
difference is not essential. Recall the remark in the Paper~I (Sec.
IV, formulae (55)--(69)) that is related to \emph{arbitrary} two
superpotentials, $\bfpotdpara$ and $\bfpotpdpara$, which differ in a
term of the type
\be \ba{l}
\Del\potdpara\mu\nu \Def \potpdpara\mu\nu - \potdpara\mu\nu\\
\quad = \lb \sna\lam \C\alp\mu\nu\lam + \C\alp{[\mu|}\rho\sig
\tor{|\nu]}\rho\sig \rb \dpara\alp + \lb \C\alp\bet\mu\nu \rb
\na\bet \dpara\alp, \ea \ee
where an \emph{arbitrary} tensor $\{ \C\alp\lam\mu\nu \}$ is totally
antisymmetric in contravariant indexes:
\be \C\alp{[\lam}\mu{\nu]} = \C\alp\lam\mu\nu. \ee
Then, such superpotentials, $\bfpotdpara$ and $\bfpotpdpara$, are
related to the \emph{same} Noether current! One can see easily
that the difference of $\bfpotpdpara$ \eqref{sec_02_05-12} and
$\bfpotdpara$ \eqref{sec_02_05-15} has just the above form with $\C\alp\lam\mu\nu =
-\G\alp{[\lam}\mu{\nu]}$.

Rather, by a simplicity, the superpotential $\bfpotpdpara$
\eqref{sec_02_05-12} could be more preferable in applications.

%
\section{Structure and interpretation of the equations of gravitational fields}\label{sec_02_05-00}

\subsection{The field equations with the total EMT and ST}\label{sec_02_06-01}
The system of the equations of motion of all the fields $\bfmet$,
$\bftor$ and $\bfffi$, as usual, is obtained by variation of the
action functional, thus
\begin{empheq}[left=\empheqlbrace]{align}
\Del I/ \Del\met\mu\nu & =  0;\label{sec_02_06-08a}\\
\Del I/ \Del\tor\lam\mu\nu & =  0;\label{sec_02_06-07a}\\
\Del I/ \Del\ffiA & =  0.\label{sec_02_06-27a}
\end{empheq}
Combining \eqref{sec_02_04-20}, \eqref{sec_02_04-39},
\eqref{sec_02_04-65} and \eqref{sec_02_04-21}, it is not difficult
to obtain
\be\label{sec_02_06-06} { 2 \DIDmet\mu\nu \eq  \semiu\mu\nu -
\sna\lam\beliu\lam\mu\nu + \DIDffiA \DbrfuAB\mu\nu \ffiB. } \ee
Then, again turning to \eqref{sec_02_04-21}, one finds that the
system \eqref{sec_02_06-08a} - \eqref{sec_02_06-27a} is equivalent
to
\begin{empheq}[left=\empheqlbrace]{align}
\semiu\mu\nu & =  0;\label{sec_02_06-08}\\
\beliud\nu\mu\lam & =  0;\label{sec_02_06-07}\\
\Del I/ \Del\ffiA & =  0.\label{sec_02_06-27}
\end{empheq}
Remark that, by the identity \eqref{sec_02_04-14}, the equation
\eqref{sec_02_06-07} induces the equation
\be\label{sec_02_06-09} \spii\pi\rho\sig = 0, \ee
and conversely. Recall also that the dynamic characteristics of the
physical system $\bfsemi$ and $\bfspii$ are total because are
related to the total action of the system.

A direct interpretation of the equations of the gravitational fields
that follows from the visible presentations of \eqref{sec_02_06-08}
and \eqref{sec_02_06-09} is evident: In an arbitrary metric-torsion theory
of gravity without background structures, both the
\emph{total} modified EMT $\bfsemi$ and the \emph{total} modified
canonical $\bfspii$ are equal to nil. The claim that the total
dynamic characteristics in a gravitational theory have to be equal
to zero is not new. In GR, it has being defended by Lorentz
\cite{Lorentz_1917_a_en, Lorentz_1917_a_en_rep} and Levi-Civita
\cite{Levi-Civita_1917, Levi-Civita_1917_rep}, later by Soriau
\cite{Souriau_1957}. Comparatively recent, Szabados
\cite{Szabados_1991, Szabados_1992} has approved this result,
examining the Belinfante procedure. In the works by Logunov and
Folomeshkin \cite{Logunov_Folomeshkin_1977_a_en,
Logunov_Folomeshkin_1977_b_en, Logunov_Folomeshkin_1977_c_en,
Logunov_Folomeshkin_1977_d_en} this claim is treated as unavoidable
conclusion in a pure metric theory of gravity.

However, under a more detailed consideration such an interpretation
meets serious objections, which lead to a necessity to reject it.
The first who was against is Einstein \cite{Einstein_1918_c,
Einstein_1918_c_en}. Replying the Lorentz work
\cite{Lorentz_1917_a_en, Lorentz_1917_a_en_rep}, he noted that there
is no a logic argument against the Lorentz interpretation. But,
basing on the equation \eqref{sec_02_06-08}, one cannot to obtain
conclusions that usually follow from the conservation laws. Indeed,
due to \eqref{sec_02_06-08}, the components of the total energy
tensor everywhere during all the time are equal to zero, that is the
total energy of the system from the beginning is equal to zero.
However, the conservation of ``zero'' does not require the next
existence of the system: one ``permits'' a disappearance of the
physical system at all. Such a conclusion looks at as extremely
non-physical. Of course, the Einstein arguments can be applied to
discuss the total ST $\bfspii$.

\subsection{Pure gravitational and matter parts of the physical system}\label{sec_02_06-02}
Recall the basis of constructing the GR and other metric theories.
One of the main requirements is that the dynamic physical picture is
postulated as follows. A bend of a curved space-time, in which the
matter propagate, is provided by the matter itself. Then, by a
natural way it turns out that the physical system is presented as a
union of divided the pure gravitational part and the matter part. As
it will be shown in the Paper~III, in the last one of the series of
the works, the problem of defining physically sensible conserved
quantities can be solved just in the framework of such a
presentation. Below, in subsection \ref{sec_02_06-04}, we give the
other arguments supporting the split presentation of the equations
and against the nil interpretation of the total dynamic
characteristics of the system.  Now, we give and discuss the main
formulae and relations for the split presentation in the framework
of the manifestly generally covariant metric-torsion theories
given in the Riemann-Cartan space.

Keeping in mind the above, represent the Lagrangian
\eqref{sec_02_01-06} as a sum of the pure gravitational $\lagG$ and
matter $\lagM$ parts:
\be\label{sec_02_05-20} \ba{rl} \lag & = \lag (\bfmet,\bfcur;
\;\bftor,\bfnator,\bfnanator; \;\bfffi,\bfnaffi,\bfnanaffi)\\
 & \Def \lag (\bfmet,\bfcur; \;\bfjfi,\bfnajfi,\bfnanajfi)= \lagG + \lagM,
\ea \ee
where
\be\label{sec_02_06-11} \lagG = \lagG(\bfmet,\bfcur) \Def \lag
(\bfmet,\bfcur; 0,0,0); \ee
\be\label{sec_02_06-51} \lagM = \lagM (\bfmet,\bfcur;
\;\bfjfi,\bfnajfi,\bfnanajfi) \Def \lag - \lagG. \ee
Remark that in the definition \eqref{sec_02_06-11}, of course, the
connection $\bfcon$ (with the use of that the curvature tensor
$\bfcur$ is constructed) continues to depend on the torsion
$\bftor$. Thus, the gravitational Lagrangian $\lagG$ not explicitly
(through $\bfcur$) depends on $\bftor$ also.

However, in all the cases, it is not possible to define $\lagG$ as
in \eqref{sec_02_06-11}, for example, in the scalar-tensor
Jordan-Brans-Dicke theories \cite{Jordan_1959, Brans_Dicke_1961},
or, in more general theories gravity with dilaton
\cite{Metsaev_Tseytlin_1987, Gross_Sloan_1987}, in ``sting''
presentation (see Refs. \cite{Papoyan_2003_en,
Frolov_Novikov_1998}). At the same time, in the ``Einstein''
presentation such a splitting can be provided easily. Then it is
necessary to define clearly what unusual fields (except of metric
and torsion ones) are related to gravitational fields.

It is evidently that the splitting of the Lagrangian
\eqref{sec_02_05-20} leads to a correspondent splitting of the
action functional:
\bse \ba{rl} I & = \intb\dome\rmet\lag = \intb\dome\rmet\lagG +
\intb\dome\rmet\lagM\\
 & \Def I^G + I^M.
\ea \ese
Of course, the Lagrangian of the vacuum system $\lagG$ has to be
\emph{generally covariant  scalar}, and then the matter Lagrangian
$\lagM$ is, like this, also. Therefore all the above results and
conclusions related to the total Lagrangian $\lag$ are left valid
for each of the Lagrangians $\lagG$ and $\lagM$ themselves.

Define next \emph{matter tensors}.
\begin{empheq}{flalign}
\spiM\pi\rho\sig & \Def \ld\spi\pi\rho\sig\ogM\label{sec_02_06-46}\\ & \mbox{(the canonical ST of matter)};\nonumber\\
\spiMa\pi\rho\sig & \Def \ld\spia\pi\rho\sig\ogM\\  & \mbox{(the additional ST of matter)};\nonumber\\
\spiMi\pi\rho\sig & \Def
\ld\spii\pi\rho\sig\ogM\label{sec_02_06-20}\\ & \mbox{(the modified
canonical ST of matter)}.\nonumber
\end{empheq}
\begin{empheq}{flalign}
\belMu\gam\bet\alp & \Def \ld\belu\gam\bet\alp\ogM\\ & \mbox{(the Belinfante tensor for ST $\bfspiM$)};\nonumber\\
\belMau\gam\bet\alp & \Def \ld\belau\gam\bet\alp\ogM\\ & \mbox{(the Belinfante tensor for ST $\bfspiMa$)};\nonumber\\
\belMiu\gam\bet\alp & \Def
\ld\beliu\gam\bet\alp\ogM\label{sec_02_05-21}\\ & \mbox{(the
Belinfante tensor for ST $\bfspiMi$)}.\nonumber
\end{empheq}
\begin{empheq}{flalign}
\semM\mu\nu & \Def \ld\sem\mu\nu\ogM\\  & \mbox{(the canonical EMT of matter)};\nonumber\\
\semMa\mu\nu & \Def \ld\sema\mu\nu\ogM\\ & \mbox{(the additional EMT of matter)};\nonumber\\
\semMi\mu\nu & \Def \ld\semi\mu\nu\ogM\label{sec_02_06-19}\\ & \mbox{(the modified canonical EMT of matter)};\nonumber\\
\semMsu\mu\nu & \Def \ld\semsu\mu\nu\ogM\\ & \mbox{(the symmetrized EMT of matter)};\nonumber\\
\semMmu\mu\nu & \Def \ld\semmu\mu\nu\ogM\label{sec_02_06-47}\\ &
\mbox{(the metrical EMT of matter)}.\nonumber
\end{empheq}
For the above defined matter tensors, relations analogous to those
between the total tensors take a place. In particular, analogously
to \eqref{sec_02_04-21} and  \eqref{sec_02_06-06}, one has
\be\label{sec_02_06-13} \boxed{ \DIMDtor\alp\bet\gam = \frac12
\belMiud\gam\bet\alp; } \ee
\be\label{sec_02_06-14}
\boxed{ \ba{l} 2 \DIMDmet\mu\nu = \semMmu\mu\nu\\
\quad \eq \semMiu\mu\nu - \sna\lam \belMiu\lam\mu\nu + \DIMDffiA
\DbrfuAB\mu\nu \ffiB. \ea} \ee

Now, define the \emph{Cartan tensor} $\bfCar \Def \{
\Carud{[\gam}{\bet]}\alp = \Carud\gam\bet\alp \}$ and the
\emph{(generalized) Einstein tensor} $\bfEin \Def \{
\Einu{(\mu}{\nu)} \neq \Einu\mu\nu \}$:
\bea
\bfrac{-1}{2k} \Carud\gam\bet\alp \Def \DIGDtor\alp\bet\gam = \bfrac12 \ld \belud\gam\bet\alp \ogG \label{sec_02_06-15}\\
\mbox{(the Cartan tensor)};\nonumber \eea
\bea \bfrac{-1}{2k} \lp \Einu{(\mu}{\nu)} - \sna\lam
\Caru\lam{(\mu}{\nu)} \rp \Def \DIGDmet\mu\nu = \bfrac12
\ld \semmu\mu\nu \ogG\label{sec_02_06-43}\\
\mbox{(symmetric part of the generalized Einstein tensor)};\nonumber
\eea
\bea \ba{rl} \bfrac{-1}{2k} \Ein\mu\nu & \Def
\bfrac12 \ld \sem\mu\nu \ogG\\
 & = \bfrac12 \lp \lagG \kro\mu\nu -
^{(G)}\Gu\alp\bet\gam\mu \curd\alp\bet\gam\nu \rp\ea\label{sec_02_06-23} \\
\mbox{(the generalized Einstein tensor)}.\nonumber \eea
Here,
\be ^{(G)}\G\alp\bet\gam\del \Def \ld \G\alp\bet\gam\del \ogG = 2
\frac{\pa{}\lagG}{\pa{} \cur\alp\bet\gam\del}; \ee
\be k \Def (D-1) \Ome_{(D-1)}\; \vkap, \ee
$\Ome_{(D-1)}$ is an area of $(D-1)$-dimensional unit sphere, and
$\vkap$ is the Newtonian gravitational constant in
$(D+1)$-dimensional space-time.

A restriction of the 2-nd Klein identity \eqref{sec_02_04-39} and
the definition \eqref{sec_02_04-40} to the case of the Lagrangian
$\lag = \lagG$ gives, with taking into account the definitions
\eqref{sec_02_06-15} - \eqref{sec_02_06-23}, the identity
\bse \ba{l}
-\Einu\mu\nu + \lf \sna\lam \Caru\lam\mu\nu + \bfrac12 \Carud\gam\bet\alp \Dbrtu\mu\nu\alp\bet\gam\tet\vphi\xi \tor\tet\vphi\xi \rf\\
\eq - \lp \Einu{(\mu}{\nu)} - \sna\lam \Caru\lam{(\mu}{\nu)} \rp +
\bfrac12 \Carud\gam\bet\alp \Dbrtu\mu\nu\alp\bet\gam\tet\vphi\xi
\tor\tet\vphi\xi, \ea \ese
or
\be\label{sec_02_06-12} \boxed{ \Einu{[\mu}{\nu]} \eq \sna\lam
\Caru\lam{[\mu}{\nu]}. } \ee
Thus, the \emph{antisymmetric part of the generalized Einstein
tensor is the divergence of the antisymmetric part of the Cartan
tensor}. Using the identity \eqref{sec_02_06-12}, one can represent
\eqref{sec_02_06-43} in the form:
\be\label{sec_02_06-16} \ba{rl} 2 \DIGDmet\mu\nu & = \bfrac{-1}{k}
\lp
\Einu{(\mu}{\nu)} - \sna\lam \Caru\lam{(\mu}{\nu)} \rp\\
 & \eq \bfrac{-1}{k} \lp \Einu\mu\nu - \sna\lam \Caru\lam\mu\nu \rp.
\ea \ee

\subsection{The gravitational field equations in the split form}\label{sec_02_06-03}
By the equations \eqref{sec_02_06-16} and \eqref{sec_02_06-47}, the
equations of motion of the metric field, $\Del (I^G + I^M)/
\Del\met\mu\nu = 0$,  can be rewritten as
\be \Einu{(\mu}{\nu)} - \sna\lam \Caru\lam{(\mu}{\nu)} = k
\semMmu\mu\nu, \ee
or, using the formulae \eqref{sec_02_06-13} - \eqref{sec_02_06-15},
\eqref{sec_02_06-16}, one can represent them in the equivalent form:
\be \ba{rl}\Einu\mu\nu = & k \semMiu\mu\nu + \sna\lam \lp
\Caru\lam\mu\nu - k
\belMiu\lam\mu\nu \rp\\
& + k \DIMDffiA \DbrfuAB\mu\nu \ffiB. \ea \ee
If here one takes into account the equations of motion for the
torsion field
\be\label{sec_02_06-17} \frac{\Del ( I^G + I^M )}{\Del
\tor\nu\mu\lam} = 0 \qquad \Leftrightarrow \qquad \Caru\lam\mu\nu =
k \belMiu\lam\mu\nu \ee
and the equations of motion for the $\bfffi$-fields: $\Del I^M/ \Del
\ffiA = 0$, one obtains the equation for the metric field only:
\be\label{sec_02_06-26} \Einu\mu\nu = k \semMiu\mu\nu \qquad
\mbox{(on the $\bfjfi$-equations)}. \ee
Now, turn to \eqref{sec_02_06-17}. After antisymmetrization  in
indexes $\mu$ and $\nu$, using the definitions \eqref{sec_02_05-21},
\eqref{sec_02_04-41} and \eqref{sec_02_06-20}, and the identity
\eqref{sec_02_04-14}, the equation  \eqref{sec_02_06-17} acquires an
equivalent form:
\be\label{sec_02_06-24} -2 \Caru\lam{[\mu}{\nu]} = k
\spiMiu\lam\mu\nu. \ee

Thus, the total system of the field equations acquires the form:
\begin{empheq}[left=\empheqlbrace,box=\fbox]{flalign}
\Einu\mu\nu & = k \semMiu\mu\nu & \mbox{(the $\bfmet$-equations)};\label{sec_02_06-29}\\
-2 \Car\lam{[\mu}{\nu]} & = k \spiMi\lam\mu\nu & \mbox{(the $\bftor$-equations)};\label{sec_02_06-30}\\
\Del I^M/ \Del \ffiA & = 0 & \mbox{(the
$\bfffi$-equations)}.\label{sec_02_06-31}
\end{empheq}
The interpretation of the gravitational equations of the system is
as follows.  \emph{The source of the metric field $\bfmet$ is the
modified canonical EMT of matter $\bfsemMi$, whereas the source of the
torsion field $\bftor$ is the modified canonical ST of matter $\bfspiMi$}.

\subsection{Geometrical identities and the equations of balance for the matter sources}\label{sec_02_06-04}
In the present subsection, we show why the total system of the
equations is more preferable just in the form \eqref{sec_02_06-29} -
\eqref{sec_02_06-31}. At first, let us discuss the matter part. The
identity \eqref{sec_02_05-22} for the Lagrangian $\lagM$ with taking
into account the definitions \eqref{sec_02_06-19} and
\eqref{sec_02_06-20} leads to the identity
\be \boxed{\ba{rl} \sna\mu \semMi\mu\nu & \eq - \semMi\mu\lam
\tor\lam\mu\nu + \bfrac12
\spiMi\pi\rho\sig \curud\rho\sig\pi\nu\\
 & \quad + \DIMDffiA \na\nu \ffiA.
\ea} \ee
From here the equations of balance for the matter modified
canonical EMT $\bfsemMi$ follows
\be\label{sec_02_06-21} \boxed{\ba{r}\sna\mu \semMi\mu\nu = -
\semMi\mu\lam
\tor\lam\mu\nu + \bfrac12 \spiMi\pi\rho\sig \curud\rho\sig\pi\nu\\
\mbox{(on the $\bfffi$-equations)}. \ea} \ee
It is important to note: in order the equation \eqref{sec_02_06-21}
to take a place \emph{it is necessary only} that the
$\bfffi$-equations hold, it is not necessary to take into account
the $\bfmet$- and $\bftor$-equations. Besides, the equation
\eqref{sec_02_06-21} is related only to $\lagM$, it does not relate
to $\lagG$. If, analogously to \eqref{sec_02_06-19} one
defines the pure ``gravitational EMT'' as $\ld \bfsemi \ogG$, then
(in the covariant sense) both the matter and gravitational EMTs,
each itself will satisfy \emph{its own} equation of balance. By the
Teitelboim terminology \cite{Teitelboim_1970}, they are
\emph{dynamically independent}. Moreover, the restriction of the
identity \eqref{sec_02_05-22} to the Lagrangian $\lagG$ leads to the
identity
\be\label{sec_02_06-22} \ba{rl} \sna\mu \ld \semi\mu\nu \ogG & \eq -
\ld
\semi\mu\lam \ogG \tor\lam\mu\nu\\
& + \bfrac12 \ld \spii\pi\rho\sig \ogG \curud\rho\sig\pi\nu, \ea \ee
that holds \emph{without any equations of motion}. The identity
\eqref{sec_02_06-22} reflects the fact only that the Lagrangian
$\lagG$ is a generally covariant scalar.

In order to fill the real sense of the identity
\eqref{sec_02_06-22}, one has to find concrete expressions, to which
the quantities $\ld\bfsemi\ogG$ and $\ld\bfspii\ogG$ correspond.
After using the definition \eqref{sec_02_04-54} for the Lagrangian
$\lagG$, the definitions \eqref{sec_02_06-23} and
\eqref{sec_02_06-15}, and the formula
\bse 2\Caru\lam{[\mu}{\nu]} = k \ld \spiiu\lam\mu\nu \ogG, \ese
which is carried out from \eqref{sec_02_06-15}, the identity
\eqref{sec_02_06-22} is rewritten in the form:
\be\label{sec_02_06-25} \boxed{ \sna\mu \Ein\mu\nu \eq \Ein\mu\lam
\tor\lam\mu\nu - \Car\pi\rho\sig \curud\rho\sig\pi\nu. } \ee
As is seen, it is the \emph{pure geometrical differential identity},
which connects the divergence of the Einstein tensor $\bfEin$ with
the Cartan tensor $\bfCar$.

So, the fact that the ``equation of balance'' for the
``gravitational EMT'' $\ld\bfsemi\ogG$ is satisfied identically is
the direct consequent of the diffeomorphism invariance of the pure
geometrical action. Therefore, one has to conclude that in background independent metric-torsion
theories of gravity, there are no
generally covariant expressions for EMTs and STs defined
\emph{classically} for the properly gravitational fields. This claim
can be stated not only for the gravitational fields, but it has an
universal character. The claim takes a place in an arbitrary gauge
invariant (in the sense of the definition in Introduction) theory:
\emph{There is no a gauge invariant expression for a current namely of the gauge field
because the theory is gauge invariant}.

At the end, let us discuss the role of the identity
\eqref{sec_02_06-25}. Namely its existence defines the fact that the
form of the gravitational equations \eqref{sec_02_06-29} -
\eqref{sec_02_06-30} is more preferable. Indeed, substituting the
Einstein $\bfEin$ and Cartan $\bfCar$ tensors with the use of the
$\bfmet$- and $\bftor$-equations \eqref{sec_02_06-29} and
\eqref{sec_02_06-30}, respectively, into the identity
\eqref{sec_02_06-25}, one obtains the equation of balance for the
matter modified EMT $\bfsemMi$:
\be \ba{r} \sna\mu \semMi\mu\nu = - \semMi\mu\lam \tor\lam\mu\nu +
\bfrac12
\spiMi\pi\rho\sig \curud\rho\sig\pi\nu\\
\mbox{(on the $\bfmet$- and $\bftor$-equations)}. \ea \ee
Recall that the copy of this equation, namely \eqref{sec_02_06-21},
conversely, has been carried out without using the gravitational
equations, but only with the use of the $\bfffi$-equations.
Therefore, one concludes that the role of the identity
\eqref{sec_02_06-25} is to \emph{state the self-consistence of the
total system of the field equations \eqref{sec_02_06-29} -
\eqref{sec_02_06-31}}. Just in this sense we generalize the
interpretation of the gravitational equations in ECT
\cite{Kibble_1961, Sciama_1962, Sciama_1964_a, Sciama_1964_b,
Trautman_1972_a, Trautman_1972_b, Trautman_1973_b, Trautman_1980,
Hehl_1973, Hehl_1974, Hehl_Heyde_Kerlick_Nester_1976}:
\begin{empheq}[left=\empheqlbrace]{align}
E^{\mu\nu} & = k\, \semMu\mu\nu;\label{sec_02_06-52}\\
\stor\lam\mu\nu & = k\, \spiM\lam\mu\nu;\label{sec_02_06-53},
\end{empheq}
where
\bse E^{\mu\nu} \Def R^{\mu\nu} - \frac12 \met\mu\nu R; \quad
\stor\lam\mu\nu \Def \tor\lam\mu\nu + \kro\lam\mu T_\nu -
\kro\lam\nu T_\mu. \ese
Here, twice contracted the Bianchi identity
\be\label{sec_02_06-54} \sna\mu E^\mu{}_\nu \equiv - E^\mu{}_\lam
\tor\lam\mu\nu + \frac12 \stor\pi\rho\sig \curud\rho\sig\pi\nu \ee
is treated as a \emph{dynamic conservation of the source}. Recall
the Wheeler words \cite{Misner_Thorn_Wheeler_1973} related to GR:
the ``gravitational field watch for the conservation of its
sources''. We see that the same can be repeated also for the
\emph{general metric-torsion theories of gravity}. Moreover,
this statement is not related to  gravitational theories only, but
has an universal character and is related to an arbitrary gauge
invariant theory. Namely, the \emph{gauge field watch for the
conservation of its matter sources}. Returning to the identity
\eqref{sec_02_06-25}, one sees that it is, thus, the generalization
of twice contracted the Bianchi identity \eqref{sec_02_06-54}.

At last, notice that in the case of the Lagrangians considered in
the works \cite{Trautman_1972_a, Trautman_1972_b, Trautman_1973_b,
Trautman_1980, Hehl_1973, Hehl_1974, Hehl_Heyde_Kerlick_Nester_1976}, the general system
of the gravitational equations \eqref{sec_02_06-29} and
\eqref{sec_02_06-30} exactly is simplified to the equations
\eqref{sec_02_06-52} and \eqref{sec_02_06-53} obtained in these works.


\section*{Acknowledgments}

The authors are very grateful to the referee for useful recommendations and D.I. Bondar for correcting English.


\appendix

%
\section{The tensors $\{ \Dd\alp\bet\gam\lam\mu\nu \}$, $\{ \Du\alp\bet\gam\lam\mu\nu \}$ and their properties}\label{app_02_a-00}

In the main text, for a significant simplification of expressions we
use the tensor:
\be\label{app_02_a-01} \boxed{ \Dd\alp\bet\gam\lam\mu\nu \Def
\frac{1}{2} \lp \kro\alp\mu \kro\bet\lam \kro\gam\nu + \kro\alp\nu
\kro\bet\lam \kro\gam\mu - \kro\alp\lam \kro\bet\mu \kro\gam\nu \rp.
} \ee
It is obtained by  differentiating the connection
$\cond\lam\mu\nu = g_{\lam\veps}\con\veps\mu\nu$ with respect to
derivatives of metric $\pa\alp \met\bet\gam $. Thus, the use of
\eqref{app_02_a-01} leads to the compact presentation:
\be \cond\lam\mu\nu = \Dd\alp\bet\gam\lam\mu\nu (\pa\alp
\met\bet\gam + \tord\alp\bet\gam ).\ee
The tensor
\be \boxed{ \Du\alp\bet\gam\lam\mu\nu \Def \frac{1}{2}  \lp
\kro\bet\lam \kro\alp\mu \kro\gam\nu + \kro\gam\lam \kro\alp\mu
\kro\bet\nu - \kro\alp\lam \kro\bet\mu \kro\gam\nu \rp } \ee
with the converse symmetry of the co- and contravariant indexes is
also useful. It is easily to obtain that
\be\label{app_02_a-02} \Dd\alp{(\bet}{\gam)}\nu\mu\lam = \frac{1}{2}
\lf \kro\alp\lam \kro{(\bet}\mu \kro{\gam)}\nu - \kro{(\gam}\lam
\krob{\bet)}\alp\mu\nu \rf, \ee
where
\be \krob\alp\bet\mu\nu \Def \kro\alp\mu \kro\bet\nu - \kro\bet\mu
\kro\alp\nu = \kro\alp\mu \kro\bet\nu - \kro\alp\nu \kro\bet\mu \ee
is the \emph{generalized Kronecker symbol}. As a consequence of
\eqref{app_02_a-02} one has
\be \Dd\alp{(\bet}{\gam)}{[\nu}{\mu]}\lam = -\frac12 \kro{(\gam}\lam
\krob{\bet)}\alp\mu\nu. \ee
The next formulae are also valid:
\be\label{sec_02_04-09} \Dd\alp{[\bet}{\gam]}\nu\mu\lam =
\Du\gam\bet\alp\lam{[\mu}{\nu]}; \ee
\be \Du{[\gam}{\bet]}\alp\lam\mu\nu = -\frac12 \krob\gam\bet\lam\mu
\kro\alp\nu - \frac14 \kro\alp\lam \krob\bet\gam\mu\nu; \ee
\be\label{sec_02_04-14} \Du\gam{[\bet}{\alp]}\lam\mu\nu = -\frac14
\kro\gam\lam \krob\bet\alp\mu\nu; \ee
\be\label{sec_02_01-33} \Dd{(\alp|}\bet{|\gam)}\sig\rho\pi +
\Dd{(\alp}{\gam)}\bet\sig\rho\pi = \kro\bet\sig \kro\gam{(\rho}
\kro\alp{\pi)}; \ee
\be\label{sec_02_01-50} \Dd\alp{(\bet}{\gam)}\sig\rho\pi -
\frac{1}{2} \kro\alp\pi \kro\bet\rho \kro\gam\sig =
\Du\alp\bet\gam\pi{[\rho}{\sig]}; \ee
\be\label{sec_02_01-46} \Dd{(\alp}{\bet)}\gam\sig\rho\pi =
\frac{1}{2} \kro\alp{(\sig} \kro\bet{\pi)} \kro\gam\rho; \ee
\be\label{sec_02_01-47} \Dd{[\alp|}\bet{|\gam]}\sig\rho\pi =
-\frac{1}{2} \kro\bet\rho \kro\alp{[\sig} \kro\gam{\pi]}. \ee
%

%
\section{The general variations of fields functions}\label{app_02_b-00}

\subsection{The functional and total variations}\label{app_02_b-01}

Let a set of variables, tensor fields,  $\bfgfi(x) = \{ \gfiA(x);\,
A=\overline{1,N} \}$, be given in a spacetime. Let a result of an
infinitesimal transformations be as
\be\label{app_02_b-06} \lf\ba{ccc}
x & \rightarrow & x';\\
\bfgfi(x) & \rightarrow & \bfgfi'(x'). \ea\rd \ee
The transformation \eqref{app_02_b-06} we will call as the
\emph{active} transformation. Then under its action a spacetime
point with coordinates $x$ transforms into a \emph{new} point with
coordinates $x'$, and a function (physical field) $\bfgfi(x)$
transforms to a \emph{new} function $\bfgfi'(x')$. At the same time.
the coordinate system is \emph{fixed/the same}.

The \emph{total variation} $\bdbfgfi(x)$ of field functions
$\bfgfi(x)$ appears as a result of a comparison of a new function
$\bfgfi'(x')$, calculated in a new point $x'$, with the initial
function  $\bfgfi(x)$, calculated in the initial point $x$:
\be\label{app_02_c-04} \bdbfgfi(x) \Def \bfgfi'(x') - \bfgfi(x). \ee
Unlike this, comparing the new and the old functions calculated in
the same (initial) point  $x$, one obtains the perturbation defined
as
\be \dbfgfi(x) \Def \bfgfi'(x) - \bfgfi(x). \ee
This perturbation, in fact, is the \emph{functional variation }
$\dbfgfi(x)$ of a field function $\bfgfi(x)$. Unlike the total
variation, it commutes with partial derivatives, and up to a sign
coincides with the Lie derivative, which appears under an
infinitesimal mapping a spacetime onto itself.

\subsection{The variation of the connection}
To obtain an explicitly covariant variation of the connection $\{
\dcon\lam\mu\nu \}$ by a more economical way one has to use the
metric compatible condition \eqref{app_01_a-02}. One obtains
\bse \na\lam (\dmet\mu\nu) = \met\mu\alp \dcon\alp\nu\lam +
\met\nu\alp \dcon\alp\mu\lam = 2\met{(\mu|}\alp
\dcon\alp{|\nu)}\lam. \ese
Now, using \eqref{app_02_a-01} and the formula \eqref{app_02_b-10}
after variating
\be \label{app_02_b-10d} \delta\con\lam{[\mu}{\nu]}=- \frac{1}{2}
\delta\tor\lam\mu\nu, \ee
one has
\bse \Dd\alp\bet\gam\lam\mu\nu \na\alp (\dmet\bet\gam) =
\met\lam\alp \dcon\alp{(\mu}{\nu)} - \met{(\mu|}\alp
\dtor\alp\lam{|\nu)}. \ese
This gives
\be\label{app_02_b-08} \dcon\lam{(\mu}{\nu)} = \metu\lam\pi
\Dd\alp\bet\gam\pi\mu\nu \na\alp  (\dmet\bet\gam) + \metu\lam\pi
\met{(\mu|}\alp \dtor\alp\pi{|\nu)}. \ee
Substituting the last formula and \eqref{app_02_b-10d} into the
evident equality $ \dcon\lam\mu\nu = \dcon\lam{(\mu}{\nu)} +
\dcon\lam{[\mu}{\nu]}$, one obtains finally
\be\label{sec_02_01-12} \boxed{ \dcon\lam\mu\nu = \metu\lam\pi
\Dd\alp\bet\gam\pi\mu\nu \lp \na\alp \dmet\bet\gam + \met\alp\sig
\dtor\sig\bet\gam \rp. } \ee

\subsection{The variation of the curvature tensor}
Varying the relation \eqref{app_01_a-02b}
\bse \cur\kap\lam\mu\nu = \pa\mu\con\kap\lam\nu -
\pa\nu\con\kap\lam\mu + \con\kap\alp\mu \con\alp\lam\nu -
\con\kap\alp\nu \con\alp\lam\mu, \ese
keeping in mind that $\{ \dcon\lam\mu\nu \}$ is a tensor and taking
into account \eqref{app_02_b-10}, one finds
\bse \dcur\kap\lam\mu\nu = 2\na{[\mu|} \dcon\kap\lam{|\nu]} +
\tor\tau\mu\nu \dcon\kap\lam\tau \ese
or
\be\label{sec_02_01-09} \boxed{ \dcur\kap\lam\mu\nu = \lp
\tor\tau\mu\nu + \krob\pi\tau\mu\nu \na\pi \rp \dcon\kap\lam\tau. }
\ee

\subsection{The variation of the $1$-st covariant derivative}
Varying the definition of the covariant derivative
\be \na\mu\jfiA = \pa\mu\jfiA + \con\kap\lam\mu \DbrjAB\lam\kap
\jfiB, \ee
where $\{ \DbrjAB\lam\kap \}$ are the Belinfante-Rosenfeld symbols
(see Appendix~\ref{app_02_c-01}), one gets
\be \del(\na\mu\jfiA) = \pa\mu (\djfiA) + \con\kap\lam\mu
\DbrjAB\lam\kap (\djfiB) + \DbrjAB\lam\kap \jfiB \dcon\kap\lam\mu.
\ee
From here one has
\be\label{app_02_b-09} \boxed{ \del(\na\mu\jfiA) = \na\mu (\djfiA) +
\DbrjAB\lam\kap \jfiB\; \dcon\kap\lam\mu. } \ee

\subsection{The variation of the $2$-nd covariant derivative}
Let us calculate $\del(\na\mu \na\nu \jfiA)$. For the sake of
simplicity, temporarily denote $\jfiA{}_\nu \Def \na\nu \jfiA$.
Then, taking into account the fact that the tensor $\{ \jfiA{}_\nu
\}$ has for a one index more than the tensor $\{ \jfiA \}$ and using
\eqref{app_02_b-09}, one obtains
\bw
\bse \ba{l}
\del(\na\mu \jfiA{}_\nu) = \na\mu \djfiA{}_\nu + \DbrjAB\lam\kap \jfiB{}_\nu \dcon\kap\lam\mu - \jfiA{}_\lam \dcon\lam\nu\mu\\
\quad = \na\mu \lb \na\nu \djfiA + \DbrjAB\lam\kap \jfiB\;
\dcon\kap\lam\nu \rb + \lb \DbrjAB\lam\kap \na\nu \jfiB - (\na\kap
\jfiA) \kro\lam\nu \rb \dcon\kap\lam\mu. \ea \ese
Consequently,
\be\label{sec_02_01-10} \boxed{ \del(\na\mu \na\nu \jfiA) = \na\mu
\na\nu \djfiA + \lb \DbrjAB\lam\kap \na\nu \jfiB - (\na\kap \jfiA)
\kro\lam\nu \rb \dcon\kap\lam\mu + \na\nu \lb \DbrjAB\lam\kap
\jfiB\; \dcon\kap\lam\nu \rb. } \ee
\ew
%

%
\section{The Belinfante-Rosenfeld symbols}\label{app_02_c-00}

\subsection{The definition and properties}\label{app_02_c-01}
Let the field functions $\bfgfi(x) = \{ \gfiA(x) \} = \{
\ten^{\mu_1\mu_2\dots\mu_r}{}_{\nu_1\nu_2\dots\nu_s}(x) \}$ present
generally covariant tensor of the rank $\lp r\atop s \rp$. Then
under a diffeomorphism

\be x^\mu \quad \rightarrow \quad x'^\mu = x'^\mu(x) \ee
its components are transformed by the known rule:
\be \ten^{\mu_1\mu_2\dots\mu_r}{}_{\nu_1\nu_2\dots\nu_s}(x) \;
\rightarrow \;
\ten'^{\mu_1\mu_2\dots\mu_r}{}_{\nu_1\nu_2\dots\nu_s}(x'); \ee
\bw
\be\label{app_02_c-05a}
\ten'^{\mu_1\mu_2\dots\mu_r}{}_{\nu_1\nu_2\dots\nu_s}(x') =
\frac{\pax'^{\mu_1}}{\pax^{\alp_1}}
\frac{\pax'^{\mu_2}}{\pax^{\alp_2}} \cdot{}\dots{}\cdot
\frac{\pax'^{\mu_r}}{\pax^{\alp_r}} \cdot
\frac{\pax^{\bet_1}}{\pax'^{\nu_1}}
\frac{\pax^{\bet_2}}{\pax'^{\nu_2}} \cdot{}\dots{}\cdot
\frac{\pax^{\bet_s}}{\pax'^{\nu_s}} \;
\ten^{\alp_1\alp_2\dots\alp_r}{}_{\bet_1\bet_2\dots\bet_s}(x). \ee
\ew
For the infinitesimal diffeomorphism
\be\label{app_02_c-05} x^\mu \quad \rightarrow \quad x'^\mu = x^\mu
+ \dpara\mu(x), \ee
induced by the vector field $\dbfpara(x) = \{ \dpara\mu(x) \}$, up
to the first order in $\dpara\mu$, the converse diffeomorphism has
the form:
\be x'^\bet \quad \rightarrow \quad x^\bet \approx x'^\bet -
\dpara\bet(x'). \ee
Thus, one has

\be \frac{\pax'^\mu}{\pax^\alp} = \kro\mu\alp + \pa\alp
\dpara\mu(x); \ee
\be \frac{\pax^\bet}{\pax'^\nu} \approx \kro\bet\nu - \partial'_\nu
\dpara\bet(x') \approx \kro\bet\nu - \pa\nu \dpara\bet(x). \ee
Then
\bw
\bse
\ba{l}
\ten'^{\mu_1\mu_2\dots\mu_r}{}_{\nu_1\nu_2\dots\nu_s}(x')\\
= \lp \kro{\mu_1}{\alp_1} + \pa{\alp_1} \dpara{\mu_1} \rp \lp \kro{\mu_2}{\alp_2} + \pa{\alp_2} \dpara{\mu_2} \rp \cdot{}\dots{}\cdot \lp \kro{\mu_r}{\alp_r} + \pa{\alp_r} \dpara{\mu_r} \rp\\
\quad\times \lp \kro{\bet_1}{\nu_1} - \pa{\nu_1} \dpara{\bet_1} \rp \lp \kro{\bet_2}{\nu_2} - \pa{\nu_2} \dpara{\bet_2} \rp \cdot{}\dots{}\cdot \lp \kro{\bet_s}{\nu_s} - \pa{\nu_s} \dpara{\bet_s} \rp \, \ten^{\alp_1\alp_2\dots\alp_r}{}_{\bet_1\bet_2\dots\bet_s}(x)\\
\\
\approx \ten^{\mu_1\mu_2\dots\mu_r}{}_{\nu_1\nu_2\dots\nu_s}(x)\\
\quad + \lp \pa{\alp_1}\dpara{\mu_1} \rp \lp \kro{\mu_2}{\alp_2} \kro{\mu_3}{\alp_3} \cdot{}\dots{}\cdot \kro{\mu_r}{\alp_r} \cdot \kro{\bet_1}{\nu_1} \kro{\bet_2}{\nu_2} \cdot{}\dots{}\cdot \kro{\bet_s}{\nu_s} \rp \ten^{\alp_1\alp_2\dots\alp_r}{}_{\bet_1\bet_2\dots\bet_s}(x)\\
\quad + \lp \pa{\alp_2}\dpara{\mu_2} \rp \lp \kro{\mu_1}{\alp_1} \kro{\mu_3}{\alp_3} \cdot{}\dots{}\cdot \kro{\mu_r}{\alp_r} \cdot \kro{\bet_1}{\nu_1} \kro{\bet_2}{\nu_2} \cdot{}\dots{}\cdot \kro{\bet_s}{\nu_s} \rp \ten^{\alp_1\alp_2\dots\alp_r}{}_{\bet_1\bet_2\dots\bet_s}(x)\\
\quad + \dots\\
\quad + \lp \pa{\alp_r}\dpara{\mu_r} \rp \lp \kro{\mu_1}{\alp_1} \kro{\mu_2}{\alp_2} \cdot{}\dots{}\cdot \kro{\mu_{r-1}}{\alp_{r-1}} \cdot \kro{\bet_1}{\nu_1} \kro{\bet_2}{\nu_2} \cdot{}\dots{}\cdot \kro{\bet_s}{\nu_s} \rp \ten^{\alp_1\alp_2\dots\alp_r}{}_{\bet_1\bet_2\dots\bet_s}(x)\\
 \\
\quad - \lp \pa{\nu_1}\dpara{\bet_1} \rp \lp \kro{\mu_1}{\alp_1} \kro{\mu_2}{\alp_2} \cdot{}\dots{}\cdot \kro{\mu_r}{\alp_r} \cdot \kro{\bet_2}{\nu_2} \kro{\bet_3}{\nu_3} \cdot{}\dots{}\cdot \kro{\bet_s}{\nu_s} \rp \ten^{\alp_1\alp_2\dots\alp_r}{}_{\bet_1\bet_2\dots\bet_s}(x)\\
\quad - \lp \pa{\nu_2}\dpara{\bet_2} \rp \lp \kro{\mu_1}{\alp_1} \kro{\mu_2}{\alp_2} \cdot{}\dots{}\cdot \kro{\mu_r}{\alp_r} \cdot \kro{\bet_1}{\nu_1} \kro{\bet_3}{\nu_3} \cdot{}\dots{}\cdot \kro{\bet_s}{\nu_s} \rp \ten^{\alp_1\alp_2\dots\alp_r}{}_{\bet_1\bet_2\dots\bet_s}(x)\\
\quad - \dots\\
\quad - \lp \pa{\nu_s}\dpara{\bet_s} \rp \lp \kro{\mu_1}{\alp_1} \kro{\mu_2}{\alp_2} \cdot{}\dots{}\cdot \kro{\mu_r}{\alp_r} \cdot \kro{\bet_1}{\nu_1} \kro{\bet_2}{\nu_2} \cdot{}\dots{}\cdot \kro{\bet_{s-1}}{\nu_{s-1}} \rp \ten^{\alp_1\alp_2\dots\alp_r}{}_{\bet_1\bet_2\dots\bet_s}(x)\\
\\
\ea
\ese
\be
\ba{l}
=  \ten^{\mu_1\mu_2\dots\mu_r}{}_{\nu_1\nu_2\dots\nu_s}(x)\\
\quad + \pa\bet\dpara\alp \lb (\kro{\mu_1}\alp \kro\bet{\alp_1}) \kro{\mu_2}{\alp_2} \kro{\mu_3}{\alp_3} \cdot{}\dots{}\cdot \kro{\mu_r}{\alp_r} \cdot \kro{\bet_1}{\nu_1} \kro{\bet_2}{\nu_2} \cdot{}\dots{}\cdot \kro{\bet_s}{\nu_s} \rd\\
\qquad\qquad\;\, + \kro{\mu_1}{\alp_1} (\kro{\mu_2}\alp \kro\bet{\alp_2}) \kro{\mu_3}{\alp_3} \cdot{}\dots{}\cdot \kro{\mu_r}{\alp_r} \cdot \kro{\bet_1}{\nu_1} \kro{\bet_2}{\nu_2} \cdot{}\dots{}\cdot \kro{\bet_s}{\nu_s}\\
\qquad\qquad\;\, + \dots\\
\qquad\qquad\;\, +  \kro{\mu_1}{\alp_1} \kro{\mu_2}{\alp_2} \cdot{}\dots{}\cdot \kro{\mu_{r-1}}{\alp_{r-1}} (\kro{\mu_r}\alp \kro\bet{\alp_r}) \cdot \kro{\bet_1}{\nu_1} \kro{\bet_2}{\nu_2} \cdot{}\dots{}\cdot \kro{\bet_s}{\nu_s}\\
 \\
\qquad\qquad\;\, - \kro{\mu_1}{\alp_1} \kro{\mu_2}{\alp_2} \cdot{}\dots{}\cdot \kro{\mu_r}{\alp_r} \cdot (\kro\bet{\nu_1} \kro{\bet_1}\alp) \kro{\bet_2}{\nu_2} \kro{\bet_3}{\nu_3} \cdot{}\dots{}\cdot \kro{\bet_s}{\nu_s}\\
\qquad\qquad\;\, - \kro{\mu_1}{\alp_1} \kro{\mu_2}{\alp_2} \cdot{}\dots{}\cdot \kro{\mu_r}{\alp_r} \cdot \kro{\bet_1}{\nu_1} (\kro\bet{\nu_2} \kro{\bet_2}\alp) \kro{\bet_3}{\nu_3} \cdot{}\dots{}\cdot \kro{\bet_s}{\nu_s}\\
\qquad\qquad\;\, - \dots\\
\qquad\qquad\;\, - \ld\kro{\mu_1}{\alp_1} \kro{\mu_2}{\alp_2} \cdot{}\dots{}\cdot \kro{\mu_r}{\alp_r} \cdot \kro{\bet_1}{\nu_1} \kro{\bet_2}{\nu_2} \cdot{}\dots{}\cdot \kro{\bet_{s-1}}{\nu_{s-1}} (\kro\bet{\nu_s} \kro{\bet_s}\alp) \rb \ten^{\alp_1\alp_2\dots\alp_r}{}_{\bet_1\bet_2\dots\bet_s}(x).
\ea
\ee
Consequently, the total variation \eqref{app_02_c-04} for the tensor of rank $\lp r\atop s\rp$ has the form
\be\label{app_02_c-06} \bdel
\ten^{\mu_1\mu_2\dots\mu_r}{}_{\nu_1\nu_2\dots\nu_s}(x)
 = \lp
\Del^\bet{}_\alp\rp\ld^{\mu_1\mu_2\dots\mu_r}{}_{\nu_1\nu_2\dots\nu_s}
\rv_{\alp_1\alp_2\dots\alp_r}{}^{\bet_1\bet_2\dots\bet_s}\;
\ten^{\alp_1\alp_2\dots\alp_r}{}_{\bet_1\bet_2\dots\bet_s}(x) \;
\pa\bet\dpara\alp(x), \ee
where
\be\label{app_02_c-08} \boxed{ \ba{l}
\lp \Del^\bet{}_\alp\rp\ld^{\mu_1\mu_2\dots\mu_r}{}_{\nu_1\nu_2\dots\nu_s} \rv_{\alp_1\alp_2\dots\alp_r}{}^{\bet_1\bet_2\dots\bet_s}\\
\quad \Def \lb (\kro{\mu_1}\alp \kro\bet{\alp_1})
\kro{\mu_2}{\alp_2} \kro{\mu_3}{\alp_3} \cdot{}\dots{}\cdot
\kro{\mu_r}{\alp_r} + \kro{\mu_1}{\alp_1} (\kro{\mu_2}\alp
\kro\bet{\alp_2}) \kro{\mu_3}{\alp_3} \cdot{}\dots{}\cdot
\kro{\mu_r}{\alp_r} + \dots + \kro{\mu_1}{\alp_1}
\kro{\mu_2}{\alp_2} \cdot{}\dots{}\cdot \kro{\mu_{r-1}}{\alp_{r-1}}
(\kro{\mu_r}\alp \kro\bet{\alp_r}) \rb  \kro{\bet_1}{\nu_1}
\kro{\bet_2}{\nu_2}
\cdot{}\dots{}\cdot \kro{\bet_s}{\nu_s}\\
\qquad -
\kro{\mu_1}{\alp_1} \kro{\mu_2}{\alp_2} \cdot{}\dots{}\cdot \kro{\mu_r}{\alp_r} \lb (\kro\bet{\nu_1} \kro{\bet_1}\alp) \kro{\bet_2}{\nu_2} \kro{\bet_3}{\nu_3} \cdot{}\dots{}\cdot \kro{\bet_s}{\nu_s} + \kro{\bet_1}{\nu_1} (\kro\bet{\nu_2} \kro{\bet_2}\alp) \kro{\bet_3}{\nu_3} \cdot{}\dots{}\cdot \kro{\bet_s}{\nu_s} + \dots + \kro{\bet_1}{\nu_1} \kro{\bet_2}{\nu_2} \cdot{}\dots{}\cdot \kro{\bet_{s-1}}{\nu_{s-1}} (\kro\bet{\nu_s} \kro{\bet_s}\alp) \rb.\\
\ea }\ee
\ew

The functional variation $\dparabfgfi$ of the field function
$\bfgfi$ induced by the diffeomorphism \eqref{app_02_c-05} is
connected with the total variation  $\bdbfgfi$ by the evident
relation:
\be \bdgfiA(x) = \lp \pa\alp \gfiA(x) \rp \dpara\alp +
\dparagfiA(x). \ee
\bw
Then, using \eqref{app_02_c-06}, one obtains for the tensor of rank
$\lp r\atop s\rp$:
\be \ba{l}
\del_\xi \ten^{\mu_1\mu_2\dots\mu_r}{}_{\nu_1\nu_2\dots\nu_s}(x)\\
\quad = - \pa\alp
\ten^{\mu_1\mu_2\dots\mu_r}{}_{\nu_1\nu_2\dots\nu_s}(x)\; \dpara\alp
+ \lp
\Del^\bet{}_\alp\rp\ld^{\mu_1\mu_2\dots\mu_r}{}_{\nu_1\nu_2\dots\nu_s}
\rv_{\alp_1\alp_2\dots\alp_r}{}^{\bet_1\bet_2\dots\bet_s}\;
\ten^{\alp_1\alp_2\dots\alp_r}{}_{\bet_1\bet_2\dots\bet_s}(x)\;
\pa\bet\dpara\alp(x), \ea \ee
\ew
Returning to the collective indexes
$A=\,^{\mu_1\mu_2\dots\mu_r}{}_{\nu_1\nu_2\dots\nu_s}$ and
$B=\,^{\alp_1\alp_2\dots\alp_r}{}_{\bet_1\bet_2\dots\bet_s}$, the
last formula is rewritten in the compact form:
\be\label{app_02_c-07} \boxed{ \dparagfiA = -\pa\alp \gfiA
\dpara\alp + \DbrgAB\alp\bet \gfiB \pa\bet \dpara\alp. } \ee
The formulae of the type \eqref{app_02_c-07} can be used also both
for the tensor densities of an arbitrary weight and for the spinors
of an arbitrary rank $\lp {k\atop l} | {{\bar{m}}\atop{\bar{n}}}
\rp$. As we know, firstly the symbols $\Dbrg\bet\alp{A}{B}$ where
introduced in the works by Belinfante \cite{Belinfante_1939} and by
Rosenfeld \cite{Rosenfeld_1940, Rosenfeld_1940_en}. Therefore we
call them as the \emph{Belinfante-Rosenfeld symbols}.

In our consideration, the tensors of the ranks $\lp 0\atop 2\rp$ and
$\lp 1\atop 2\rp$ are more important, for them one has to use
\be \Dbrm\bet\alp\mu\nu\veps\kap = - (\kro\veps\alp \kro\bet\mu)
\kro\kap\nu - \kro\veps\mu (\kro\kap\alp \kro\bet\nu); \ee
\be \Dbrt\bet\alp\lam\mu\nu\pi\rho\sig = \lp \kro\lam\alp
\kro\bet\pi \rp \kro\rho\mu \kro\sig\nu - \kro\lam\pi \lb
(\kro\rho\alp \kro\bet\mu) \kro\sig\nu + \kro\rho\mu (\kro\sig\alp
\kro\bet\nu) \rb. \ee
Thus, for the metric tensor $\bfmet = \{\met\mu\nu \}$ and for the
torsion tensor $\bftor = \{\tor\lam\mu\nu \}$ one finds,
respectively,
\be\label{app_02_c-10} \boxed{ \Dbrm\bet\alp\mu\nu\veps\kap
\met\veps\kap = -2\met\alp{(\mu} \kro\bet{\nu)} } \ee
and
\be\label{app_02_c-11} \boxed{ \Dbrt\bet\alp\lam\mu\nu\pi\rho\sig
\tor\pi\rho\sig = \kro\lam\alp \tor\bet\mu\nu + 2\tor\lam\alp{[\mu}
\kro\bet{\nu]}. } \ee

\subsection{The covariant derivative}\label{app_02_c-02}
With the use of the Belinfante-Rosenfeld symbols \eqref{app_02_c-08}
the covariant derivative of the tensor of the rank $\lp r\atop s\rp$
\bw
\be \ba{l}
\na\lam \ten^{\mu_1\mu_2\dots\mu_r}{}_{\nu_1\nu_2\dots\nu_s} = \pa\lam \ten^{\mu_1\mu_2\dots\mu_r}{}_{\nu_1\nu_2\dots\nu_s}\\
\quad + \con{\mu_1}\bet\lam\; \ten^{\bet\mu_2\mu_3\dots\mu_r}{}_{\nu_1\nu_2\dots\nu_s} + \con{\mu_2}\bet\lam\; \ten^{\mu_1\bet\mu_3\dots\mu_r}{}_{\nu_1\nu_2\dots\nu_s} + \dots + \con{\mu_r}\bet\lam\; \ten^{\mu_1\mu_2\dots\mu_{r-1}\bet}{}_{\nu_1\nu_2\dots\nu_s}\\
\quad - \con\alp{\nu_1}\lam\;
\ten^{\mu_1\mu_2\dots\mu_r}{}_{\alp\nu_2\nu_3\dots\nu_s} -
\con\alp{\nu_2}\lam\;
\ten^{\mu_1\mu_2\dots\mu_r}{}_{\nu_1\alp\nu_3\dots\nu_s} - \dots -
\con\alp{\nu_s}\lam\;
\ten^{\mu_1\mu_2\dots\mu_r}{}_{\nu_1\nu_2\dots\nu_{s-1}\alp} \ea \ee
can be presented in the form:
\be \na\lam \ten^{\mu_1\mu_2\dots\mu_r}{}_{\nu_1\nu_2\dots\nu_s} =
\pa\lam \ten^{\mu_1\mu_2\dots\mu_r}{}_{\nu_1\nu_2\dots\nu_s} +
\con\alp\bet\lam  \lp
\Del^\bet{}_\alp\rp\ld^{\mu_1\mu_2\dots\mu_r}{}_{\nu_1\nu_2\dots\nu_s}
\rv_{\gam_1\gam_2\dots\gam_r}{}^{\del_1\del_2\dots\del_s}\;
\ten^{\gam_1\gam_2\dots\gam_r}{}_{\del_1\del_2\dots\del_s}, \ee
\ew
or, returning to the collective indexes $A$ and $B$, as
\be\label{app_02_c-09} \boxed{ \na\lam \gfiA = \pa\lam \gfiA +
\con\alp\bet\lam \DbrgAB\bet\alp \gfiB. } \ee

\subsection{The covariant form of the variation $\dparabfgfi$}\label{app_02_c-03}
In the Paper~I, Sec. I,  for the functional variation $\dparabfgfi$
the generalized formula (15) has been stated, due to which the
formula \eqref{app_02_c-07} has to acquire the form:
\be\label{app_02_c-12} \dparagfiA = \gfiaA\alp \dpara\alp +
\gfibA\alp\bet \na\bet \dpara\alp. \ee
Define the explicit expressions for the coefficients in this
formula. Using \eqref{app_02_c-09}, find the quantity $\{ \pa\lam
\gfiA \}$, and using
\be \na\bet \dpara\alp = \pa\bet \dpara\alp + \con\alp\gam\bet
\dpara\gam \ee
find the quantity $\{ \pa\bet \dpara\alp \}$. Next, substitute the
results into \eqref{app_02_c-07} and obtain
\bse \ba{rl} \dparagfiA & = \lf -\na\alp \gfiA +
2\con\gam{[\bet}{\alp]}
\DbrgAB\bet\gam \gfiB \rf \dpara\alp\\
& \quad + \lf \DbrgAB\bet\alp \gfiB \rf \na\bet \dpara\alp. \ea \ese
Substituting here
$ 2\con\gam{[\bet}{\alp]} = -\tor\gam\bet\alp,$
obtain finally
\be \boxed{ \ba{rl} \dparagfiA & = -\lf \na\alp \gfiA +
\tor\gam\bet\alp
\DbrgAB\bet\gam \gfiB \rf \dpara\alp\\
 & \quad + \lf \DbrgAB\bet\alp \gfiB \rf
\na\bet \dpara\alp. \ea} \ee
Comparing this formula with \eqref{app_02_c-12} one finds
\be\label{sec_02_01-20} \boxed{ \gfiaA\alp = -\lf \na\alp \gfiA +
\tor\gam\bet\alp \DbrgAB\bet\gam \gfiB\rf; } \ee
\be\label{sec_02_01-21} \boxed{ \gfibA\alp\bet = \DbrgAB\bet\alp
\gfiB. } \ee
In particular, for the metric tensor $\bfmet$, using the metric compatible
condition $\na\alp \met\bet\gam = 0$ and the formula \eqref{app_02_c-10}, one
gets

\be\label{sec_02_01-24} \boxed{ \meta\alp\bet\gam = 2
\tord{(\bet}{\gam)}\alp; } \ee
\be \boxed{ \metb\alp\bet\kap\lam = -2 \met\alp{(\kap}
\kro\bet{\lam)}. } \ee
Analogously, using the formula \eqref{app_02_c-11}, one has for the
torsion tensor $\bftor$:
\be \boxed{\ba{l} \tora\alp{\veps}\bet\gam = -\na\alp
\tor\veps\bet\gam\\
\quad - (\tor\veps\kap\alp \tor\kap\bet\gam + \tor\veps\kap\bet
\tor\kap\gam\alp + \tor\veps\kap\gam \tor\kap\alp\bet); \ea} \ee
\be\label{sec_02_01-25} \boxed{ \torb\alp\bet\veps\kap\lam =
\kro\veps\alp \tor\bet\kap\lam + 2\tor\veps\alp{[\kap}
\kro\bet{\lam]}. } \ee

\subsection{Commutator of the covariant derivatives}
With the use of the Belinfante-Rosenfeld symbols \eqref{app_02_c-08}
the commutator of the covariant derivatives of the tensor of the
rank $\lp r\atop
s\rp$
\bw
\be \ba{l}
(\na\rho \na\sig - \na\sig \na\rho )\ten^{\mu_1\mu_2\dots\mu_r}{}_{\nu_1\nu_2\dots\nu_s} = -\tor\lam\rho\sig \na\lam \ten^{\mu_1\mu_2\dots\mu_r}{}_{\nu_1\nu_2\dots\nu_s}\\
\quad + \cur{\mu_1}\lam\rho\sig \ten^{\lam\mu_2\mu_3\dots\mu_r}{}_{\nu_1\nu_2\dots\nu_s} + \cur{\mu_2}\lam\rho\sig \ten^{\mu_1\lam\mu_3\dots\mu_r}{}_{\nu_1\nu_2\dots\nu_s} + \dots + \cur{\mu_r}\lam\rho\sig \ten^{\mu_1\mu_2\dots\mu_{r-1}\lam}{}_{\nu_1\nu_2\dots\nu_s}\\
\quad - \cur\kap{\nu_1}\rho\sig
\ten^{\mu_1\mu_2\dots\mu_r}{}_{\kap\nu_2\nu_3\dots\nu_s} -
\cur\kap{\nu_2}\rho\sig
\ten^{\mu_1\mu_2\dots\mu_r}{}_{\nu_1\kap\nu_3\dots\nu_s} - \dots -
\cur\kap{\nu_s}\rho\sig
\ten^{\mu_1\mu_2\dots\mu_r}{}_{\nu_1\nu_2\dots\nu_{s-1}\kap} \ea \ee
can be presented in the form:
\be \ba{rl}
(\na\rho \na\sig - \na\sig \na\rho )\ten^{\mu_1\mu_2\dots\mu_r}{}_{\nu_1\nu_2\dots\nu_s} & = -\tor\lam\rho\sig \na\lam \ten^{\mu_1\mu_2\dots\mu_r}{}_{\nu_1\nu_2\dots\nu_s}\\
\\
 & \quad + \cur\kap\lam\rho\sig  \lp
\Del^\lam{}_\kap\rp\ld^{\mu_1\mu_2\dots\mu_r}{}_{\nu_1\nu_2\dots\nu_s}
\rv_{\gam_1\gam_2\dots\gam_r}{}^{\del_1\del_2\dots\del_s}
\ten^{\gam_1\gam_2\dots\gam_r}{}_{\del_1\del_2\dots\del_s}, \ea \ee
%
%
Returning to the collective indexes $A$, $B$, \dots, it is presented
as
\be\label{sec_02_02-01} \boxed{ (\na\rho \na\sig - \na\sig \na\rho )
\gfiA = -\tor\lam\rho\sig \na\lam \gfiA + \cur\kap\lam\rho\sig
\DbrgAB\lam\kap \gfiB. } \ee
\ew

%
\section{The transformation of the expression $\lp \frac{-1}{2} \belud\gam\bet\alp \na\nu \tor\alp\bet\gam \rp$}\label{app_02_d-00}

Let $\{ \belu\gam\bet\alp \} \Def \{ \Du\gam\bet\alp\pi\rho\sig
\spiu\pi\rho\sig \}$, where $\{ \spiu\pi{[\rho}{\sig]} =
\spiu\pi\rho\sig \}$ be an \emph{arbitrary} tensor with such a symmetry.
Then $\belu{[\gam}{\bet]}\alp = \belu\gam\bet\alp$. Basing on this,
transform the expression $\lp \frac{-1}{2} \belud\gam\bet\alp \na\nu
\tor\alp\bet\gam \rp$ as follows.
\bn
\item Substituting the Ricci identity in the form
\bw
\bse \ba{rl}
\na\nu \tor\alp\bet\gam & \eq \cur\alp\nu\bet\gam + \cur\alp\bet\gam\nu + \cur\alp\gam\nu\bet - \lp \na\bet \tor\alp\gam\nu + \na\gam \tor\alp\nu\bet + \tor\alp\lam\nu \tor\lam\bet\gam + \tor\alp\lam\bet \tor\lam\gam\nu + \tor\alp\lam\gam \tor\lam\nu\bet \rp\\
& = \cur\alp\nu\bet\gam + 2\cur\alp{[\bet}{\gam]}\nu - 2\na{[\bet}
\tor\alp{\gam]}\nu -  \tor\alp\lam\nu \tor\lam\bet\gam -
2\tor\alp\lam{[\bet} \tor\lam{\gam]}\nu, \ea \ese
\ew
one obtains
\be\label{app_02_d-01} \ba{l}
-\bfrac12 \belud\gam\bet\alp \na\nu \tor\alp\bet\gam \eq -\bfrac12 \belud\gam\bet\alp \cur\alp\nu\bet\gam - \belud\gam\bet\alp \cur\alp\bet\gam\nu\\
\quad + \belud\gam\bet\alp \na\bet \tor\alp\gam\nu + \bfrac12
\belud\gam\bet\alp \tor\alp\lam\nu \tor\lam\bet\gam +
\belud\gam\bet\alp \tor\alp\lam\bet \tor\lam\gam\nu. \ea \ee
\item \label{app_02_d-02} Turn to the first term on the
right  hand side of \eqref{app_02_d-01}. Then, recall the identity
(C2) in the Paper~I, Appendix C.1:
\bse \sna\mu \lb \sna\eta \potb\nu\mu\eta +
\bfrac{1}{2}\potb\nu\rho\sig \tor\mu\rho\sig \rb \eq -\bfrac{1}{2}
\cur\lam\nu\rho\sig \potb\lam\rho\sig, \ese
change here $\potb\nu\mu\eta = \belud\mu\eta\nu$ and obtain for this
term:
\bse -\bfrac12 \belud\gam\bet\alp \cur\alp\nu\bet\gam = -\sna\mu \lb
\sna\eta \belud\mu\eta\nu + \bfrac12 \belud\veps\eta\nu
\tor\mu\veps\eta\rb. \ese
\item The second term on the
right  hand side of \eqref{app_02_d-01} is equal to
\bse \ba{l}
-\belud\gam\bet\alp \cur\alp\bet\gam\nu = -\belu\gam\bet\alp \curd\alp\bet\gam\nu = -\Du\gam\bet\alp\pi\rho\sig \spiu\pi\rho\sig \curd\alp\bet\gam\nu\\
\quad = -\bfrac12 \lp \spiu\bet\gam\alp + \spiu\alp\gam\bet - \spiu\gam\bet\alp \rp \curd\alp\bet\gam\nu\\
\quad = \lp \spiu{(\alp}{\bet)}\gam - \frac12 \spiu\gam\alp\bet \rp
\curd\alp\bet\gam\nu = -\bfrac12 \spiu\pi\rho\sig
\curd\rho\sig\pi\nu; \ea \ese
\item Using the differentiation by part in the third term on the
right  hand side of \eqref{app_02_d-01}, one finds
\bse \belud\gam\bet\alp \na\bet \tor\alp\gam\nu = -\sna\mu \lp
\belud\mu\bet\alp \tor\alp\bet\nu \rp - \lp \sna\eta
\belud\mu\eta\lam \rp \tor\lam\mu\nu; \ese

\item\label{app_02_d-03} At last, one rewrites fourth and fifth terms on the
right  hand side of \eqref{app_02_d-01}, respectively,
as
\bse \frac12 \belud\gam\bet\alp \tor\alp\lam\nu \tor\lam\bet\gam =
-\frac12 \lp \belud\veps\eta\lam \tor\mu\veps\eta \rp
\tor\lam\mu\nu. \ese
and
\bse \belud\gam\bet\alp \tor\alp\lam\bet \tor\lam\gam\nu = - \lp
\belud\mu\bet\alp \tor\alp\bet\lam \rp \tor\lam\mu\nu; \ese
\en
Combining the results of the points \ref{app_02_d-02} --
\ref{app_02_d-03} in the formula \eqref{app_02_d-01}, one obtains
the search identity:
\bw
\be\label{sec_02_03-07} \boxed{ \ba{rl}
-\bfrac12 \belud\gam\bet\alp \na\nu \tor\alp\bet\gam & \eq - \sna\mu \lb \sna\eta \belud\mu\eta\nu + \bfrac12 \belud\veps\eta\nu \tor\mu\veps\eta + \belud\mu\bet\alp \tor\alp\bet\nu \rb \\
 & \quad - \lb \sna\eta \belud\mu\eta\lam + \bfrac12 \belud\veps\eta\lam \tor\mu\veps\eta + \belud\mu\bet\alp \tor\alp\bet\lam \rb \tor\lam\mu\nu - \bfrac12 \spi\pi\rho\sig \curud\rho\sig\pi\nu.
\ea } \ee
\ew
%

%
\bibliography{Lompay_Petrov_-_Part_2_[Bibliography]}

\end{document}